\documentclass[11pt]{article}
\usepackage{tikz}
\usetikzlibrary{decorations.pathreplacing,calligraphy}
\usetikzlibrary{calc}
\usetikzlibrary{shapes.geometric}
\usetikzlibrary{arrows.meta}
\usepackage[utf8]{inputenc}
\usepackage[T1]{fontenc}
\usepackage{stmaryrd}
\usepackage{textcomp,amssymb,amsmath,amsthm}
\usepackage{enumitem}
\usepackage[retainorgcmds]{IEEEtrantools}
\usepackage{optidef}
\usepackage[margin=2cm]{geometry}
\usepackage{hyperref}
\usepackage{xcolor}
\hypersetup{
    colorlinks,
    linkcolor={red!70!black},
    citecolor={blue!50!black},
    urlcolor={blue!80!black}
  }
\usepackage[capitalize,nameinlink]{cleveref}
\usepackage[final]{showlabels} 
\usepackage{bbm}

\crefname{equation}{}{}

\usepackage[backend=biber,sorting=nyt,maxnames=10,style=numeric]{biblatex}
\tikzset{sttriangle/.style={
    draw=black,
    fill=white,
    regular polygon,
    regular polygon sides=3,
    minimum size = 0pt,
    inner sep=1pt
}}
\tikzset{stcircle/.style={
    fill=white,
    draw=black,
    circle,inner sep=1pt,
    minimum size = 1.5em
}}
\tikzset{rednode/.style={
    fill=white,
    draw=black,
    circle,
    inner sep=1pt,
    minimum size = 1.5em
}}
\tikzset{blacknode/.style={
    text=white,
    fill=black,
    draw=black,
    circle,
    inner sep=1pt,
    minimum size = 0.5em
}}
\tikzset{splitter/.style={
    fill=black,
    circle,
    inner sep=0pt,
    minimum size = 4pt
}}
\tikzset{term/.style={
    draw=black,
    fill=black,
    rectangle,
    inner sep=0pt,
    minimum size=4pt
}}

\tikzset{chooser/.style={
    draw=black,
    fill=white,
    circle,
    inner sep=0pt,
    minimum size = 4pt
}}
\tikzset{saturated/.style={very thick}}
\tikzset{fluid/.style={}}
\tikzset{OutPrio/.tip = {||.}}

\newcommand{\bull}{\raisebox{0.05em}{\tikz{\draw (0,0) -- (0.3em,0.15em) -- (0,0.3em) -- cycle;}}}

\theoremstyle{plain}
\newtheorem{theorem}{Theorem}[section]
\newtheorem{proposition}[theorem]{Proposition}
\newtheorem{corollary}[theorem]{Corollary}
\newtheorem{lemma}[theorem]{Lemma}
\newtheorem{claim}[theorem]{Claim}
\theoremstyle{definition}
\newtheorem{definition}[theorem]{Definition}

\newcommand*{\disunion}{\uplus}
\newcommand*{\eqdef}{:=}
\newcommand*{\interval}[2]{\left\llbracket #1, #2 \right\rrbracket}

\newcommand*{\splitnetwork}{G = (I \disunion S \disunion O ,E)}

\newcommand*{\update}[1]{\overline{#1}}

\newcommand*{\optt}{{t^\star}}

\newcommand*{\allone}{\mathbbm{1}}

\newcommand*{\Cequal}{\mathcal{C}^{=}}
\newcommand*{\stationary}{\mathrm{sd}}
\newcommand*{\binary}[2][]{\mathrm{binary}_{#1}\left({#2}\right)}
\newcommand*{\symmetry}[1]{\overleftarrow{#1}}

\newcommand*{\up}{\uparrow}
\newcommand*{\down}{\downarrow}

\newcommand\restr[2]{{
  \left.\kern-\nulldelimiterspace 
  #1 
  \right|_{#2} 
  }}

\title{The steady-states of splitter networks} 
\author{
  Basile Couëtoux\footnote{Aix-Marseille Université, CNRS, LIS, Marseille, France,
    \href{mailto:basile.couetoux@univ-amu.fr}{basile.couetoux@univ-amu.fr}}, 
  Bastien Gastaldi\footnote{Télécom SudParis, Institut Polytechnique de Paris, Evry, France,
    \href{mailto:bastien.gastaldi@telecom-sudparis.eu}{bastien.gastaldi@telecom-sudparis.eu}}, 
  Guyslain Naves\footnote{Aix-Marseille Université, CNRS, LIS, Marseille, France,
    \href{mailto:guyslain.naves@univ-amu.fr}{guyslain.naves@univ-amu.fr}, corrresponding author.}
}

\begin{document}

\begin{center}
{\Large\bfseries The steady-states of splitter networks}\medskip

Basile Couëtoux\footnote{Aix-Marseille Université, CNRS, LIS, Marseille, France,
  \href{mailto:basile.couetoux@univ-amu.fr}{basile.couetoux@univ-amu.fr}}, 
Bastien Gastaldi\footnote{Télécom SudParis, Institut Polytechnique de Paris, Evry, France,
  \href{mailto:bastien.gastaldi@telecom-sudparis.fr}{bastien.gastaldi@telecom-sudparis.eu}}, 
Guyslain Naves\footnote{Aix-Marseille Université, CNRS, LIS, Marseille, France,
  \href{mailto:guyslain.naves@univ-amu.fr}{guyslain.naves@univ-amu.fr}, corrresponding author.}

\end{center}\medskip

\noindent{\bfseries Abstract:} We introduce splitter networks, which
abstract the behavior of conveyor belts found in the video game
Factorio. Based on this definition, we show how to compute the
steady-state of a splitter network. Then, leveraging insights from the
players community, we provide multiple designs of splitter networks
capable of load-balancing among several conveyor belts, and prove that
any load-balancing network on $n$ belts must have $\Omega(n \log n)$
nodes. Incidentally, we establish connections between splitter
networks and various concepts including flow algorithms, flows with
equality constraints, Markov chains and the Knuth-Yao theorem about
sampling over rational distributions using a fair coin.

\section{Introduction}

The transportation of materials or data within various networks
represents an inexhaustible source of mathematical problems, which has
lead to almost as many solutions, theories and algorithms. These
advancements have brought about significant improvements across
diverse fields including supply chain management, logistics, network
optimization. Transportation also serves as a central component in
numerous games, as evidenced by the transportation category on
BoardGameGeek which lists almost two thousand games~\cite{bgg}. In
Factorio~\cite{factorio}, a video game published in 2020 by Wube
Software, players must mine natural resources to feed a
rocket-building factory on an hostile planet. A major part of the
gameplay involves the movement of resources within the factory,
employing various mechanism: robotic arms, conveyor belts, drones or
trains.

In this work, we study the conveyor belts of Factorio. An item placed
on a belt will move at a constant speed toward the end of the belt,
until it reaches that end, or is blocked by an item preceding it.
Belts in Factorio can be combined using a splitter, connecting one or
two incoming belts to one or two outgoing belts. A splitter takes
items from the incoming belts and places them on the outgoing belts,
trying to split the flow as fairly as possible between the incident
belts, while maximizing the throughput. Given the scale of a typical
Factorio game, players frequently encounter the need to balance the
loads across multiple belts, and the community has devised numerous
efficient networks to address this load-balancing problem.

An intriguing aspect of Factorio is its encouragement for players to
construct vast systems of automation, requiring intensive planning and
optimization. Ultimately, the limiting factor arises from the CPU load
generated by game state updates. Consequently, players are
incentivized to prioritize resource efficiency, particularly concerning
gameplay elements that entail frequent computations such as splitters.
This motivates the minimization of the number of splitters in
load-balancing networks.

Our goal is two-fold: first we model the steady-state of a network of
splitters. The network of conveyor belts is abstracted as a directed
graph, with nodes corresponding to splitters and arcs to belts. A
steady-state is a throughput function on the arcs; a circulation with
additional constraints to capture the fact that splitters are fair and
locally optimizing. We present two polynomial-time algorithms for
computing a steady-state in a splitter network. An analogy is made
with two classical maximum-flow algorithms: the blocking-flow
algorithm~\cite{dinits1973method} and the push-relabel
algorithm~\cite{goldberg1988new}. In contrast to maximum flows, the
primary challenge arises when a belt reaches full capacity, as its
supplying splitter may no longer stay both fair and maximizing. In
that case, the splitter is allowed to become unfair, but that decision
changes the constraints applied to the flow, making the problem
fundamentally non-convex. In a second part, we showcase various
load-balancing network designs sourced from the Internet, formalizing
concepts defined by the players community. Furthermore, we prove that
those designs approach optimality. Specifically, we prove that any
balancing network on $n$ belts must have $\Omega(n \log n)$ splitters,
by exhibiting a relation with the problem of sampling the uniform
distribution over a set of $n$ elements using only a fair coin. The
core design is the Beneš network, a circuit-switching network
well-known in the field of
telecommunication~\cite{benevs1962rearrangeable,benevs1964permutation}.

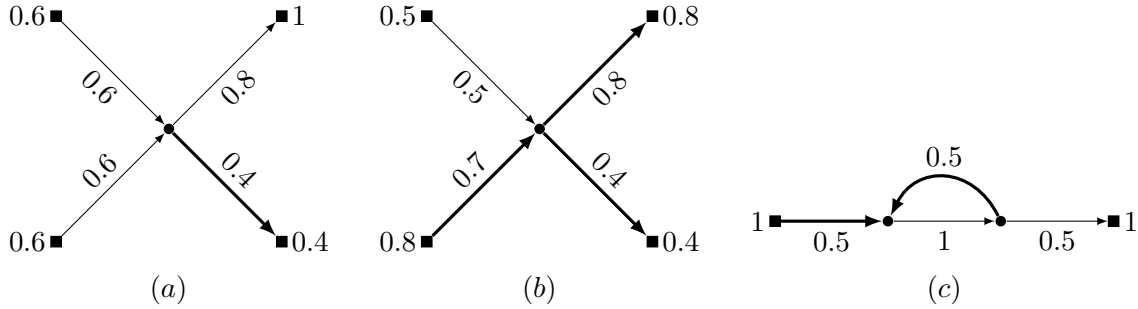
\begin{figure}
  \begin{tabular}{ccc}
    \begin{tikzpicture}[x=1.5cm,y=1.5cm,>=latex]
      \node[splitter] (s) at (0,0) {};
      \foreach \n/\y/\c in {1/-1/0.6,2/1/0.6} {
        \node[term] (i\n) at (-1,\y) {};
        \draw (-1,\y) node[anchor=east] {$\c$};
      }
      \foreach \n/\y/\c in {1/-1/0.4,2/1/1} {
        \node[term] (o\n) at (1,\y) {};
        \draw (1,\y) node[anchor=west] {$\c$};
      }
      \draw[fluid,->] (i1) -- (s) node[midway,sloped,above] {$0.6$};
      \draw[fluid,->] (i2) -- (s) node[midway,sloped,below] {$0.6$};
      \draw[saturated,->] (s) -- (o1) node[midway,sloped,above] {$0.4$};
      \draw[fluid,->] (s) -- (o2) node[midway,sloped,below] {$0.8$};
    \end{tikzpicture}
    &
    \begin{tikzpicture}[x=1.5cm,y=1.5cm,>=latex]
      \node[splitter] (s) at (0,0) {};
      \foreach \n/\y/\c in {1/-1/0.8,2/1/0.5} {
        \node[term] (i\n) at (-1,\y) {};
        \draw (-1,\y) node[anchor=east] {$\c$};
      }
      \foreach \n/\y/\c in {1/-1/0.4,2/1/0.8} {
        \node[term] (o\n) at (1,\y) {};
        \draw (1,\y) node[anchor=west] {$\c$};
      }
      \draw[saturated,->] (i1) -- (s) node[midway,sloped,above] {$0.7$};
      \draw[fluid,->] (i2) -- (s) node[midway,sloped,below] {$0.5$};
      \draw[saturated,->] (s) -- (o1) node[midway,sloped,above] {$0.4$};
      \draw[saturated,->] (s) -- (o2) node[midway,sloped,below] {$0.8$};
    \end{tikzpicture} 
    &
    \begin{tikzpicture}[x=1.5cm,y=1.5cm,>=latex]
      \node[splitter] (s1) at (0,0) {};
      \node[splitter] (s2) at (1,0) {};
      \node[term] (i) at (-1,0) {};
      \draw (-1,0) node[anchor=east] {$1$};
      \node[term] (o) at (2,0) {};
      \draw (2,0) node[anchor=west] {$1$};
  
      \draw[saturated,->] (i) -- (s1) node[midway,sloped,below] {$0.5$};
      \draw[fluid,->] (s1) -- (s2) node[midway,sloped,below] {$1$};
      \draw[fluid,->] (s2) -- (o) node[midway,sloped,below] {$0.5$};
      \draw[saturated,->] (s2) .. controls (0.75,0.5) and (0.25,0.5)
        ..  (s1) node[midway,sloped,above] {$0.5$};
    \end{tikzpicture} 
    \\
    $(a)$ & $(b)$ & $(c)$
  \end{tabular}
  \caption{3 splitter networks given with capacities and associated
    steady-states. Splitter will be represented by circle vertices,
    terminals by square vertices. Each terminal is tagged by its
    capacity, and each arc by its throughput. Saturated arcs are
    bolder than fluid arcs.}
  \label{fig:fair-example}
\end{figure}

The blocking-flow-like algorithm relies on finding circulations with
equality constraints. A \emph{circulation} on a directed graph is a
flow without any excess at any vertex. Given a directed graph $(G,A)$,
we denote $\delta^+(v)$ and $\delta^-(v)$ the sets of outgoing and
incoming arcs incident to a vertex $v$. Let $\Cequal$ be a partition
of $A$ such that for each part $C \in \Cequal$, there is some vertex
$v$ with $C \subseteq \delta^+(v)$. The $\Cequal$-circulation problem
is to decide whether there is a non-zero circulation $f$ that is
constant within each part. While this problem can easily be solved
using linear programming, we require a good characterization of graphs
admitting a $\Cequal$-circulation, Additionaly a polynomial-time
algorithm is needed to either construct a $\Cequal$-circulation or
identify an obstacle that prevents its existence. The algorithm relies
on the computation of a stationary distribution of an auxiliary graph.
In contrast, solving maximum integral flow problems with additional
equality constraints is known to be NP-hard~\cite{meyers2009}, even
when the partition is exactly the sets of leaving arcs of each
vertex~\cite{srinathan2002theory,parmar2007integer}.

Sorting networks~\cite{knuth1998art} and Beneš networks have
topologies similar to splitter networks, with nodes of in-degree and
out-degree 2. In microfluidics, mixing graphs are used to produce
droplets of specific concentration, using devices that produces two
identical droplets from two droplets of any
concentration~\cite{COVIELLOGONZALEZ202098}. The concentration values
on the arcs are subject to equality constraints similar to those of
splitter networks, but without a maximizing constraint. The topology
of splitter networks is nonetheless more general than these examples,
as splitter networks may have directed cycles, those being necessary
in particular to achieve load-balancing with an arbitrary number of
outputs.

In an answer to a question on the mathematics section of
\texttt{stackexchange}, David Ketcheson attempted to model and compute
the throughputs of splitter networks~\cite{ketcheson}. Rather than
binary categorizing each belt as full or not, each arc is assigned a
density and a velocity. The density will be monotonically increasing,
and the velocity monotonically decreasing during the run of the
algorithm, until a steady-state is reached. In fact the velocity
increases only after the density reaches its maximum at one. Therefore
this description is equivalent to our solution, which involves a
throughput function and a set of full belts. Unfortunately his
algorithm does not always terminate, and its solutions do not satisfy
that splitters use their incoming belts fairly. Ketcheson also gave a
procedure, albeit non-polynomial, to determine whether a network (not
necessarily load-balancing) may limit throughput. Hovewer, this
procedure is applicable only to networks without directed cycle.
In~\cite{leue2021verification}, Leue modeled splitter networks using
Petri nets, and uses model checking to check the load-balancing
properties of some small networks.

The Factorio community is very active and creative. Players have
designed load-balancing networks of various sizes, with efficient
embeddings into the grid while respecting the constraints of the game.
Additionally, they have developed general methods for constructing
arbitrary large load-balancing networks. They introduced the concept
of balancing networks, along with the more robust properties of being
throughput unlimited or universal, and subsequently designed networks
that exhibit these characterics. A notable example is the universal
balancer presented by \emph{pocarski}~\cite{pocarski}, although it
uses non-fair splitters too; our universal balancer only uses fair
splitters. They also discovered the relationship with Beneš networks.
Factorio-SAT~\cite{factoriosat} is a project that uses a SAT-solver to
find optimal embeddings of splitter networks in the grid. The project
\emph{VeriFactory} uses a SAT-solver to check various load-balancing
properties of splitter networks~\cite{verifactory}. Factorio belts are
actually sufficiently complex to be
Turing-complete~\cite{beltturingcomplete}. There are many
implementations of various devices inside Factorio, ranging from
raytracers to programming language interpreters, using the diverse set
of available gameplay mechanisms. Factorio has been the inspiration
for several other academic
works~\cite{reid2021factory,patterson2023towards,boardman2021simulation,covello2023using,duhan2019factory}.

The rest of this section presents an overview of the main concepts and
results of this work. Then \Cref{sec:cequal-problem} is dedicated to
the $\Cequal$-circulation problem, which is a requirement for the next
\Cref{sec:algo}, that focuses on the two algorithms to compute a
steady-state. \Cref{sec:balancers} contains the constructions and
proofs of the load-balancing designs. \Cref{sec:lowerbounds}
formalizes the proof for the lower bound on the number of splitters in
a load-balancing network. In \Cref{sec:capacity}, we define networks
that simulate arcs of arbitrary rational capacities. They will be used
in \Cref{sec:uniformSSS}, where we will prove that splitter networks
have unique throughputs. In \Cref{sec:priority}, we investigate
splitter networks when each splitter may receive priorities,
prioritizing one incident arc over the others. Using priorities, in
\Cref{sec:saturating-balancer} we will define a balancer with a number
of fair splitters closer to our lower bound. Finally in
\Cref{sec:perspectives} we will present some perspectives.

\subsection{Splitter networks and their steady-states}

We start by modeling networks of conveyor belts and splitters by
directed graphs, where each single belt is an arc, and each splitter is
a node (thus abstracting the length of the belts).

\begin{definition}
  A \emph{splitter network} is a directed graph $G$ (with possible
  loops or parallel arcs) whose vertex set can be partitioned into
  three sets $V(G) = I \disunion S \disunion O$ where
  \begin{enumerate}[label=(\roman*),nosep]
  \item $I$ is the set of \emph{inputs}, and $d^+(i) = 1$,
    $d^-(i) = 0$ for any input $i$;
  \item $O$ is the set of \emph{outputs}, and $d^-(o) = 1$, $d^+(o) = 0$ for
    any output $o$;
  \item $S$ is the set of \emph{splitters}, and $d^-(s) = d^+(s) = 2$
    for any splitter $s$.
  \end{enumerate}
\end{definition}

We will use the word \emph{flow} to informally describe the material
transported by the network, and \emph{throughput} for the amount of
flow going through the arcs. Our work aims to understand the
throughputs inside a splitter network at steady state, when some
maximum throughputs are forced on its inputs and its outputs, which
are respectively the sources and sinks of the flow passing through the
network. To this end we will consider capacity functions on the input
and output. A capacity $c$ on an input means that the input has an
incoming flow of throughput $c$. The input will try to push that much
into the network, but no more. A capacity $c$ on an output means that
the output will accept a maximum throughput of $c$. We consider that
the maximum throughput of any arc is $1$, with all belts being
identical.

A splitter can be described using two operational rules. The first
rule, which takes precedence, is to maximize the amount of flow that
goes through it. The secondary rule is to be fair. A splitter is fair
relatively to its outgoing arcs: it tries to push as much flow onto
each of them. It is also fair relatively to its incoming arcs: it
tries to pull as much flow from each of them. As the maximization rule
takes precedence, it will not be fair when being unfair leads to
higher throughput. For instance, consider the network in
\Cref{fig:fair-example} $(a)$, depicting a network with a single
splitter. As one of the output has a lower capacity, it pushes more
flow toward the other output, thereby maximizing the total throughput,
while still being as fair as possible as it minimizes the difference
of throughputs on its outgoing arcs.

The throughput of an arc may reach a limit when its head is an output
with a low capacity. For example in \Cref{fig:fair-example} $(a)$, an
output of capacity $0.4$ acts as a bottleneck. In other cases the head
of an arc is a splitter, which itself is limited by what its outgoing
arcs can accept. For example in \Cref{fig:fair-example} $(b)$, as all
the outputs have reached their capacities, the splitter cannot accept
more flow, even if the bottom input could provide even more flow. In
terms of conveyor belts, some belts will initially receive more items
that they can deliver, causing them to fill up. Once full, they can
only accept from upstream as much as they deliver downstream, which
may in turn limit throughputs upstream. We say that such belts are
\emph{saturated}.

The output capacities are not the only factor that limit the total
throughput and create bottlenecks. This can be observed in
\Cref{fig:fair-example} $(c)$. There, the rightmost splitter tries to
be fair and send some of the flow back to the left. The leftmost
splitter also tries to be fair, thus accept the flow coming from the
right. This results in the stabilization into the given throughputs.
This example illustrates that the throughput is not globally
maximized, contrary to the expectation of a total throughput of $1$
for this network. Instead, it is only $0.5$.

The following definition formalizes the notions of capacity,
throughput and saturations, as well as the behaviour of splitters
related to the flow going through the network in a steady state.

\begin{definition}
  Let $G = (I \disunion S \disunion O, E)$ be a splitter network, and
  let $c : I \cup O \to [0,1]$ be the maximal \emph{capacities} of each input
  and output node. A \emph{steady-state} for $(G,c)$ is a pair $(t,F)$
  where
  \begin{enumerate}[label=R\arabic*,nosep]
  \item \label{eqn:throughput} $t : E \to [0,1]$ is the \emph{throughput} function;
  \item \label{eqn:fluid-arcs} $F \subseteq E$ is the set of
    \emph{fluid arcs}, $E \setminus F$ is the set of \emph{saturated}
    arcs;
  \item \label{eqn:input-capacity} for each $i \in I$ with $\delta^+(i) = \{e\}$,
    $t(e) \leq c(i)$ and moreover if $e \in F$ then $t(e) = c(i)$;
  \item \label{eqn:output-capacity} for each $o \in O$ with $\delta^-(o) = \{e\}$,
    $t(e) \leq c(o)$ and moreover if $e \notin F$ then $t(e) = c(o)$;
  \item \label{eqn:conservation} for each $s \in S$, with
    $\delta^-(s) = \{ e_1, e_2 \}$ and $\delta^+(s) = \{e_3, e_4\}$,
    $t(e_1) + t(e_2) = t(e_3) + t(e_4)$;
  \item \label{eqn:incoming-rule} for any $e_1,e_2 \in E$ with
    $\{e_1, e_2\} = \delta^-(s)$ and $e_1 \notin F$,
    $ t(e_1) \geq t(e_2)$;
  \item \label{eqn:outgoing-rule} for any $e_1, e_2 \in E$ with
    $\{e_1, e_2\} = \delta^+(s)$ and $e_1 \in F$, $t(e_1) \geq t(e_2)$;
  \item \label{eqn:maximization} for any $uv \in E \setminus F$ and
    $vw \in F$, $t(uv) = 1$ or $t(vw) = 1$.
  \end{enumerate}
\end{definition}

Rules~\ref{eqn:input-capacity} and~\ref{eqn:output-capacity} say that
the throughputs are limited at each input and each ouput, and
moreover, an input pushes as much flow as allowed by its capacity on a
fluid arc. Similarly an output absorbs as much flow as allowed by its
capacity from a saturated arc. Rule~\ref{eqn:conservation} imposes the
conservation of flow. Rules~\ref{eqn:incoming-rule}
and~\ref{eqn:outgoing-rule} enforce the fairness constraints: a
splitter consumes no less flow from a saturated arc than from another
incoming arc. A saturated arc represents a belt that is full.
Therefore, the splitter is not limited in how much flow it can pull
from that arc, and thus cannot pull less than from the other incoming
arc. Similarly it produces no less flow in a fluid outgoing arc than
in another outgoing arc. In particular, if both incoming arcs are
saturated, or if both outgoing arcs are fluid, they must have equal
throughput, suggesting the following definition.

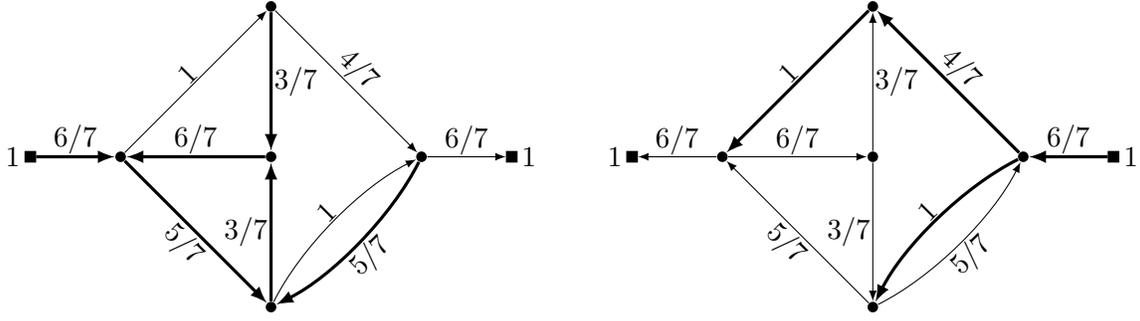
\begin{figure}
  \centering
  \begin{tikzpicture}[x=2cm,y=2cm,>=latex]
    \foreach \i/\x/\y in {1/1/0,2/2/1,3/2/0,4/2/-1,5/3/0} {
      \node[splitter] (s\i) at (\x,\y) {};
    }
    \node[term] (i) at (0.4,0) {};
    \draw (0.4,0) node[anchor=east] {$1$};
    \node[term] (o) at (3.6,0) {};
    \draw (3.6,0) node[anchor=west] {$1$};
  
    \draw[saturated,->] (i) -- (s1) node[midway,sloped,above=-3pt] {$6/7$};
    \draw[fluid,->] (s1) -- (s2) node[midway,sloped,above=-3pt] {$1$};
    \draw[saturated,->] (s1) -- (s4) node[midway,sloped,below=-3pt] {$5/7$};
    \draw[saturated,->] (s2) -- (s3) node[midway,right=-3pt] {$3/7$};
    \draw[fluid,->] (s2) -- (s5) node[midway,sloped,above=-3pt] {$4/7$};
    \draw[saturated,->] (s3) -- (s1) node[midway,sloped,above=-3pt] {$6/7$};
    \draw[saturated,->] (s4) -- (s3) node[midway,left=-3pt] {$3/7$};
    \draw[fluid,->] (s5) -- (o) node[midway,sloped,above=-3pt] {$6/7$};
    \draw[fluid,->] (s4) .. controls (2.2,-0.6) and (2.6,-0.2) ..  (s5) node[midway,sloped,above=-3pt] {$1$};
    \draw[saturated,->] (s5) .. controls (2.8,-0.4) and (2.4,-0.8) .. (s4) node[midway,sloped,below=-3pt] {$5/7$};

    \begin{scope}[xshift=8cm]
      \foreach \i/\x/\y in {1/1/0,2/2/1,3/2/0,4/2/-1,5/3/0} {
        \node[splitter] (s\i) at (\x,\y) {};
      }
      \node[term] (i) at (0.4,0) {};
      \draw (0.4,0) node[anchor=east] {$1$};
      \node[term] (o) at (3.6,0) {};
      \draw (3.6,0) node[anchor=west] {$1$};
      
      \draw[fluid,<-] (i) -- (s1) node[midway,sloped,above=-3pt] {$6/7$};
      \draw[saturated,<-] (s1) -- (s2) node[midway,sloped,above=-3pt] {$1$};
      \draw[fluid,<-] (s1) -- (s4) node[midway,sloped,below=-3pt] {$5/7$};
      \draw[fluid,<-] (s2) -- (s3) node[midway,right=-3pt] {$3/7$};
      \draw[saturated,<-] (s2) -- (s5) node[midway,sloped,above=-3pt] {$4/7$};
      \draw[fluid,<-] (s3) -- (s1) node[midway,sloped,above=-3pt] {$6/7$};
      \draw[fluid,<-] (s4) -- (s3) node[midway,left=-3pt] {$3/7$};
      \draw[saturated,<-] (s5) -- (o) node[midway,sloped,above=-3pt] {$6/7$};
      \draw[saturated,<-] (s4) .. controls (2.2,-0.6) and (2.6,-0.2) ..  (s5) node[midway,sloped,above=-3pt] {$1$};
    \draw[fluid,<-] (s5) .. controls (2.8,-0.4) and (2.4,-0.8) .. (s4) node[midway,sloped,below=-3pt] {$5/7$};

    \end{scope}
  \end{tikzpicture}
  \caption{An example of steady-state in a moderately small network,
    and the reverse network with its steady-state obtained by reversal.
    Notice that the reversed steady-state satisfies
    rule~\ref{eqn:maximization} but not
    rule~\ref{eqn:strong-maximization}.}
  \label{fig:67-example}
\end{figure}

\begin{definition}
  Given a splitter network $\splitnetwork$, and a set
  $F \subseteq E$ of fluid arcs, we say that two arcs $e, e' \in E$
  are 
  \begin{itemize}[label=\bull,nosep]
  \item \emph{in-coupled} if $e, e' \notin F$ and there is a splitter
    vertex $v \in S$ with $\delta^-(v) = \{e, e'\}$,
  \item \emph{out-coupled} if $e, e' \in F$ and there is a splitter
    vertex $v \in S$ with $\delta^+(v) = \{e, e'\}$,
  \item \emph{coupled} if they are in-coupled or out-coupled.
  \end{itemize}
\end{definition}

Finally rule~\ref{eqn:maximization} imposes the maximization of the
throughput by each splitter. Indeed, a saturated arc can provide more
flow, while a fluid arc can absorb more flow. Thus, a steady-state
cannot contain a saturated arc followed by a fluid arc. The only
exception is when one of them already has a throughput of 1.

We will prove in \Cref{sec:symmetry} that the definitions of splitter
networks and steady-states exhibit a remarkable symmetry. By reversing
each arc, exchanging the role of inputs and outputs, and complementing
the set of fluid arcs, a steady-state is transformed into a
steady-state of the reverse graph, as seen in \Cref{fig:67-example}.

For convenience, when defining or representing splitter networks, we
will allow splitters with in-degree one or out-degree one (see
\Cref{fig:67-example} for instance). This is justified by the fact
that if a splitter $s$ has in-degree one, we can add a dummy input
node $i$ with capacity $c(i) = 0$. An arc from $i$ to $s$ can then be
added, that will always remain fluid. Similarly if $s$ has out-degree
one, we can add a dummy output node $o$ with capacity $c(o) = 0$, and
an always-saturated arc from $s$ to $o$. The throughputs on those arcs
are forced to be $0$. Therefore it does not induce any new constraint
on the non-dummy arcs as rules~\ref{eqn:incoming-rule}
and~\ref{eqn:outgoing-rule} are clearly true for those arcs.

Additionally, for convenience, for any input $i \in I$ with outgoing
arc $e$, we note $t(i) \eqdef t(e)$, and similarly for any output
$o \in O$ with incoming arc $e$, $t(o) \eqdef t(e)$. We also extend
the capacities to arcs by setting $c(e)$ to be either $c(i)$ if
$e \in \delta^+(i), i \in I$, or $c(o)$ is
$e \in \delta^-(o), o \in O$, or $1$ otherwise.

\subsection{Existence and computation of steady-states}

Let $F$ be a fixed set of fluid arcs. Then the set of possible
throughput functions $t$ of a steady-state $(t,F)$ can be described as
a polyhedron. Indeed, each of the rules~\ref{eqn:throughput},
\ref{eqn:input-capacity}, \ref{eqn:output-capacity},
\ref{eqn:conservation}, \ref{eqn:incoming-rule},
\ref{eqn:outgoing-rule} can be encoded by linear inequations.
Rule~\ref{eqn:maximization} is non-convex, but we will later introduce
its slight strengthening, rule~\ref{eqn:strong-maximization}. That
stronger rule admits an encoding as a family of linear inequations.
Thanks to linear programming, finding a steady-state thus reduces to
finding a set of fluid arcs that admits a steady-state. Nevertheless,
we still need to find $F$. We propose two algorithms to compute a
steady-state, which relates to two families of maximum flow algorithm:
\begin{itemize}[label=\bull,nosep]
\item a push-relabel-like algorithm, where we relax the conservation
  rule~\ref{eqn:conservation}, thus defining a \emph{pre-steady-state}
  by analogy with \emph{pre-flows}. Given a set $F$, we use a linear
  program to compute an optimal pre-steady-state $(t,F)$ (for some
  well-chosen objective), and prove that either $(t,F)$ is a
  steady-state, or there is an arc $e \in F$ such that
  $(t, F \setminus e)$ is also a (non-optimal) pre-steady-state. Then
  after at most $|E|$ steps we get a steady-state;
\item a blocking-flow-like algorithm, where we relax the
  rule~\ref{eqn:input-capacity} on input capacities, removing the
  requirement that an input whose throughput is less than its capacity
  must have a saturated outgoing arc. This defines the notion of
  \emph{sub-steady-state}. Given a set $F$, we solve a linear system
  to find a sub-steady-state $t$, and prove once again that either $(t,F)$
  is a steady-state or there is an arc $e \in F$ such that
  $(t,F \setminus e)$ is a sub-steady-state.
\end{itemize}
The pre-steady-state algorithm is technically simpler but requires an
LP-solver. The sub-steady-state only requires an algorithm to compute
stationary distributions in directed graphs. We defer a complete
presentation and proof of these algorithms to
\Cref{sec:push-relabel,sec:blocking-flow-algo}, and focus for now on
explaining the sub-steady-state algorithm.
\begin{definition}
  Given $\splitnetwork$ a splitter network with capacities
  $c: I \disunion O \to [0,1]$, a \emph{sub-steady-state} for $(G,c)$
  is a pair $(t,F)$ satisfying~\ref{eqn:throughput},
  \ref{eqn:fluid-arcs}, \ref{eqn:output-capacity},
  \ref{eqn:conservation}, \ref{eqn:incoming-rule},
  \ref{eqn:outgoing-rule} and the strong maximization
  rule~\ref{eqn:strong-maximization}, and for any $i \in I$ and
  $e \in \delta^+(i)$, $t(e) \leq c(i)$.
\end{definition}

The algorithm starts with the trivial sub-steady-state
$(t : e \to 0, E)$, and will improve it iteratively until reaching a
steady-state. At each iteration of the algorithm, we will be trying to
increase the throughputs of the arcs without violating any rule.
Unlike in maximum flows, we do not have the choice of which leaving
arc to increase the flow on. Furthermore, rule~\ref{eqn:maximization}
forces each splitter to send as much flow forward as possible. A
non-obvious consequence is that, when increasing the input capacities,
throughputs can only increase on fluid arcs, and can only decrease on
saturated arcs. This suggests a definition of the residual graph for
the sub-steady-state $(t,F)$. Its vertex set is $\{z\} \cup S$, where
$z$ is obtained by identifying all the inputs and outputs into a
single node. Its edge set contains some fluid arcs and the reverses of
some saturated arcs.

\begin{figure}
  \centering
  \small
  \begin{tikzpicture}[x=1.3cm,y=0.75cm,>=latex]
    \begin{scope}[xshift=0cm]
      \node[splitter] (s) at (0,0) {};
      \draw[->,fluid] (-1,1) -- (s) node[midway,sloped,above] {$+\varepsilon$}; 
      \draw (-1,1) node[anchor=east] {$e_1$};
      \draw[->,fluid] (-1,-1) -- (s); 
      \draw (-1,-1) node[anchor=east] {$e_2$}; 
      \draw[<-,fluid] (1,1) -- (s) node[midway,sloped,above] {$+\frac{\varepsilon}{2}$}; 
      \draw (1,1) node[anchor=west] {$e_3$};
      \draw[<-,fluid] (1,-1) -- (s) node[midway,sloped,below] {$+\frac{\varepsilon}{2}$}; 
      \draw (1,-1) node[anchor=west] {$e_4$};
      \draw (0,-1.5) node {$(a)$};
    \end{scope}
    \begin{scope}[xshift=4cm]
      \node[splitter] (s) at (0,0) {};
      \draw[->,fluid] (-1,1) -- (s) node[midway,sloped,above] {$+\varepsilon$}; 
      \draw (-1,1) node[anchor=east] {$e_1$};
      \draw[->,fluid] (-1,-1) -- (s); 
      \draw (-1,-1) node[anchor=east] {$e_2$}; 
      \draw[<-,saturated] (1,1) -- (s); 
      \draw (1,1) node[anchor=west] {$e_3$};
      \draw[<-,fluid] (1,-1) -- (s) node[midway,sloped,below] {$+\varepsilon$}; 
      \draw (1,-1) node[anchor=west] {$e_4$};
      \draw (0,-1.5) node {$(b)$};
    \end{scope}
    \begin{scope}[xshift=8cm]
      \node[splitter] (s) at (0,0) {};
      \draw[->,fluid] (-1,1) -- (s) node[midway,sloped,above] {$+\varepsilon$}; 
      \draw (-1,1) node[anchor=east] {$e_1$};
      \draw[->,saturated] (-1,-1) -- (s) node[midway,sloped,below] {$-\varepsilon$}; 
      \draw (-1,-1) node[anchor=east] {$e_2$}; 
      \draw[<-,saturated] (1,1) -- (s); 
      \draw (1,1) node[anchor=west] {$e_3$};
      \draw[<-,saturated] (1,-1) -- (s); 
      \draw (1,-1) node[anchor=west] {$e_4$};
      \draw (0,-1.5) node {$(c)$};

    \end{scope}
    \begin{scope}[xshift=12cm]
      \node[splitter] (s) at (0,0) {};
      \draw[->,saturated] (-1,1) -- (s) node[midway,sloped,above] {$-\frac{\varepsilon}{2}$}; 
      \draw (-1,1) node[anchor=east] {$e_1$};
      \draw[->,saturated] (-1,-1) -- (s) node[midway,sloped,below] {$-\frac{\varepsilon}{2}$}; 
      \draw (-1,-1) node[anchor=east] {$e_2$}; 
      \draw[<-,saturated] (1,1) -- (s) node[midway,sloped,above] {$-\varepsilon$}; 
      \draw (1,1) node[anchor=west] {$e_3$};
      \draw[<-,saturated] (1,-1) -- (s); 
      \draw (1,-1) node[anchor=west] {$e_4$};
      \draw (0,-1.5) node {$(d)$};
    \end{scope}
  \end{tikzpicture}
    \caption{Four examples of throughput changes at a single splitter,
    depending on which arcs are fluid.}
  \label{fig:routing-flow}
\end{figure}
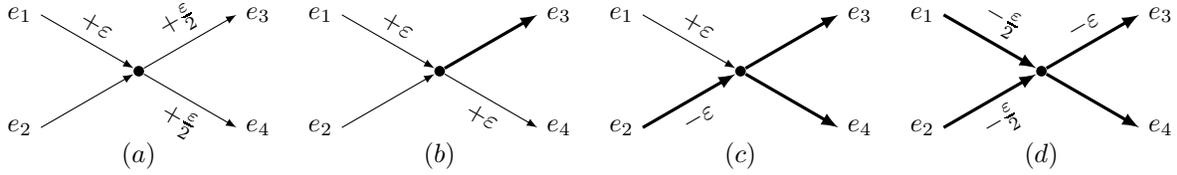

Consider the splitters in \Cref{fig:routing-flow}. We examine what happens
when we increase the throughput on edge $e_1$ by $+\varepsilon$, or in
case $(d)$ when we decrease $t(e_3)$ by $\varepsilon$. In case $(a)$,
by rule~\ref{eqn:outgoing-rule}, the throughputs on the two leaving
arcs must stay equal, hence both increases by $\varepsilon/2$. In case
$(b)$, only the throughput of the fluid leaving arc $e_4$ can
increase. In case $(c)$, both leaving arc are saturated, the splitter
cannot push more flow downward, hence it is forced to push back flow
through its incoming saturated arcs. Thus $t(e_2)$ decreases while
$t(e_1)$ increases, by no more than $(t(e_2) - t(e_1))/2$ because of
rule~\ref{eqn:incoming-rule}. Finally in case $(d)$, if we decrease
$t(e_3)$, then $t(e_1)$ and $t(e_2)$ must decrease by half as much.

Case $(c)$ presents a challenge due to rule~\ref{eqn:incoming-rule},
which imposes $t(e_1) \leq t(e_2)$. When $t(e_1) = t(e_2)$, the
throughput of $e_1$ cannot increase, and the throughput of $e_2$
cannot decrease. We say that $e_1$ and $e_2$ are \emph{tight}. In such
a case, removing $e_1$ from $F$ is allowed by
rule~\ref{eqn:incoming-rule}. Fluid arcs $e$ with $t(e) = c(e)$ or
saturated arc with $t(e) = 0$ are also \emph{tight}, since we cannot
modify their throughput further. Then we define the edge-set of the
residual graph to only contain non-tight fluid arcs and reverses of
non-tight saturated arcs.

\begin{figure}
  \centering
  \begin{tikzpicture}[x=1cm,y=1cm,>=latex]
    \draw (-1.5,0) node[anchor=east] {Network:};
    \draw (-1.5,-2) node[anchor=east] {Residual graph:};
    \begin{scope}[xshift=0cm]
      \node[splitter] (s) at (0,0) {};
      \draw[->,fluid] (-1,0.5) -- (s);
      \draw[->,fluid] (-1,-0.5) -- (s);
      \draw[<-,fluid] (1,0.5) -- (s);
      \draw[<-,fluid] (1,-0.5) -- (s);
      \draw[->,double] (0,-0.5) -- (0,-1.5);
      \node[splitter] (r) at (0,-2) {};
      \draw[->,fluid] (-1,-1.5) -- (r);
      \draw[->,fluid] (-1,-2.5) -- (r);
      \draw[<-,fluid,red] (1,-1.5) -- (r);
      \draw[<-,fluid,red] (1,-2.5) -- (r);
    \end{scope}      
    \begin{scope}[xshift=2.5cm]
      \node[splitter] (s) at (0,0) {};
      \draw[->,fluid] (-1,0.5) -- (s);
      \draw[->,fluid] (-1,-0.5) -- (s);
      \draw[<-,saturated] (1,0.5) -- (s);
      \draw[<-,fluid] (1,-0.5) -- (s);
      \draw[->,double] (0,-0.5) -- (0,-1.5);
      \node[splitter] (r) at (0,-2) {};
      \draw[->,fluid] (-1,-1.5) -- (r);
      \draw[->,fluid] (-1,-2.5) -- (r);
      \draw[->,fluid] (1,-1.5) -- (r);
      \draw[<-,fluid,red] (1,-2.5) -- (r);
    \end{scope}      
    \begin{scope}[xshift=5cm]
      \node[splitter] (s) at (0,0) {};
      \draw[->,fluid] (-1,0.5) -- (s);
      \draw[->,fluid] (-1,-0.5) -- (s);
      \draw[<-,saturated] (1,0.5) -- (s);
      \draw[<-,saturated] (1,-0.5) -- (s);
      \draw[->,double] (0,-0.5) -- (0,-1.5);
      \node[splitter] (r) at (0,-2) {};
      \draw[->,fluid] (-1,-1.5) -- (r);
      \draw[->,fluid] (-1,-2.5) -- (r);
      \draw[->,fluid] (1,-1.5) -- (r);
      \draw[->,fluid] (1,-2.5) -- (r);
    \end{scope}      
    \begin{scope}[xshift=7.5cm]
      \node[splitter] (s) at (0,0) {};
      \draw[->,fluid] (-1,0.5) -- (s);
      \draw[->,saturated] (-1,-0.5) -- (s);
      \draw[<-,saturated] (1,0.5) -- (s);
      \draw[<-,saturated] (1,-0.5) -- (s);
      \draw[->,double] (0,-0.5) -- (0,-1.5);
      \node[splitter] (r) at (0,-2) {};
      \draw[->,fluid] (-1,-1.5) -- (r);
      \draw[<-,fluid,red] (-1,-2.5) -- (r);
      \draw[->,fluid] (1,-1.5) -- (r);
      \draw[->,fluid] (1,-2.5) -- (r);
    \end{scope}      
    \begin{scope}[xshift=10cm]
      \node[splitter] (s) at (0,0) {};
      \draw[->,saturated] (-1,0.5) -- (s);
      \draw[->,saturated] (-1,-0.5) -- (s);
      \draw[<-,saturated] (1,0.5) -- (s);
      \draw[<-,saturated] (1,-0.5) -- (s);
      \draw[->,double] (0,-0.5) -- (0,-1.5);
      \node[splitter] (r) at (0,-2) {};
      \draw[<-,fluid,red] (-1,-1.5) -- (r);
      \draw[<-,fluid,red] (-1,-2.5) -- (r);
      \draw[->,fluid] (1,-1.5) -- (r);
      \draw[->,fluid] (1,-2.5) -- (r);
    \end{scope}      
   \end{tikzpicture}  
  \caption{Configurations of splitters and the corresponding vertex in
    the residual graph. The outgoing arcs from a vertex of the
    residual graphs are highlighted in red: notice that in a
    sub-steady-state, the throughputs on these arcs must be equal.}
  \label{fig:configuration-residual}
\end{figure}
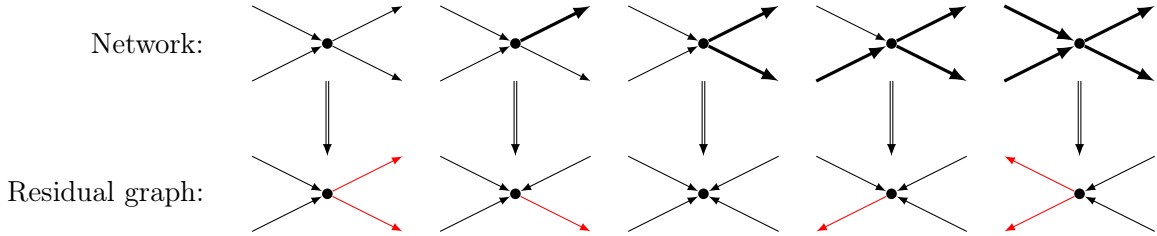

\begin{figure}
  \centering
  \begin{tikzpicture}[x=2cm,y=2cm,>=latex]
    \node[term] (i) at (0.4,1) {};
    \node[term] (o) at (3.6,1) {};
    \draw (0.4,1) node[anchor=east] {$1$};
    \draw (3.6,1) node[anchor=west] {$1$};
    \foreach \n/\x/\y in {a/1/1,b/2/2,c/2/1,d/2/0,e/3/1} {
      \node[splitter] (s\n) at (\x,\y) {};
    }
    \draw[fluid,->] (i) -- (sa) node[midway,sloped,above=-3pt] {$0$};
    \draw[fluid,->] (sa) -- (sb) node[midway,sloped,above=-3pt] {$0$};
    \draw[fluid,->] (sa) -- (sd) node[midway,sloped,below=-3pt] {$0$};
    \draw[fluid,->] (sb) -- (sc) node[pos=0.6,right=-3pt] {$0$};
    \draw[fluid,->] (sb) -- (se) node[midway,sloped,above=-3pt] {$0$};
    \draw[fluid,->] (sc) -- (sa) node[midway,sloped,above=-3pt] {$0$};
    \draw[fluid,->] (sd) -- (sc) node[pos=0.6,left=-3pt] {$0$};
    \draw[fluid,->] (se) -- (o) node[midway,sloped,above=-3pt] {$0$};
    \draw[fluid,->] (sd) .. controls (2.2,0.4) and (2.6,0.8) ..  (se)
    node[midway,sloped,above=-3pt] {$0$};
    \draw[fluid,->] (se) .. controls (2.8,0.6) and (2.4,0.2) .. (sd)
    node[midway,sloped,below=-3pt] {$0$};
    \begin{scope}[xshift=8cm]
      \node[splitter] (i) at (0.4,1) {};
      \node[splitter] (o) at (3.6,1) {};
      \draw (0.4,1) node[anchor=east] {$z$};
      \draw (3.6,1) node[anchor=west] {$z$};
      \foreach \n/\x/\y in {a/1/1,b/2/2,c/2/1,d/2/0,e/3/1} {
        \node[splitter] (s\n) at (\x,\y) {};
      }
      \draw[->] (i) -- (sa) node[midway,sloped,above=-3pt] {$1/12$};
      \draw[->] (sa) -- (sb) node[midway,sloped,above=-3pt] {$1/8$};
      \draw[->] (sa) -- (sd) node[midway,sloped,below=-3pt] {$1/8$};
      \draw[->] (sb) -- (sc) node[pos=0.6,right=-3pt] {$1/16$};
      \draw[->] (sb) -- (se) node[midway,sloped,above=-3pt] {$1/16$};
      \draw[->] (sc) -- (sa) node[midway,sloped,above=-3pt] {$1/6$};
      \draw[->] (sd) -- (sc) node[pos=0.6,left=-3pt] {$5/48$};
      \draw[->] (se) -- (o) node[midway,sloped,above=-3pt] {$1/12$};
      \draw[->] (sd) .. controls (2.2,0.4) and (2.6,0.8) ..  (se)
      node[midway,sloped,above=-3pt] {$5/48$};
      \draw[->] (se) .. controls (2.8,0.6) and (2.4,0.2) .. (sd)
      node[midway,sloped,below=-3pt] {$1/12$};
      \end{scope}
      
  \end{tikzpicture}
  \caption{Starting from a trivial sub-steady-state, we compute a
    residual graph and a stationary circulation in this graph (the two
    vertices marked $z$ should be identified). Then we increase the
    throughputs accordingly, as much as possible without violating a
    sub-steady-state rule, by adding $\lambda = 6$ times the
    circulation at which point some edge reaches its capacity (see
    \Cref{fig:algo-round2}).}
  \label{fig:algo-round1}
\end{figure}
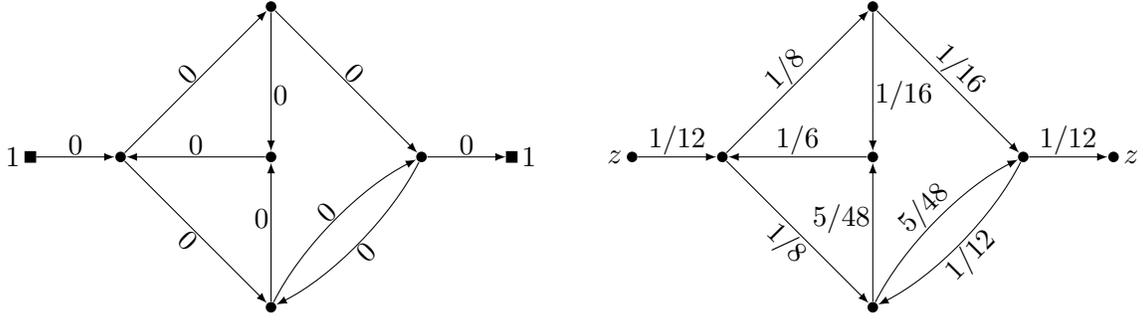

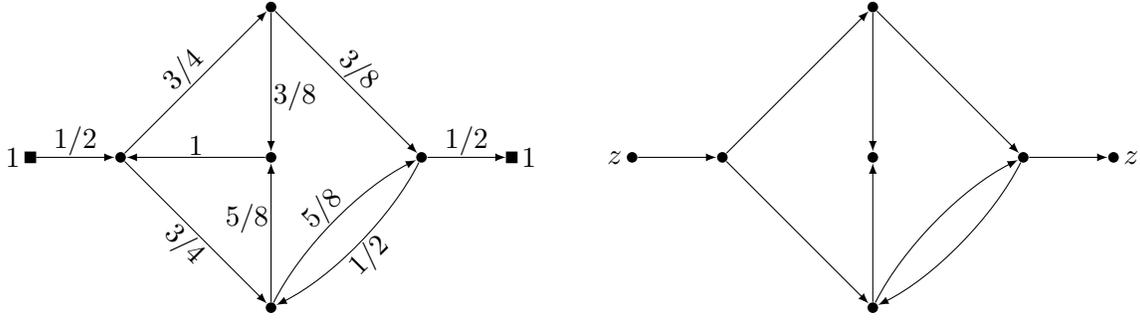
\begin{figure}
  \centering
  \begin{tikzpicture}[x=2cm,y=2cm,>=latex]
    \node[term] (i) at (0.4,1) {};
    \node[term] (o) at (3.6,1) {};
    \draw (0.4,1) node[anchor=east] {$1$};
    \draw (3.6,1) node[anchor=west] {$1$};
    \foreach \n/\x/\y in {a/1/1,b/2/2,c/2/1,d/2/0,e/3/1} {
      \node[splitter] (s\n) at (\x,\y) {};
    }
    \draw[fluid,->] (i) -- (sa) node[midway,sloped,above=-3pt] {$1/2$};
    \draw[fluid,->] (sa) -- (sb) node[midway,sloped,above=-3pt] {$3/4$};
    \draw[fluid,->] (sa) -- (sd) node[midway,sloped,below=-3pt] {$3/4$};
    \draw[fluid,->] (sb) -- (sc) node[pos=0.6,right=-3pt] {$3/8$};
    \draw[fluid,->] (sb) -- (se) node[midway,sloped,above=-3pt] {$3/8$};
    \draw[fluid,->] (sc) -- (sa) node[midway,sloped,above=-3pt] {$1$};
    \draw[fluid,->] (sd) -- (sc) node[pos=0.6,left=-3pt] {$5/8$};
    \draw[fluid,->] (se) -- (o) node[midway,sloped,above=-3pt] {$1/2$};
    \draw[fluid,->] (sd) .. controls (2.2,0.4) and (2.6,0.8) ..  (se)
    node[midway,sloped,above=-3pt] {$5/8$};
    \draw[fluid,->] (se) .. controls (2.8,0.6) and (2.4,0.2) .. (sd)
    node[midway,sloped,below=-3pt] {$1/2$};
    \begin{scope}[xshift=8cm]
      \node[splitter] (i) at (0.4,1) {};
      \node[splitter] (o) at (3.6,1) {};
      \draw (0.4,1) node[anchor=east] {$z$};
      \draw (3.6,1) node[anchor=west] {$z$};
      \foreach \n/\x/\y in {a/1/1,b/2/2,c/2/1,d/2/0,e/3/1} {
        \node[splitter] (s\n) at (\x,\y) {};
      }
      \draw[->] (i) -- (sa);
      \draw[->] (sa) -- (sb);
      \draw[->] (sa) -- (sd);
      \draw[->] (sb) -- (sc);
      \draw[->] (sb) -- (se);
      \draw[->] (sd) -- (sc);
      \draw[->] (se) -- (o);
      \draw[->] (sd) .. controls (2.2,0.4) and (2.6,0.8) ..  (se);
      \draw[->] (se) .. controls (2.8,0.6) and (2.4,0.2) .. (sd);
      \end{scope}
      
  \end{tikzpicture}
  \caption{We compute a new residual graph, which does not contain the
    arc with throughput $1$, since this arc cannot increase. Then the
    existence of a sink prevents us to find a stationary circulation
    in this residual graph (the two vertices marked $z$ should be
    identified). We remove from $F$ the incoming arc to the sink with
    highest throughput, and go to the next iteration.}
  \label{fig:algo-round2}
\end{figure}

Due to the conservation rule~\ref{eqn:conservation}, any iterative
change to the throughputs of the network must be in accordance with a
circulation of the residual graph. Because of
rules~\ref{eqn:incoming-rule} and~\ref{eqn:outgoing-rule}, some arcs
are constrained to have the same throughput. Therefore the chosen
circulation itself has similar constraints. This is illustrated in
\Cref{fig:configuration-residual}, where the arcs that have equal
throughput are highlighted in the residual graph. As may be readily
checked, those constraints are exactly set on the leaving arcs in the
residual graph of each vertex corresponding to a splitter. As for the
special vertex $z$, obtained from the identification of the inputs and
the outputs, we may non-deterministically select one of its leaving
arc. Then we force all other arcs leaving $z$ to have zero flow, by
removing those arcs from the residual graph. From the residual graph,
we compute a circulation satisfying each equality constraint. First
compute a stationary distribution of a random walk on the residual
graph. Then assign to each arc the probability of being the next arc
in a random walk from that distribution. This results in a so-called
\emph{stationary circulation} (see \Cref{fig:algo-round1}). One must
be careful if the residual graph is not strongly connected. Then
either we can find a strongly connected subgraph induced by the
leaving arcs of some subset of vertices, or the residual graph
contains a sink (as in \Cref{fig:algo-round2}). In the former case we
can still find a circulation, while in the latter case, we will be
able to remove some arc from $F$.

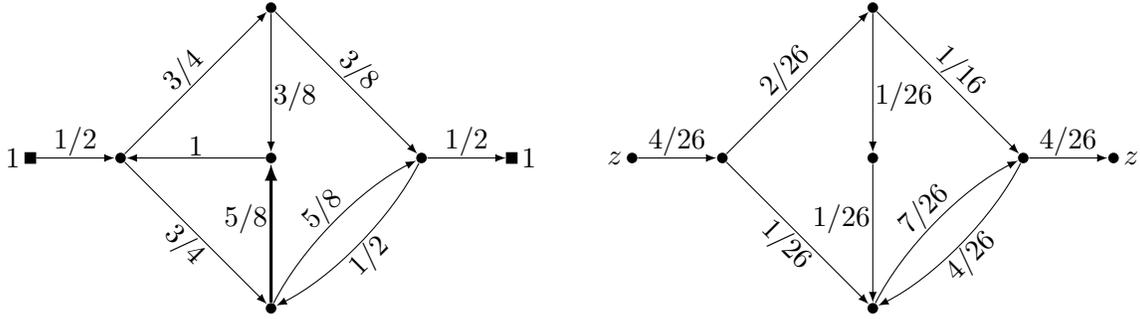
\begin{figure}
  \centering
  \begin{tikzpicture}[x=2cm,y=2cm,>=latex]
    \node[term] (i) at (0.4,1) {};
    \node[term] (o) at (3.6,1) {};
    \draw (0.4,1) node[anchor=east] {$1$};
    \draw (3.6,1) node[anchor=west] {$1$};
    \foreach \n/\x/\y in {a/1/1,b/2/2,c/2/1,d/2/0,e/3/1} {
      \node[splitter] (s\n) at (\x,\y) {};
    }
    \draw[fluid,->] (i) -- (sa) node[midway,sloped,above=-3pt] {$1/2$};
    \draw[fluid,->] (sa) -- (sb) node[midway,sloped,above=-3pt] {$3/4$};
    \draw[fluid,->] (sa) -- (sd) node[midway,sloped,below=-3pt] {$3/4$};
    \draw[fluid,->] (sb) -- (sc) node[pos=0.6,right=-3pt] {$3/8$};
    \draw[fluid,->] (sb) -- (se) node[midway,sloped,above=-3pt] {$3/8$};
    \draw[fluid,->] (sc) -- (sa) node[midway,sloped,above=-3pt] {$1$};
    \draw[saturated,->] (sd) -- (sc) node[pos=0.6,left=-3pt] {$5/8$};
    \draw[fluid,->] (se) -- (o) node[midway,sloped,above=-3pt] {$1/2$};
    \draw[fluid,->] (sd) .. controls (2.2,0.4) and (2.6,0.8) ..  (se)
    node[midway,sloped,above=-3pt] {$5/8$};
    \draw[fluid,->] (se) .. controls (2.8,0.6) and (2.4,0.2) .. (sd)
    node[midway,sloped,below=-3pt] {$1/2$};
    \begin{scope}[xshift=8cm]
      \node[splitter] (i) at (0.4,1) {};
      \node[splitter] (o) at (3.6,1) {};
      \draw (0.4,1) node[anchor=east] {$z$};
      \draw (3.6,1) node[anchor=west] {$z$};
      \foreach \n/\x/\y in {a/1/1,b/2/2,c/2/1,d/2/0,e/3/1} {
        \node[splitter] (s\n) at (\x,\y) {};
      }
      \draw[->] (i) -- (sa) node[midway,sloped,above=-3pt] {$4/26$};
      \draw[->] (sa) -- (sb) node[midway,sloped,above=-3pt] {$2/26$};
      \draw[->] (sa) -- (sd) node[midway,sloped,below=-3pt] {$2/26$};
      \draw[->] (sb) -- (sc) node[pos=0.6,right=-3pt] {$1/26$};
      \draw[->] (sb) -- (se) node[midway,sloped,above=-3pt] {$1/26$};
      \draw[<-] (sd) -- (sc) node[pos=0.6,left=-3pt] {$1/26$};
      \draw[->] (se) -- (o) node[midway,sloped,above=-3pt] {$4/26$};
      \draw[->] (sd) .. controls (2.2,0.4) and (2.6,0.8) ..  (se)
      node[midway,sloped,above=-3pt] {$7/26$};
      \draw[->] (se) .. controls (2.8,0.6) and (2.4,0.2) .. (sd)
      node[midway,sloped,below=-3pt] {$4/26$};
    \end{scope}  
  \end{tikzpicture}
  \caption{We compute a new residual graph, including the reverse of
    the newly saturated arc. Then we compute a stationary circulation
    in this residual graph. This leads to an improved sub-steady-state
    (taking $\lambda = \frac{3 \cdot 26}{4 \cdot 14}$, when another
    edge reaches throughput 1, see \Cref{fig:algo-round4})}
  \label{fig:algo-round3}
\end{figure}

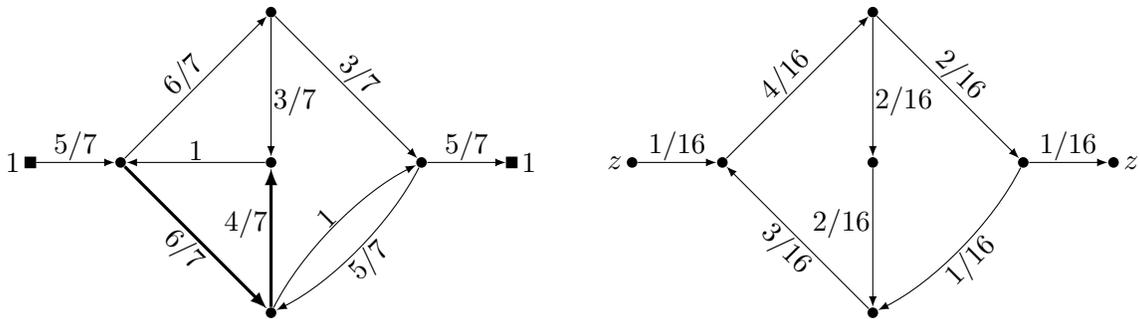
\begin{figure}
  \centering
  \begin{tikzpicture}[x=2cm,y=2cm,>=latex]
    \node[term] (i) at (0.4,1) {};
    \node[term] (o) at (3.6,1) {};
    \draw (0.4,1) node[anchor=east] {$1$};
    \draw (3.6,1) node[anchor=west] {$1$};
    \foreach \n/\x/\y in {a/1/1,b/2/2,c/2/1,d/2/0,e/3/1} {
      \node[splitter] (s\n) at (\x,\y) {};
    }
    \draw[fluid,->] (i) -- (sa) node[midway,sloped,above=-3pt] {$5/7$};
    \draw[fluid,->] (sa) -- (sb) node[midway,sloped,above=-3pt] {$6/7$};
    \draw[saturated,->] (sa) -- (sd) node[midway,sloped,below=-3pt] {$6/7$};
    \draw[fluid,->] (sb) -- (sc) node[pos=0.6,right=-3pt] {$3/7$};
    \draw[fluid,->] (sb) -- (se) node[midway,sloped,above=-3pt] {$3/7$};
    \draw[fluid,->] (sc) -- (sa) node[midway,sloped,above=-3pt] {$1$};
    \draw[saturated,->] (sd) -- (sc) node[pos=0.6,left=-3pt] {$4/7$};
    \draw[fluid,->] (se) -- (o) node[midway,sloped,above=-3pt] {$5/7$};
    \draw[fluid,->] (sd) .. controls (2.2,0.4) and (2.6,0.8) ..  (se)
    node[midway,sloped,above=-3pt] {$1$};
    \draw[fluid,->] (se) .. controls (2.8,0.6) and (2.4,0.2) .. (sd)
    node[midway,sloped,below=-3pt] {$5/7$};
    \begin{scope}[xshift=8cm]
      \node[splitter] (i) at (0.4,1) {};
      \node[splitter] (o) at (3.6,1) {};
      \draw (0.4,1) node[anchor=east] {$z$};
      \draw (3.6,1) node[anchor=west] {$z$};
      \foreach \n/\x/\y in {a/1/1,b/2/2,c/2/1,d/2/0,e/3/1} {
        \node[splitter] (s\n) at (\x,\y) {};
      }
      \draw[->] (i) -- (sa) node[midway,sloped,above=-3pt] {$1/16$};
      \draw[->] (sa) -- (sb) node[midway,sloped,above=-3pt] {$4/16$};
      \draw[<-] (sa) -- (sd) node[midway,sloped,below=-3pt] {$3/16$};
      \draw[->] (sb) -- (sc) node[pos=0.6,right=-3pt] {$2/16$};
      \draw[->] (sb) -- (se) node[midway,sloped,above=-3pt] {$2/16$};
      \draw[<-] (sd) -- (sc) node[pos=0.6,left=-3pt] {$2/16$};
      \draw[->] (se) -- (o) node[midway,sloped,above=-3pt] {$1/16$};
      \draw[fluid,->] (se) .. controls (2.8,0.6) and (2.4,0.2) .. (sd)
      node[midway,sloped,below=-3pt] {$1/16$};
    \end{scope}  
  \end{tikzpicture}
  \caption{The next residual graph contains a sink, thus we remove one
    more arc from $F$. This yields the residual graph on the right. We
    compute the stationary circulation in this residual graph. Then we
    imporve the sub-steady-state according to this circulation (here
    $\lambda = \frac{4}{7}$, when the incoming arcs of the central
    vertex become tight, see \Cref{fig:algo-round5})}
  \label{fig:algo-round4}
\end{figure}

\begin{figure}
  \centering
  \begin{tikzpicture}[x=2cm,y=2cm,>=latex]
    \node[term] (i) at (0.4,1) {};
    \node[term] (o) at (3.6,1) {};
    \draw (0.4,1) node[anchor=east] {$1$};
    \draw (3.6,1) node[anchor=west] {$1$};
    \foreach \n/\x/\y in {a/1/1,b/2/2,c/2/1,d/2/0,e/3/1} {
      \node[splitter] (s\n) at (\x,\y) {};
    }
    \draw[saturated,->] (i) -- (sa) node[midway,sloped,above=-3pt] {$6/7$};
    \draw[fluid,->] (sa) -- (sb) node[midway,sloped,above=-3pt] {$1$};
    \draw[saturated,->] (sa) -- (sd) node[midway,sloped,below=-3pt] {$5/7$};
    \draw[saturated,->] (sb) -- (sc) node[pos=0.6,right=-3pt] {$3/7$};
    \draw[fluid,->] (sb) -- (se) node[midway,sloped,above=-3pt] {$4/7$};
    \draw[saturated,->] (sc) -- (sa) node[midway,sloped,above=-3pt] {$6/7$};
    \draw[saturated,->] (sd) -- (sc) node[pos=0.6,left=-3pt] {$3/7$};
    \draw[fluid,->] (se) -- (o) node[midway,sloped,above=-3pt] {$6/7$};
    \draw[fluid,->] (sd) .. controls (2.2,0.4) and (2.6,0.8) ..  (se)
    node[midway,sloped,above=-3pt] {$1$};
    \draw[saturated,->] (se) .. controls (2.8,0.6) and (2.4,0.2) .. (sd)
    node[midway,sloped,below=-3pt] {$5/7$};
    \begin{scope}[xshift=8cm]
      \node[splitter] (i) at (0.4,1) {};
      \node[splitter] (o) at (3.6,1) {};
      \draw (0.4,1) node[anchor=east] {$z$};
      \draw (3.6,1) node[anchor=west] {$z$};
      \foreach \n/\x/\y in {a/1/1,b/2/2,c/2/1,d/2/0,e/3/1} {
        \node[splitter] (s\n) at (\x,\y) {};
      }
      \draw[<-] (i) -- (sa);
      \draw[<-] (sc) -- (sa);
      \draw[<-] (sa) -- (sd);
      \draw[<-] (sb) -- (sc);
      \draw[->] (sb) -- (se);
      \draw[<-] (sd) -- (sc);
      \draw[->] (se) -- (o);
      \draw[<-] (se) .. controls (2.8,0.6) and (2.4,0.2) .. (sd);
    \end{scope}  
  \end{tikzpicture}
  \caption{After removing one more arc from $F$, $z$ becomes a
    sink in the residual graph. It implies that $(t,F)$ is a
    steady-state. The algorithm stops.}
  \label{fig:algo-round6}
\end{figure}
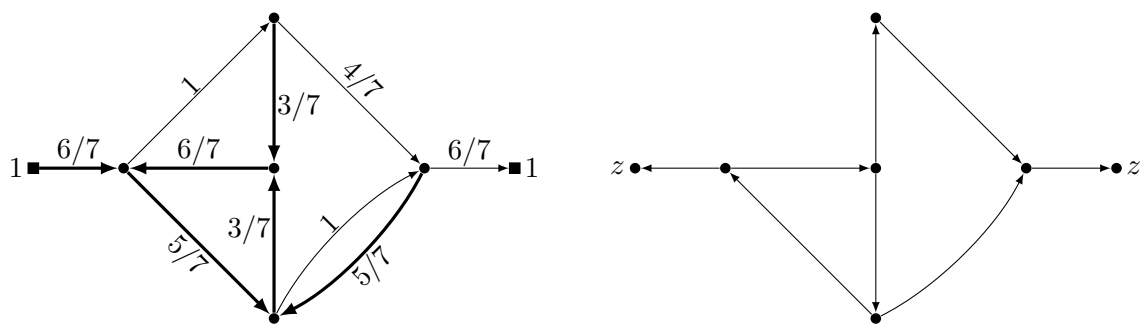

Once a circulation is found, we increase the throughput as much as
possible. This process will result in the creation of at least one
sink in the updated residual graph. We show that when the residual
graph contains a sink, some arc can be safely removed from $F$ and
becomes saturated. This bounds the number of steps until the algorithm
stops, when $z$ itself becomes a sink. At this point, any arc leaving
an input node is either at full capacity or is saturated. Hence
rule~\ref{eqn:input-capacity} is satisfied, $(t,F)$ is a steady-state.
Some additional details, not covered in this presentation, are
provided in \Cref{sec:blocking-flow-algo}. An example of run of the
algorithm on a simple network is given in
\Cref{fig:algo-round1,fig:algo-round2,fig:algo-round3,fig:algo-round4,fig:algo-round5,fig:algo-round6}.
Summarizing the discussion, we get:

\begin{theorem}\label{thm:equi-algo}
  There is an algorithm that given a splitter network $\splitnetwork$
  with capacities $c: I \disunion O \to [0,1]$, finds a steady-state
  $(t,F)$ in time $O(|S|^2 + |S| \ \stationary(G_z))$, where $G_z$ is the
  graph obtained by identifying $I \cup O$ into a single vertex $z$,
  and $\stationary(G_z)$ denotes the time to compute a stationary
  distribution on any orientation of a subgraph of $G_z$.
\end{theorem}

\begin{figure}
  \centering
  \begin{tikzpicture}[x=1cm,y=1cm,>=latex]
    \node[term] (i) at (0,0) {}; \draw (0,0) node[anchor=east] {$1$};
    \node[term] (o) at (12,0) {}; \draw (12,0) node[anchor=west] {$1$};
    \foreach \x in {1,2,3,4,5,7,8,9,10,11} {
      \node[splitter] (s\x) at (\x,0) {};
    }
    \draw[->,saturated] (i) -- (s1) node[midway,above=-3pt,sloped] {$1/4$};
    \draw[->,saturated] (s1) -- (s2) node[midway,above=-3pt,sloped] {$1/2$};
    \draw[->,saturated] (s2) -- (s3) node[midway,above=-3pt,sloped] {$1$};
    \draw[->,saturated] (s3) .. controls (2.7,-0.3) and (2.3,-0.3) .. (s2) 
    node[midway,below=-3pt,sloped] {$1/2$};
    \draw[->,fluid] (s3) -- (s4) node[midway,above=-3pt,sloped] {$1/2$};
    \draw[->,saturated] (s4) .. controls (3.5,-1.5) and (1.5,-1.5) .. (s1)
    node[midway,below=-3pt,sloped] {$1/4$};
    \draw[->,fluid] (s4) -- (s5) node[midway,above=-3pt,sloped] {$1/4$};
    \draw[->,fluid] (s5) .. controls (5.4,0.4) and (6.6,0.4) .. (s7)
    node[midway,above=-3pt,sloped] {$1/2 + \varepsilon$};
    \draw[->,fluid] (s7) .. controls (6.6,-0.4) and (5.4,-0.4) .. (s5)
    node[midway,below=-3pt,sloped] {$1/4 + \varepsilon$};
    \draw[->,saturated] (s7) -- (s8) node[midway,above=-3pt,sloped] {$1/4$};
    \draw[->,saturated] (s8) -- (s9) node[midway,above=-3pt,sloped] {$1/2$};
    \draw[->,saturated] (s9) -- (s10) node[midway,above=-3pt,sloped] {$1$};
    \draw[->,fluid] (s10) -- (s11) node[midway,above=-3pt,sloped] {$1/2$};
    \draw[->,fluid] (s11) -- (o) node[midway,above=-3pt,sloped] {$1/4$};
    \draw[->,saturated] (s10) .. controls (9.7,-0.3) and (9.3,-0.3) .. (s9) 
    node[midway,below=-3pt,sloped] {$1/2$};
    \draw[->,saturated] (s11) .. controls (10.5,-1.5) and (8.5,-1.5) .. (s8)
    node[midway,below=-3pt,sloped] {$1/4$};

  \end{tikzpicture}
  \caption{A network having several steady-states. Any value for
    $\varepsilon$ between $0$ and $\frac{1}{2}$ gives a steady-state.}
  \label{fig:non-unique}
\end{figure}
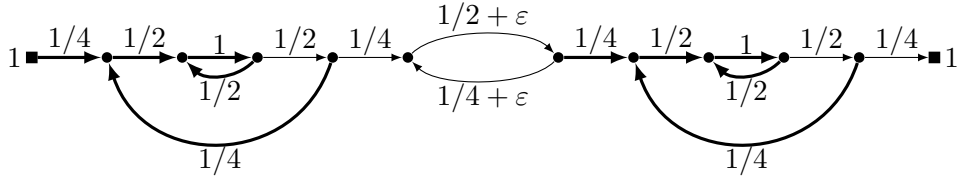

Steady-states are not unique: a directed cycle with no input or output
can have any constant throughput on all its arcs.
\Cref{fig:non-unique} showcases a more interesting network, having one
input, one output, and many possible steady-states. However, in this
example, all steady-states have the same throughputs on the inputs and
outputs. Is there a network with two steady-states significantly
different steady-states? We will answer this question negatively in
\Cref{sec:unique}, proving that all the steady-states induce the same
throughputs on the inputs and outputs of the splitter network.

\subsection{Balancers}\label{sec:intro-balancer}

We now define load-balancing networks and their properties. The goal
of a load-balancing network is to divide some input flow evenly
between several output belts. In the simplest case, the output belts
can receive an arbitrarily large flow (up to the capacity of the
belt). In more general cases, some outputs may be restricted but we
still want the flow to be divided as evenly as possible, without
limiting the total throughput available. We distinguish three
properties of load-balancing networks. The first of these properties
considers networks where the output capacities are not constrained.

\begin{definition}
  A splitter network $\splitnetwork$ is a \emph{balancer} if for any
  $c : I \disunion O \to [0,1]$ such that for each output $o \in O$,
  $c(o) = 1$, there is a steady-state $(t,F)$ for $(G,c)$ with
  $t$ constant on $\delta^-(O)$. An $(n,p)$-balancer is defined as a balancer
  with $|I| = n$ inputs and $|O| = p$ outputs.
\end{definition}

\begin{figure}
  \begin{center}
    \small
    \begin{tikzpicture}[x=1.8cm,y=1.3cm,>=latex]
      \foreach \xa/\ya/\xb/\yb in {
        0.7/-0.6/3.2/3.6,
        0.8/-0.4/2.2/1.4,
        0.8/1.6/2.2/3.4,
        0.9/-0.2/1.1/0.2,
        0.9/0.8/1.1/1.2,
        0.9/1.8/1.1/2.2,
        0.9/2.8/1.1/3.2}
      {
        \fill[rounded corners=3pt,nearly transparent,blue] (\xa,\ya) 
          rectangle (\xb,\yb);
      }

      \foreach \n/\y/\c in {0/-0.25/1,1/0.25/1,2/0.75/1,3/1.25/1,4/1.75/1,5/2.25/1,6/2.75/1,7/3.25/0} {
        \node[term] (i\n) at (0,\y) {};
        \node[term] (o\n) at (4,\y) {};
        \draw (0,\y) node[anchor=east] {$0.5$};
        \draw (4,\y) node[anchor=west] {$\c$};
      }
      \foreach \y in {0,1,2,3} {
        \foreach \x in {1,2,3} {
          \node[splitter] (s\x\y) at (\x,\y) {};
        }
      }
      \draw[fluid,->] (i0) -- (s10) node[sloped,midway,below] {$0.5$};
      \draw[fluid,->] (i1) -- (s10) node[sloped,midway,above] {$0.5$};
      \draw[fluid,->] (i2) -- (s11) node[sloped,midway,below] {$0.5$};
      \draw[fluid,->] (i3) -- (s11) node[sloped,midway,above] {$0.5$};
      \draw[fluid,->] (i4) -- (s12) node[sloped,midway,below] {$0.5$};
      \draw[fluid,->] (i5) -- (s12) node[sloped,midway,above] {$0.5$};
      \draw[fluid,->] (i6) -- (s13) node[sloped,midway,below] {$0.5$};
      \draw[fluid,->] (i7) -- (s13) node[sloped,midway,above] {$0.5$};

      \draw[fluid,->] (s10) -- (s20) node[midway,below] {$0.5$};
      \draw[fluid,->] (s11) -- (s21) node[midway,above] {$0.5$};
      \draw[fluid,->] (s12) -- (s22) node[midway,below] {$0.5$};
      \draw[fluid,->] (s13) -- (s23) node[midway,above] {$0.5$};
      \draw[fluid,->] (s10) -- (s21) node[sloped,pos=0.2,above] {$0.5$};
      \draw[fluid,->] (s11) -- (s20) node[sloped,pos=0.7,above] {$0.5$};
      \draw[fluid,->] (s12) -- (s23) node[sloped,pos=0.2,above] {$0.5$};
      \draw[fluid,->] (s13) -- (s22) node[sloped,pos=0.7,above] {$0.5$};

      \draw[fluid,->] (s20) -- (s30) node[midway,below] {$0.5$};
      \draw[fluid,->] (s21) -- (s31) node[midway,above] {$0.5$};
      \draw[fluid,->] (s22) -- (s32) node[midway,below] {$0.5$};
      \draw[fluid,->] (s23) -- (s33) node[midway,above] {$0.5$};
      \draw[fluid,->] (s20) -- (s32) node[sloped,pos=0.15,above] {$0.5$};
      \draw[fluid,->] (s21) -- (s33) node[sloped,pos=0.9,below] {$0.5$};
      \draw[fluid,->] (s22) -- (s30) node[sloped,pos=0.8,above] {$0.5$};
      \draw[fluid,->] (s23) -- (s31) node[sloped,pos=0.2,below] {$0.5$};

      \draw[fluid,->] (s30) -- (o0) node[sloped,midway,below] {$0.5$};
      \draw[fluid,->] (s30) -- (o1) node[sloped,midway,above] {$0.5$};
      \draw[fluid,->] (s31) -- (o2) node[sloped,midway,below] {$0.5$};
      \draw[fluid,->] (s31) -- (o3) node[sloped,midway,above] {$0.5$};
      \draw[fluid,->] (s32) -- (o4) node[sloped,midway,below] {$0.5$};
      \draw[fluid,->] (s32) -- (o5) node[sloped,midway,above] {$0.5$};
      \draw[fluid,->] (s33) -- (o6) node[sloped,midway,below] {$1$};
      \draw[saturated,->] (s33) -- (o7) node[sloped,midway,above] {$0$};
      
      \begin{scope}[xshift=9cm,yshift=1.3cm]
        \foreach \n/\y/\c in {0/-0.25/0,1/0.25/0,2/0.75/1,3/1.25/1} {
          \node[term] (i\n) at (0,\y) {};
          \node[term] (o\n) at (3,\y) {};
          \draw (0,\y) node[anchor=east] {$\c$};
          \draw (3,\y) node[anchor=west] {$\c$};
        }
        \foreach \x/\y in {1/0,1/1,2/0,2/1} {
          \node[splitter] (s\x\y) at (\x,\y) {};
        }
        \draw[fluid,->] (i0) -- (s10) node[midway,sloped,below] {$0$};
        \draw[fluid,->] (i1) -- (s10) node[midway,sloped,above] {$0$};
        \draw[fluid,->] (i2) -- (s11) node[midway,sloped,below] {$0.5$};
        \draw[fluid,->] (i3) -- (s11) node[midway,sloped,above] {$0.5$};
        \draw[fluid,->] (s10) -- (s20) node[midway,below] {$0$};
        \draw[fluid,->] (s10) -- (s21) node[pos=0.2,sloped,above] {$0$};
        \draw[fluid,->] (s11) -- (s20) node[pos=0.8,sloped,above] {$0$};
        \draw[fluid,->] (s11) -- (s21) node[midway,above] {$1$};
        \draw[fluid,->] (s20) -- (o0) node[midway,sloped,below] {$0$};
        \draw[fluid,->] (s20) -- (o1) node[midway,sloped,above] {$0$};
        \draw[fluid,->] (s21) -- (o2) node[midway,sloped,below] {$0.5$};
        \draw[fluid,->] (s21) -- (o3) node[midway,sloped,above] {$0.5$};
        
      \end{scope}
    \end{tikzpicture}
  \end{center}
  \caption{On the left, the simple balancer of order $3$, with a steady-state that
    is not balanced when some output capacity is not $1$. The capacity
    of each input (resp. output) is given at their left (resp.
    right). On the right, a simple balancer of order $2$, with a steady-state with
    total throughput less than both the total input capacity and the total output
    capacity.}
  \label{fig:simple-balancer}
\end{figure}
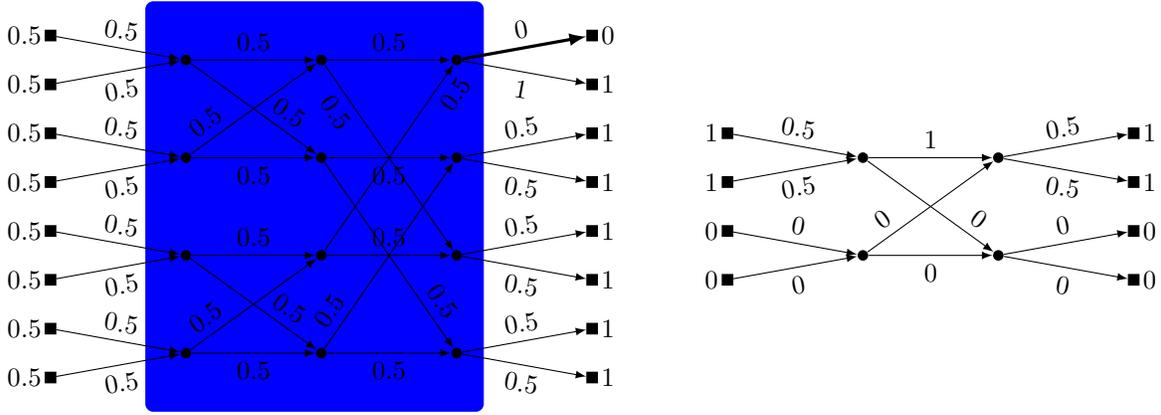

When $|I| = |O| = 2^k$, the \emph{simple balancer of order $k$} is a
balancer network. It can be defined recursively: a simple balancer of
order $k+1$ is made from two simple balancers of order $k$ in
parallel. We identify each pair of outputs with equal index from the
two balancers, creating a new splitter whose leaving arcs go to new
output nodes. The recursive process is highlighted by blue boxes in
\Cref{fig:simple-balancer}. A drawback of the simple balancer occurs
when the output capacities are not uniformly $1$. Then the balancing
property is lost, as can be seen on the network in the left side of
\Cref{fig:simple-balancer}.

Another limitation of simple balancers is that the total throughput at
steady-state is not as much as we could expect. A simple upper bound
on the total throughput is $\min \{ c(I), c(O)\}$. It is reasonable to
expect from a load-balancing network to always reach that bound.
However, simple balancers do not have this property, as shown by the
example on the right side of \Cref{fig:simple-balancer}. Improving
over the definition of simple balancer, the concept of
throughput-unlimited balancer imposes a maximized global throughput.

\begin{definition}
  A balancer $\splitnetwork$ is \emph{throughput-unlimited} if for any
  $c : I \disunion O \to [0,1]$, there is a steady-state $(t,F)$ for
  $(G,c)$ such that total throughput $t(I) = t(O)$ is maximized at
  $\min \{ c(I), c(O)\}$.
\end{definition}

\begin{figure}
  \centering
  \begin{tikzpicture}[x=0.7cm,y=0.7cm,>=latex]
    \fill[rounded corners=3pt,nearly transparent, blue] (0.9,-0.1) rectangle (7.1,7.1);
    \fill[rounded corners=3pt,nearly transparent, blue] (1.9,-0.1) rectangle (6.1,3.1);
    \fill[rounded corners=3pt,nearly transparent, blue] (1.9,3.9) rectangle (6.1,7.1);
    \foreach \y in {0,2,4,6} {
      \fill[rounded corners=3pt,nearly transparent, blue] ($(2.9,-0.1) + (0,\y)$) rectangle ($(5.1,1.1) + (0,\y)$);
    }
    \foreach \y in {0,1,...,7} {
      \fill[rounded corners=3pt,nearly transparent, blue] ($(3.9,-0.1) + (0,\y)$) rectangle ($(4.1,0.1) + (0,\y)$);
    }
    \foreach \y in {0,1,...,15} {
      \node[term] (i\y) at ($1/2*(0,\y) + (0,-0.25)$) {}; 
      \node[term] (o\y) at ($1/2*(0,\y) + (8,-0.25)$) {};
    }
    \foreach \x in {1,2,...,7} {
      \foreach \y in {0,1,...,7} {
        \node[splitter] (s\x\y) at (\x,\y) {};
      }
    }
    \foreach \y/\z in {0/0,1/0,2/1,3/1,4/2,5/2,6/3,7/3,8/4,9/4,10/5,11/5,12/6,13/6,14/7,15/7} {
      \draw[->,fluid] (i\y) -- (s1\z);
      \draw[->,fluid] (s7\z) -- (o\y);
    }
    \foreach \x/\w in {1/2,2/3,3/4,4/5,5/6,6/7} {
      \foreach \y in {0,1,...,7} {
        \draw[->,fluid] (s\x\y) -- (s\w\y);
      }
    }
    \foreach \y/\z in {0/4,1/5,2/6,3/7,4/0,5/1,6/2,7/3} {
      \draw[->,fluid] (s1\y) -- (s2\z);
    }
    \foreach \y/\z in {0/2,1/3,2/0,3/1,4/6,5/7,6/4,7/5} {
      \draw[->,fluid] (s2\y) -- (s3\z);
    }
    \foreach \y/\z in {0/1,1/0,2/3,3/2,4/5,5/4,6/7,7/6} {
      \draw[->,fluid] (s3\y) -- (s4\z);
    }
    \foreach \y/\z in {0/1,1/0,2/3,3/2,4/5,5/4,6/7,7/6} {
      \draw[->,fluid] (s4\y) -- (s5\z);
    }
    \foreach \y/\z in {0/2,1/3,2/0,3/1,4/6,5/7,6/4,7/5} {
      \draw[->,fluid] (s5\y) -- (s6\z);
    }
    \foreach \y/\z in {0/4,1/5,2/6,3/7,4/0,5/1,6/2,7/3} {
      \draw[->,fluid] (s6\y) -- (s7\z);
    }

  \end{tikzpicture}
  \caption{A Beneš network of order $4$ with the recursive structure being made explicit.}
  \label{fig:benes-network}
\end{figure}
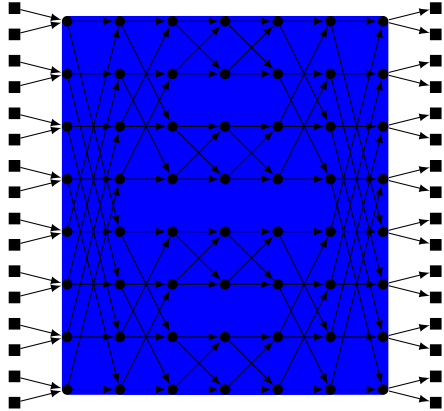

Notice that it has to be balancing only when the output capacities are
uniformly $1$. Beneš networks are throughput-unlimited networks with
$|O| = |I| = 2^k$. They can be described as gluing two simple
balancers, where the second balancer is reversed, see
\Cref{fig:benes-network}. Observe that Beneš networks are their own
reverses.

On the negative side, Beneš network are still not balancing when
output capacities are not uniformly $1$, for instance one could extend
the steady-state in the network on the left side of
\Cref{fig:simple-balancer} to a steady-state in a Beneš network with
the same throughputs. This calls for a stronger property, that a
network should be load-balancing and throughput-unlimited for any
capacity function. This is the notion of universal balancer.

\begin{definition}
  A splitter network $\splitnetwork$ is \emph{universally
    balancing} if for each capacity $c : I \disunion O \to [0,1]$,
  there is a steady-state $(t,F)$ and
  $\alpha, \beta \in \mathbb{R}_{\geq 0}$ such that 
  \begin{enumerate}[label=(\roman*),nosep]
  \item\label{ubn:input-fair} for each input $i$, $t(\delta^+(i)) = \min \{c(i), \alpha\}$,
  \item\label{ubn:output-fair} for each output $o$, $t(\delta^-(o)) = \min \{c(o), \beta\}$.
  \item\label{ubn:maximizing} the total throughput $T \eqdef t(\delta^+(I))$ equals
    $\min \{c(I), c(O)\}$.
  \end{enumerate}  
\end{definition}

In \Cref{sec:balancers}, we will show how to build a universal
balancer with $|I| = |O| = 2^k$. From such a universal balancer, by
ignoring any set of inputs and outputs (setting their capacities to
$0$), we can make balancers with arbitrary numbers of inputs and
outputs. We will also prove that every balancer presented here
contains $\Theta(n \log n)$ splitters where $n$ is the number of
inputs and outputs.

\begin{proposition}\label{prop:count-splitters}
  The number of splitters in the simple balancer, Beneš network and
  universal network of order $k$ are respectively
  $S(k) = k \cdot 2^{k-1}$, $B(k) = (2k-1) \cdot 2^{k-1}$, and
  $U(k) = (k+1)2^{k+2}$.
\end{proposition}

\subsection{Lower bounds on the number of splitters}.

Our next goal is to provide an $\Omega((n+p) \log (n+p))$ lower bound
on the number of splitters in a $(n,p)$-balancer. We begin with what
may seem as an unrelated problem: sampling in a discrete probability
distribution. Given a fair coin that can be tossed arbitrarily often,
how to choose an outcome in $\{1,\ldots,d\}$, with probabilities given
by a distribution $\pi \in [0,1]^d$? First, consider the case when
$\pi(i)$ is a rational for each $i \in \{1,\ldots,d\}$, say
$\pi(i) = p_i/q$ where $q$ is a common denominator. Then a sequence of
coin tossing can be described as a (possibly infinite) binary decision
tree, with each leaf labeled with a sampled value. Here we present a
construction of such a tree. Start from a single vertex, which serves
as the root. Grow the tree in repeated iterations. At each iteration,
add two children to every unlabelled leaf. As soon as the deepest
level of the tree contains at least $q$ leaves, label $p_i$ of these
leaves with $i$, for each $i \in \{1,\ldots,d\}$. Once labeled, each
leaf becomes definitive and will not grow anymore. The process goes on
by once again growing the unlabelled leaves, as long (possibly
infinitely) as some unlabelled leaf exists. After the tree is
completed, the tree can be optimized using a simple trick repeated
multiple times. If at any depth $d$, two leaves share a common label,
move them under a common parent, then replace these two leaves with a
single leaf at depth $d-1$ bearing the same label. This process can be
generalized to irrational probabilities, and gives a sampling
algorithm that minimizes the number of coins tossed:

\begin{theorem}[\cite{knuth1976complexity}]\label{thm:knuth-yao} Let
  $\pi \in [0,1]^d$ a discrete probability distribution (so
  $\allone \pi = 1$). Then the minimum expected number of coin tosses
  necessary to sample an element with probability distribution $\pi$
  is
  $\sum_{i=1}^d \sum_{k \in \mathbb{N}} \frac{k}{2^k} \binary[k]{\pi_i}$.
  This minimum is achieved by a binary decision tree where at each
  depth $k$ and for each $i \in \interval{1}{d}$, the number of leaves
  with label $i$ is $\binary[k]{\pi_i}$.
\end{theorem}

Consider a splitter network, and think of the flow as discrete,
arbitrarily small items. An item enters the network from some input,
then meets splitters repeatedly until reaching an output. When an item
arrives at a splitter with both outgoing arcs being fluid, it will
continue on any of the two outgoing arcs, without preference for one
over the other because the splitter is fair. It implies that, from the
perspective of this single item, the splitter network behaves like a
coin-tossing network, with each splitter corresponding to a coin toss. If
the network is a balancer, the sampled distribution is the uniform
distribution on $O$.

\begin{figure}
  \centering
  \begin{tikzpicture}[x=2cm,y=1.5cm,>=latex]
    \foreach \n in {1,2,3} {
      \node[term] (i\n) at (0,\n) {}; 
      \node[term] (o\n) at (3,\n) {};
    }
    \foreach \n/\x/\y in {a/1/0.5,b/1/2.5,c/2/0.5,d/2/2.5} {
      \node[splitter] (\n) at (\x,\y) {};
    }
    \draw[->,fluid] (i1) -- (a);
    \draw[->,fluid] (i2) -- (b);
    \draw[->,fluid] (i3) -- (b);
    \draw[->,fluid] (a) -- (c);
    \draw[->,fluid] (a) -- (d);
    \draw[->,fluid] (b) -- (c);
    \draw[->,fluid] (b) -- (d);
    \draw[->,fluid] (c) -- (o1);
    \draw[->,fluid] (d) -- (o2);
    \draw[->,fluid] (d) -- (o3);
    \draw[->,fluid] (c) .. controls (2.5,-0.33) and (2.5,-0.5) .. (2,-0.5) 
    -- (1,-0.5) .. controls (0.5,-0.5) and (0.5,-0.33) .. (a);
    \draw[blue,very thick] ($(i3.center) + (0,0.1)$) 
    -- ($(0.5,2.75) + (0,0.1)$) node[above=-3pt] {$1$}; 
    \draw[blue,line width=0.95pt] ($(0.5,2.75) + (0,0.1)$)
    .. controls ($(b.center) + (0,0.1)$) 
    .. ($(1.5,2.5) + (0,0.1)$) node[above=-3pt] {$1/2$}; 
    \draw[blue,thick] ($(1.5,2.5) + (0,0.1)$) 
    .. controls ($(d.center) + (0,0.1)$) 
    .. ($(2.5,2.75) + (0,0.1)$);
    \draw[blue,thick] ($(2.5,2.75) + (0,0.1)$) 
    -- ($(o3.center) + (0,0.1)$) node[above=-3pt,sloped,pos=0.8] {$1/4$};;
    \draw[blue,thick] ($(1.5,2.5) + (0,0.1)$) 
    .. controls ($(d.center) + (0,0.1)$) 
    .. ($(2.5,2.25) + (0,0.1)$); 
    \draw[blue,thick] ($(2.5,2.25) + (0,0.1)$) 
    -- ($(o2.center) + (0,0.1)$) node[above=-3pt,sloped,pos=0.8] {$1/4$};
    \draw[blue,line width=0.95pt] ($(0.5,2.75) + (0,0.1)$) 
    .. controls ($(b.center) + (0,0.1)$) 
    .. ($(1.5,1.5) + (0,0.1)$) node[above=-3pt,sloped,pos=0.8] {$1/2$}; 
    \draw[blue,thick] ($(1.5,1.5) + (0,0.1)$) 
    .. controls ($(c.center) + (0,0.1)$)
    .. ($(2.5,0.75) + (0,0.1)$) node[above=-3pt] {$1/4$}; 
    \draw[blue,thick] ($(2.5,0.75) + (0,0.1)$) -- ($(o1.center) + (0,0.1)$);
    \draw[blue,thick] ($(1.5,1.5) + (0,0.1)$) 
    .. controls ($(c.center) + (0,0.1)$)
    .. (2.2,0.3)
    .. controls (2.55,0.-0.23) and (2.6,-0.6) .. (2,-0.6) 
    -- (1,-0.6) node[midway,below=-3pt] {$1/4$}
    .. controls (0.4,-0.6) and (0.4,-0.23) 
    .. (0.8,0.3) [semithick] 
    .. controls ($(a.center) + (-0.05,+0.03)$)
    .. ($(1.5,1.5)$) node[pos=0.6,above=-3pt,sloped] {$1/8$};
    \draw[blue,thick] (1,-0.6)
    .. controls (0.4,-0.6) and (0.4,-0.23)
    .. (0.8,0.3) [semithick]
    .. controls (0.9,0.4) and ($(a.center) + (0.2,-0.1)$) 
    .. ($(1.5,0.5) - (0,0.1)$) node[below=-3pt] {$1/8$};
    \draw[blue,thin] ($(1.5,1.5)$)
    .. controls ($(d.center) - (0,0.1)$) 
    .. ($(2.5,2.75) - (0,0.1)$) 
    -- ($(o3.center) - (0,0.1)$) node[pos=0.8,below=-3pt,sloped] {\small $1/16$}; 
    \draw[blue,thin] ($(1.5,1.5)$)
    .. controls ($(d.center) - (0,0.1)$) 
    .. ($(2.5,2.25) - (0,0.1)$) 
    -- ($(o2.center) - (0,0.1)$) node[pos=0.8,below=-3pt,sloped] {\small $1/16$}; 
    \draw[blue,thin] ($(1.5,0.5) - (0,0.1)$)
    .. controls ($(c.center) - (0,0.1)$) 
    .. ($(2.5,0.75) - (0,0.1)$)
    -- ($(o1.center) - (0,0.1)$) node[pos=0.8,below=-3pt,sloped] {\small $1/16$};
    \draw[blue,thin] ($(1.5,0.5) - (0,0.1)$)
    .. controls (1.7,0.4) and ($(c.center) - (0,0.1)$) 
    .. (2.12,0.23)
    .. controls ($(2.24,0.06) - (0,0.1)$) and ($(2.4,-0.5) + (0,0.1)$) 
    .. ($(2,-0.5) + (0,0.1)$);
    \draw[blue,very thin,dashed] ($(2,-0.5) + (0,0.1)$)
    -- ($(1,-0.5) + (0,0.1)$) node[midway,above=-3pt,sloped] {\small $1/16$};
    \draw (o1) node[anchor=west] {c};
    \draw (o2) node[anchor=west] {b};
    \draw (o3) node[anchor=west] {a};
    \begin{scope}[xshift=-5cm,x=0.7cm,y=0.7cm]
      \foreach \n/\x/\y in {a/0/4,b/-1.5/3,c/1.5/3,g/2.5/2,h/1,i/4/1}
      {
        \node[splitter,blue] (\n) at (\x,\y) {};
      } 
      \foreach \n/\x/\y in {d/-2.5/2,e/-0.5/2,f/0.5/2,j/0/0,k/2/0,l/3/0}
      {
        \node[term,blue] (\n) at (\x,\y) {};
      } 
      \foreach \u/\v in {a/b,a/c} {
        \draw[blue,line width=0.95] (\u) -- (\v) node[sloped,midway,above=-3pt] {$1/2$};
      }
      \foreach \u/\v in {b/d,b/e,c/f,c/g} {
        \draw[blue,thick] (\u) -- (\v) node[sloped,midway,above=-3pt] {$1/4$};
      }
      \foreach \u/\v in {g/h,g/i} { 
        \draw[blue,semithick] (\u) -- (\v) node[sloped,midway,above=-3pt] {$1/8$};
      }
      \foreach \u/\v in {h/j,h/k,i/l} {
        \draw[blue,thin] (\u) -- (\v) node[sloped,midway,above=-3pt] {\small $1/16$};
      }

      \draw[blue,thin] (i) -- (4.5,0.5) [dashed] -- (5,0);
      \draw ($(d) - (0,1em)$) node[anchor=base] {a};
      \draw ($(e) - (0,1em)$) node[anchor=base] {b};
      \draw ($(f) - (0,1em)$) node[anchor=base] {c};
      \draw ($(j) - (0,1em)$) node[anchor=base] {a};
      \draw ($(k) - (0,1em)$) node[anchor=base] {b};
      \draw ($(l) - (0,1em)$) node[anchor=base] {c};
    \end{scope}
  \end{tikzpicture}
  \caption{The infinite decision tree (in blue) used to sample
    uniformly over a three-element set $\{a,b,c\}$ can be embedded
    from any input into a $(3,3)$-balancer. Moreover, the sum of the
    probabilities of the 3 trees, one from each input, will be at most
    one on any arc, which shows that this network is indeed a simple
    balancer.}
  \label{fig:embed-decision-tree}
\end{figure}
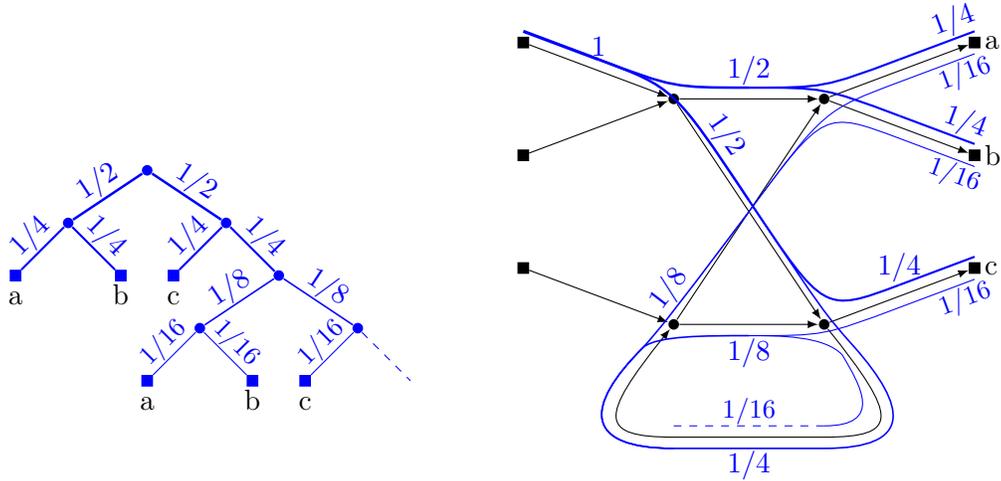

Formally, when all the arcs remains fluid, increasing a single input
capacity from $0$ to $1$ results in a non-decreasing throughput on
each arc. Because all arcs are still fluid, the sub-steady-state
algorithm performs a single iteration. Therefore the increase in
throughputs follows a single stationary circulation. As illustrated on
\Cref{fig:embed-decision-tree}, it is obtained from the embedding of a
binary decision tree $T$ onto the splitter network. The increase in
throughput on an arc $e$ is the sum of probabilities of the edges
mapped to $e$. Furthermore, in a balancer network, the increase of
throughput is the same on every output. This implies that, as we
progressively increase each input capacity from $0$ to $1$, each
binary decision tree must uniformly sample from $O$.

In each binary decision tree, label each edge $e$ with the probability
of its usage during sampling. The sum of these labels represents the
expected number of tosses, and can be bounded as shown in
\Cref{thm:knuth-yao}. When mapped into the splitter network, for an
arc $e$, the sum of these labels on each edge of the tree mapped to
$e$ is the additional throughput on $e$. By summing over all the
binary decision tree, we get that the sum of all labels is at most the
number of outgoing arcs of all splitters, that is $2|S|$. Applied on
balancers, it yields:

\begin{theorem}\label{thm:weak-lower-bound}
  Let $\splitnetwork$ be an $(n,p)$-balancer, such that when all input
  capacities are 1, the steady-state has no saturated arc. Then
\[|S| \geq \frac{1}{2} |I| |O| \sum_{k \in \mathbb{N}} \frac{k}{2^k} \binary[k]{\frac{1}{|O|}} \]
\end{theorem}

For a balancer with $|I| = |O| = 2^k$, since
$\sum_{k \in \mathbb{N}} \frac{k}{2^k} \binary[k]{\frac{1}{|O|}} = \frac{k}{2^k}$,
we get a lower bound of $k 2^{k-1}$ splitters, matching the value of
$S(k)$. Therefore the simple balancer of order $k$ is optimal among
all balancer networks without any saturated arcs in their
steady-states. By extending this argument to steady-states with
saturated arcs, we can remove that restriction, albeit at the cost of
halving the lower bound.

Consider the various configurations of fluid and saturated arcs
incident to a splitter, illustrated in \Cref{fig:routing-flow}. If a
splitter has two fluid outgoing arcs, any additional flow is evenly
distributed between the two outputs, akin to the probabilities of a
coin toss. If a splitter has two incoming saturated arcs, by
rule~\ref{eqn:maximization}, its outgoing arcs are saturated or at
full capacity. In an augmenting circulation, the throughput on those
arcs may only decrease by the same quantity by
rule~\ref{eqn:incoming-rule}: the splitter still acts as a coin toss,
but on the flow that is pushed back. Otherwise, a positive change in
throughput on an incoming arc will be followed by an increase on a
single outgoing fluid arc or a decrease on a single incoming saturated
arc. Similarly a negative change of throughput on an outgoing
saturated arc will impact only one other arc. Any additional unit of
flow entering the splitter would be routed deterministically.
Therefore, in the embedding of a binary decision tree into the
splitter network, a node cannot be mapped to such a vertex, and no
coin toss occurs here. Thus any splitter, depending on which of its
incident arcs are fluid, acts as either a coin toss or a deterministic
router. Thus, even in the presence of saturated arcs, we can embed a
binary decision tree, by mapping each edge to a directed path in the
residual graph. The inner nodes of any such path are
\emph{deterministic} splitters, while its extremities are
\emph{tossing} splitters. By \Cref{coro:monotonicity}, the throughput
on each arc increases until it becomes saturated, then decreases.
Therefore its throughput varies by at most $2$ during the whole
algorithm. This limits the extent to which an arc can be utilized by
the embeddings of binary decision trees, leading us to the following
conclusion:

\begin{theorem}\label{thm:lower-bound}
  Let $\splitnetwork$ be an $(n,p)$-balancer. Then
  \[|S| \geq \frac{1}{4} |I| |O| \sum_{k \in \mathbb{N}} \frac{k}{2^k} \binary[k]{\frac{1}{|O|}} \]
\end{theorem}

We will formalize this discussion and prove the theorem in
\Cref{sec:lowerbounds}

\section{The \texorpdfstring{$\Cequal$}{C=}-circulation
  problem}\label{sec:cequal-problem}

Let $G = (V,E)$ be a directed graph. We consider a partition
$\Cequal \subseteq 2^E$ of the arcs, such that each $C \in \Cequal$ is
a subset of the outgoing arcs $C \subseteq \delta^+(v)$ of some vertex
$v \in V$. Two arcs $e', e'' \in E$ are considered
\emph{$\Cequal$-coupled} (or just \emph{coupled} when no ambiguity
arises) if there is $C \in \Cequal$ such that $e', e''$ belongs to $C$
(an arc is coupled to itself). An arc $e'$ is \emph{single} if
$\{e'\} \in \Cequal$, indicating that it is not coupled to another
arc. The tuple $(G,e,\Cequal)$, where $e \in E$, is then an instance
of the \emph{$\Cequal$-circulation} problem: find a circulation in $G$
that is non-zero on $e$ and is constant on any set of $\Cequal$. It
can be expressed as finding a vector $x$ with $x_{e} > 0$ satisfying
the following linear system:

\begin{equation}
  \left\{\begin{array}{rcll}
        x(\delta^+(v)) - x(\delta^-(v)) & = & 0 & \qquad(v \in V)\\
        x_e - x_{e'} & = & 0 & \qquad(e, e' \in C \in \Cequal) \\
        x & \geq & 0 &
           
         \end{array}\right.
  \label{eqn:eqcirculation}
\end{equation}

\begin{theorem}\label{thm:eqcirculation}
  An instance $(G = (V,E),ts,\Cequal)$ of the $\Cequal$-circulation
  problem has a solution $x$ with $x_{ts} > 0$ if and only if there is
  no set $S \subseteq V \setminus \{t\}$ with a partition
  $S = S_0 \disunion S_1 \ldots \disunion S_k$ where $s \in S_0$
  and for any arc $e \in \delta^+(S_i)$, there is an arc $e'$ coupled
  to $e$ (possibly $e = e'$) such that $e' \in E[S_i,S_j]$ and $i < j$.
\end{theorem}

\begin{proof}
  Suppose $S = S_0 \disunion S_1 \ldots \disunion S_k$ exist and let
  $x$ denote a solution to \Cref{eqn:eqcirculation}. We prove by induction
  from $k$ to $0$ that for all $i \in \interval{0}{k}$,
  $x(\delta(S_i)) = 0$. This will imply that $x_{ts} = 0$ as
  $ts \in \delta^-(S_0)$. By assumption, $\delta^+(S_k) = \emptyset$,
  and because $x$ is a circulation, $x(\delta^-(S_k)) = 0$, proving
  the base case. Let $i \in \interval{1}{k-1}$ and suppose that we have for all
  $j \in \interval{i+1}{k}$, $x(\delta(S_j)) = 0$. Let
  $uv \in \delta^+(S_i)$, then there is an arc $uw \in E[S_i,S_j]$
  with $j > i$, where $uw$ is either equal to or coupled with $uv$, hence by
  the induction hypothesis $x_{uv} = x_{uw} = 0$. Thus
  $x(\delta^+(S_i)) = 0$. Again, by $x$ being a circulation, we get
  that $x(\delta^-(S_i)) = 0$.

  Now suppose that there is no $x \in \mathbb{R}_{\geq 0}^E$
  satisfying \Cref{eqn:eqcirculation} and $x_{ts} > 0$. By Farkas
  lemma, this implies that the following linear system has a solution
  $y \in \mathbb{R}^V$:

  \begin{equation}
    \makebox[0pt]{
      \begin{minipage}{\linewidth}
        \begin{IEEEeqnarray*}{rCll}
          y_s - y_t & \geq & 1 & \\
          \sum_{uv \in C} y_v - y_u & \geq & 0 & \qquad(C \in \Cequal) 
        \end{IEEEeqnarray*}
      \end{minipage}
    }
    \label{eqn:eqcirculation-dual}
  \end{equation}
  
  We may partition $V$ into $S_{-l}, S_{-l+1}, \ldots, S_k$, the
  equivalence classes defined by the relation $u \sim v$ if
  $y_u = y_v$, ordered based on increasing $y$-values, and such that
  $s \in S_0$. Because $y_t < y_s$,
  $t \notin S_0 \cup S_1 \ldots \cup S_k$.

  Let $i \in \interval{0}{k}$ and let $uv \in \delta^+(S_i)$. Let
  $C \in \Cequal$ with $uv \in C$. Then
  $\sum_{uw \in C} y_w - y_u \geq 0$ implies that either $y_w = y_u$
  for all $uw \in C$, which is not the case as $uv \in \delta^+(S_i)$
  hence $y_u \neq y_v$, or there is some $uw \in C$ with $y_w > y_u$.
  But then $w \in S_j$ with $j > i$, proving that the partition $S$
  satisfies the desired property.
\end{proof}

\begin{corollary}
  An instance $(G = (V,E),ts,\Cequal)$ of the $\Cequal$-circulation
  problem where $G$ is strongly connected has a solution $x$ with
  $x_{ts} > 0$.
\end{corollary}

\begin{proof}
  By contraposition, consider an instance lacking a non-zero
  circulation, and let $S_0 \cup \ldots \cup S_k$ be the partition
  given by \Cref{thm:eqcirculation}. Then $\delta^+(S_k)$ is empty,
  indicating that $G$ is not strongly connected.
\end{proof}

We can compute efficiently a $\Cequal$-circulation in the strongly
connected case. We do so by restricting the problem to a spanning
subgraph in which the classes in $\Cequal$ coincides with the leaving
arcs of each vertex.

\begin{lemma}\label{lemma:eqcirculation-SC}
  There is an algorithm that, given an instance
  $(G=(V,E), ts, \Cequal)$ of the $\Cequal$-circulation problem with
  $G$ strongly connected, find a feasible solution $x$ with
  $x_{ts} > 0$ in time $O(|E| + \stationary(G))$, where
  $\stationary(G)$ is the complexity of computing a stationary
  distribution on any subgraph of $G$.
\end{lemma}

\begin{proof}
  Let $T$ denote a maximal in-arborescence of $G$ rooted at $t$. Let
  $$E' := \{ e \in E~:~\textrm{$e$ is coupled to some arc
    in $T \cup \{ts\}$}\}.$$ By construction, $t$ is reachable from
  any vertex in $(V,E')$. Because $ts \in E'$, $s$ and $t$ are in the
  same strongly connected component of $(V,E')$. Let $H = (V_H, E_H)$
  be the strongly connected component of $(V,E')$ containing $s$ and
  $t$. Because any vertex in $T$ has out-degree at most 1, and by our
  choice of $E'$, for any $v \in V$,
  $\delta^+(v) \cap E' \in \Cequal$. Let
  $\Cequal_H := \{ C \in \Cequal : C \subseteq E_H\}$.

  We can rewrite the problem of finding a $\Cequal_H$-circulation on
  $H$, by setting $y_v \eqdef x_e$ for $e \in \delta^+(v)$:
  \begin{IEEEeqnarray*}{rCl}
    \sum_{uv \in \delta^-(v)} y_u & = & d^+(v) y_v, \\
    y & \geq & 0,
  \end{IEEEeqnarray*}
  and then setting $\pi_v = d^+(v) y_v$:
  \begin{IEEEeqnarray*}{rCl}
    \sum_{uv \in \delta^-(v)} \frac{\pi_u}{d^+(u)} & = & \pi_v, \\
    \pi & \geq & 0.
  \end{IEEEeqnarray*}
  We recognize the last system as the one defining a stationary
  distribution of a Markov chain over $H$, provided that we add the
  constraint $\allone \pi = 1$. As $H$ is strongly connected, the
  stationary distribution exists and is nowhere zero. Thus we obtain a
  non-zero $\Cequal_H$-circulation for $H$. We extend it into a
  non-zero $\Cequal$-circulation for $G$ by setting $x_e = 0$ for any
  edge not in $E_H$.
\end{proof}

The strongly connected case serves as a basis for an algorithm solving
the general problem.

\begin{lemma}\label{lemma:algo-eqcirculation}
  There is an algorithm that, given an instance
  $(G = (V,E), ts, \Cequal)$ of the $\Cequal$-circulation problem, in
  time $O(|E|\log^4 |V| + \stationary(G))$, either find a feasible
  solution $x$ with $x_{ts} > 0$ or correctly assert that none exists.
  Here $\stationary(G)$ is the complexity of computing a stationary
  distribution on any subgraph of $G$.
\end{lemma}

\begin{proof}
  We start by computing a strongly connected subgraph $H$ of $G$
  containing $ts$ and ensuring that for any $C \in \Cequal$, either
  $C \subseteq E(H)$ or $C \cap E(H) = \emptyset$. If no such $H$
  exists, then the $\Cequal$-circulation instance does not admit a
  non-zero solution. If $H$ exists, we reduce the problem of finding
  $x$ to computing a stationary distribution in $H$.

  To compute $H$, we use a dynamic decremental single-sink
  reachability algorithm with sink $t$. If ever $t$ becomes
  unreachable from $s$, then we conclude that the instance does not
  admit a non-zero solution. The algorithm keeps a queue of vertices
  to delete, initially the vertices from which $t$ is not reachable.
  While the queue is non-empty, it takes a vertex from it and delete
  its incident arcs and all the arcs coupled to them. As soon as it
  detects that some vertex $v$ cannot reach $t$, it queues that vertex
  $v$ for deletion. When the queue is empty, remove all the vertices
  not reachable from $s$ and return the remaining subgraph $H$.

  First, suppose that this algorithm fails to build $H$ because $t$
  becomes unreachable from $s$. Each time a subset of vertices becomes
  disconnected from $t$, we label it $S_i$ (with decreasing index $i$,
  adjusting the value of the starting index at the end of the
  algorithm). When such a component appears, all its leaving arcs must
  have been removed, which means they are coupled to an arc entering
  some already removed component. Hence we get the sequence
  $S_0, S_1 \ldots S_k$ proving that a non-zero $\Cequal$-circulation
  cannot exist.

  Suppose now that the algorithm returns a subgraph $H = (V_H,E_H)$.
  By construction, for any $C \in \Cequal$, either $C \subseteq E_H$
  or $C \cap E_H = \emptyset$. Indeed, if any arc of $C$ is removed
  during the main loop, then $C$ is removed in the same iteration. If
  some arc of $C$ is removed during the final phase, when vertices
  unreachable from $s$ are removed, then the vertex $v$ for which
  $C \subseteq \delta^+(v)$ must not be reachable from $s$ and thus
  $C$ was removed.

  Next we claim that $H$ is strongly connected. By construction, for
  any vertex $v$, there is a path from $t$ to $v$ starting with $ts$.
  Moreover, notice that before removing the vertices unreachable from
  $s$, there was a path from $v$ to $t$. Thus any vertex on this path
  is also reachable from $s$ and is kept during the last phase. Hence $t$
  is reachable from $v$, proving that $H$ is strongly connected. We
  then apply \Cref{lemma:eqcirculation-SC} to compute $x$, and extend
  it to $G$ by setting $x_e = 0$ for any arc not in $H$. The
  complexity follows by using the decremental single-sink reachability
  algorithm from~\cite{bernstein2019decremental}.
\end{proof}

The next lemma gives a sufficient condition for the existence of a
non-zero $\Cequal$-circulation, when we may choose any arc to be
positive instead of the specific arc $ts$ as above.

\begin{lemma}\label{lemma:circulation-or-sink}
  Given a connected directed graph $G = (V,E)$ and a partition
  $\Cequal$ of $E$ that is a refinement of the out-incidencies of $G$,
  then
  \begin{itemize}[nosep,label=\bull]
  \item either there is a non-zero $\Cequal$-circulation in $G$,
  \item or there is a vertex $v \in V$ with $\delta^+(v) = \emptyset$,
  \end{itemize}
  and we can find one or the other in time $O(|E| + \stationary(G))$.
\end{lemma}

\begin{proof}
  One can find in time $O(|E|)$ a strongly connected component $X$ of
  $G$ that is a sink component: $\delta^+(X) = \emptyset$. If
  $|X| = 1$ we are done, because $v \in X$ is a sink. Now, assume that
  $|X| > 1$.

  Because for any $C \in \Cequal$, there is a vertex $v$ with
  $C \subseteq \delta^+(v)$, for any arc $e \in E[X]$, any arc coupled
  to $e$ is also in $E[X]$. Hence, a $\Cequal$-circulation $x$ of
  $G[X]$ can be extended to a $\Cequal$-circulation on $G$ by setting
  $x_e = 0$ for each $e \notin E[X]$. Such a $\Cequal$-circulation
  on $G[X]$ exists and can be computed in time $O(|E| + \stationary(G))$ by
  \Cref{lemma:eqcirculation-SC}.  
\end{proof}

\section{Computing a steady-state}\label{sec:algo}

In this section, we give two algorithms to compute a steady-state in a
splitter network.

\subsection{A stronger maximization rule}

The algorithms use a slightly stronger property than
rule~\ref{eqn:maximization}:

\begin{enumerate}[label=\labelcref*{eqn:maximization}S]
\item\label{eqn:strong-maximization} for any
  arcs $uv \in E \setminus F$ and $vw \in F$, $t(vw) = 1$.
\end{enumerate}

Clearly rule~\ref{eqn:strong-maximization} implies
rule~\ref{eqn:maximization}. The two rules are actually almost
equivalent:

\begin{lemma}\label{lemma:equivalence-strong-maximization}
  If $(t,F)$ is a steady-state, then there is a steady-state $(t,F')$
  that satisfies rule~\ref{eqn:strong-maximization}, with
  $F \subseteq F'$.
\end{lemma}

\begin{proof}
  By induction on the number of splitters on which
  rule~\ref{eqn:strong-maximization} is not true. Let $uv \notin F$,
  $vw \in F$ with $t(vw) < 1$, by rule~\ref{eqn:maximization}
  $t(uv) = 1$. Let $u'v$ be the arc in-coupled to $uv$. If $u'v$ is
  fluid, then $(t,F \cup \{uv\})$ is a steady-state as
  rule~\ref{eqn:incoming-rule} is checked on $v$ and
  rule~\ref{eqn:outgoing-rule} is checked on $u$. If $u'v$ is
  saturated, by rule~\ref{eqn:incoming-rule}, $t(u'v) = 1$, then
  $(t,F \cup \{uv,u'v\})$ is a steady-state for similar reasons.
  Notice that in both cases, the modification preserves
  rule~\ref{eqn:strong-maximization} on any splitter for which it
  held. Hence the number of those splitters increases by one.
\end{proof}

\subsection{An push-relabel-like algorithm to compute steady-states}\label{sec:push-relabel}

In analogy with a pre-flow in a push-relabel max-flow algorithm, we
relax the conservation rule~\ref{eqn:conservation}, to define:

\begin{definition}
  Given a splitter network $\splitnetwork$ with capacities
  $c: I \disunion O \to [0,1]$, a \emph{pre-steady-state} for $(G,c)$
  is a pair $(t,F)$ satisfying~\ref{eqn:throughput},
  \ref{eqn:fluid-arcs}, \ref{eqn:input-capacity},
  \ref{eqn:output-capacity}, \ref{eqn:incoming-rule},
  \ref{eqn:outgoing-rule}, and \ref{eqn:strong-maximization}, and such
  that for each splitter $s \in S$,
  \[ t(\delta^+(s)) \leq t(\delta^-(s)).\]
\end{definition}

The next claim which can be readily checked, asserts the existence of
a simple pre-steady-state.

\begin{claim}\label{claim:starting-preeq}
  The pair $(t,E(G))$ is a pre-steady-state for $(G,c)$, where
  $$t: uv \to \left\{
    \begin{array}{ll}
      c(u) & \textrm{if $u \in I$,} \\
      0 & \textrm{otherwise}
    \end{array}\right. 
  $$
\end{claim}

We proceed by iteratively improving upon this initial pre-steady-state.
Let $\splitnetwork$ be a splitter network with capacities
$c : I \disunion O \to [0,1]$. Let $F \subseteq E$ be a set of fluid
arcs. If $(t,F)$ is a pre-steady-state, then $t$ is a feasible solution
to the following linear program:
\begin{equation}\tag{PSS}\label{eqn:presteady-state-lp}
\left|\begin{array}{ll@{\hspace{1.5cm}}ll}
  \max t(\delta^-(O)) &\textrm{subject to} & & \\
  0 \leq t \leq 1 & (e \in E) 
& t(\delta^+(s)) \leq t(\delta^-(s)) & (s \in S) \\
  t(is) \leq c(i) & (is \in \delta^+(I)) 
& t(is) = c(i) & (is \in \delta^+(I) \cap F) \\
  t(so) \leq c(o) & (so \in \delta^-(O)) 
& t(so) = c(o) & (so \in \delta^-(O) \setminus F) \\
  t(wv) \leq t(uv) & (uv \in E \setminus F, wv \in E) 
& t(uw) \leq t(uv) & (uv \in F, uw \in E) \\
  t(vw) = 1 & (uv \in E \setminus F, vw \in F)
& & 
\end{array}\right.
\end{equation}

Conversely, the following lemma is immediate:

\begin{lemma}
  Given $\splitnetwork$ a splitter network,
  $c : I \disunion O \to [0,1]$ a capacity function and
  $F \subseteq E$ a set of fluid arcs, for any feasible solution $t$
  of~\eqref{eqn:presteady-state-lp}, $(t,F)$ is a pre-steady-state of
  $(G,c)$.
\end{lemma}

Our algorithm is based on the next lemma.

\begin{lemma}\label{lemma:presteady-state-lp}
  Let $\optt$ be an optimal solution
  to~\eqref{eqn:presteady-state-lp} such that
  $\sum_{e \in F} t(e) - \sum_{e \in E \setminus F} t(e)$ is
  maximized. Then either $(\optt, F)$ is a steady-state, or there is
  an arc $e \in F$ such that $(\optt, F \setminus \{e\})$ is a
  pre-steady-state.
\end{lemma}

\begin{proof}
  Assume that $(\optt,F)$ is not a steady-state: there is a vertex
  $s \in S$ with $\optt(\delta^-(s)) - \optt(\delta^+(s)) > 0$. Let
  $\{e_1,e_2\} = \delta^-(s)$ with $\optt(e_1) \leq \optt(e_2)$, and
  $\{e_3, e_4\} = \delta^+(s)$ with $\optt(e_3) \leq \optt(e_4)$.

  \begin{enumerate}[label=Case \arabic*:,ref=\arabic*,wide]
  \item\label{case:leavingcoupled} there is a fluid leaving arc
    $e \in \{e_3,e_4\} \cap F$ such that $e$ is in-coupled to an arc
    $e' \in E \setminus F$ and $\optt(e) = \optt(e')$. Then
    $(\optt,F \setminus \{e\}$) is a pre-steady-state, as
    rules~\ref{eqn:incoming-rule} and~\ref{eqn:strong-maximization}
    are clearly still satisfied.
  \item\label{case:leavingoutput} there is a fluid leaving arc
    $e \in \{e_3,e_4\} \cap F$ such that $e \in \delta^-(o)$ for some
    output $o \in O$, and $\optt(e_3) = c(o)$. Then
    $(\optt,F \setminus \{e\}$) is a pre-steady-state, as
    rules~\ref{eqn:output-capacity} and~\ref{eqn:outgoing-rule} are
    clearly still satisfied.\medskip

    We may now assume that the fluid leaving arcs do not check the
    conditions for cases~\ref{case:leavingcoupled}
    and~\ref{case:leavingoutput}. Consequently, increasing the output
    $\optt(e)$ to $\optt(e) + \varepsilon$ for some sufficiently small
    $\varepsilon$ do not break any rule on the destination node of $e$.
  \item\label{case:fluidleavings} $e_3$ and $e_4$ are both fluid, with
    $\optt(e_3) = \optt(e_4) < 1$. Then, for some $\varepsilon > 0$
    sufficiently small, $(\optt + \varepsilon \chi_{\{e_3,e_4\}},F)$
    is a pre-steady-state, because we increase the throughput on both
    arcs uniformly, preserving rule~\ref{eqn:outgoing-rule}. achieving
    a better objective than $\optt$. The throughput of this
    pre-steady-state improves over $\optt$, a contradiction.
  \item\label{case:oneleavingfluid} exactly one of $e_3$, $e_4$ is
    fluid with throughput strictly less that one, and we may assume it
    is $e_4$ by rule~\ref{eqn:outgoing-rule}. Then, for some
    $\varepsilon > 0$ sufficiently small,
    $\optt + \varepsilon \chi_{\{e_4\}}$ is a better solution then
    $\optt$, again a contradiction.\medskip

    We may now assume that $e_3 \notin F$ or $\optt(e_3) = 1$, and that
    $e_4 \notin F$ or $\optt(e_4) = 1$.

  \item\label{case:e2fluid} $e_2 \in F$. Then
    $(\optt,F \setminus \{e_2\})$ is a pre-steady-state, as
    rules~\ref{eqn:incoming-rule} and~\ref{eqn:strong-maximization}
    are both satisfied.\medskip

    We may now assume that $e_2 \in E \setminus F$.

  \item\label{case:e1fluid} $e_1 \in F$ and $\optt(e_1) = \optt(e_2)$.
    Then $(\optt,F \setminus \{e_1\})$ is a pre-steady-state, as
    rules~\ref{eqn:incoming-rule} and~\ref{eqn:strong-maximization}
    are satisfied.

  \item\label{case:e1sat} $e_1 \in E \setminus F$. Then
    $\optt - \varepsilon \chi_{\delta^-(s)}$ is a better solution than
    $\optt$ for some $\varepsilon > 0$ sufficiently small, which leads
    to a contradiction.

  \item\label{case:final} $\optt(e_1) < \optt(e_2)$, $e_1 \in F$,
    $e_2 \notin F$. Then, for some $\varepsilon > 0$ sufficiently
    small, $\optt - \varepsilon \chi_\{e_2\}$ is a better solution
    to~\eqref{eqn:presteady-state-lp} than $\optt$, again a
    contradiction. \qedhere
  \end{enumerate}
\end{proof}

This leads to the iterative algorithm of repeatedly solving the LP and
removing an arc from $F$, that stops after at most $|E|+1$ iterations,
and solves $|E| + 1$ linear programs of polynomial size in the worst
case. 

Notice that the proof still holds with additional constraints
$t(e) \geq l_e$ for any fluid arc $e$, and $t(e) \leq u_e$ for any
saturated arc $e$. This is because the algorithm tries to increase the
throughput on fluid arcs, and decrease it on saturated arcs.
Therefore, as long as the algorithm can be initialized with a feasible
solution, we can find a steady-state that also checks these
constraints. The feasible solution is a pre-steady-state that already
satisfies those additional constraints. This implies the following
corollary.

\begin{corollary}\label{coro:mono-pre-steady}
  Given $\splitnetwork$ a splitter network,
  $c : I \disunion O \to [0,1]$ a capacity function and $(t_0,F_0)$ a
  pre-steady-state for $G,c$. Then there exists steady-state $(t,F)$
  for $G,c$ such that 
  \begin{enumerate}[label=(\roman*),nosep]
  \item $F \subseteq F_0$;
  \item for each arc $e \in F$, $t(e) \geq t_0(e)$;
  \item for each arc $e \in E \setminus F_0$, $t(e) \leq t_o(e)$.
  \end{enumerate}
\end{corollary}

\begin{proof}
  Find a steady-state with the additional constraints
  $t(e) \leq t_0(e)$ for each $e \in E \setminus F_0$, and
  $t(e) \geq t_0(E)$ as long as $e$ is a fluid arc.
\end{proof}

\subsection{A blocking-flow-like algorithm to compute steady-states}\label{sec:blocking-flow-algo}




Let $(t,F)$ be a sub-steady-state for a splitter network
$\splitnetwork$ with capacities $c : I \disunion O \to [0,1]$. Similar
to the blocking-flow algorithm, we employ a notion of residual graph.
We notice that throughput values may be bounded by
rule~\ref{eqn:incoming-rule}, which governs the throughput of arcs
entering a splitter, and~\ref{eqn:outgoing-rule}, which bounds the
throughput leaving a splitter. Let $s$ be a splitter, with
$\delta^-(s) = \{e, e'\}$, where $e \in F$ and $e' \in E \setminus F$,
and $t(e) = t(e')$, then $t(e)$ is lower-bounded and $t(e')$ is
upper-bounded by rule~\ref{eqn:incoming-rule}. We say that $e$ is
\emph{upper-tight} and $e'$ is \emph{lower-tight}. For any output
$o \in O$, with an incoming fluid arc $e \in \delta^-(o)$, if
$t(e) = c(o)$, $e$ is also \emph{upper-tight}. Furthermore,
rules~\ref{eqn:incoming-rule} and~\ref{eqn:outgoing-rule} imposes some
arcs to have equal throughput: out-coupled fluid arcs, or in-coupled
saturated arcs. Any modification to such an arc imposes a modification
to its coupled arc. Therefore, we say that a fluid arc is \emph{loose}
if none of its out-coupled fluid arcs is upper-tight. A saturated arc
is \emph{loose} if none of its in-coupled saturated arcs is
lower-tight.

\begin{definition}
  Let $\splitnetwork$ be a splitter network, with capacities
  $c : I \disunion O \to [0,1]$ and a sub-steady-state $(t,F)$. The
  \emph{residual graph} for $(t,F)$ is the pair $(H, \Cequal)$ where
  $H = (V_H,E_H)$ is a graph defined by
  \begin{itemize}[nosep,label=\bull]
  \item $V_H \eqdef \{z\} \cup S$ where $z$ is a new vertex.
  \item for each fluid arc $uv \in F$ with $t(uv) < c(e)$, $\phi(u)\phi(v) \in E_H$,
  \item for each saturated arc $uv \in E \setminus F$ with
    $t(uv) > 0$, $\phi(v)\phi(u) \in E_H$,
  \end{itemize}
  where $\phi(u) = z$ if $u \in I \cup O$, $\phi(u) = u$ otherwise.

  We denote $\rho$ the natural bijection from the arcs in $E$ to
  $E_H$. For each $uv \in E$, let
  $
  C_{uv} \eqdef \{ \rho(u'v')~:~ u'v' \in E \textrm{ coupled with } uv\} 
  $
  and $\Cequal \eqdef \{ C_{uv}~:~uv \in E \}$.
  Notice that $\Cequal$ forms a partition of $E_H$, and for all
  $C \in \Cequal$, there is a vertex $v \in V_H$ with
  $C \subseteq \delta^+_H(v)$. The connected component of $H$
  containing $z$ will be called its \emph{main component}.
\end{definition}

Coupled arcs must have the same throughput, thus their throughput can
only be changed by an equal amount. Therefore $\Cequal$ is the
partition of the arcs into the sets of coupled arcs. We first ensure
that all arcs of the residual graph correspond to loose arcs.

\begin{lemma}\label{lemma:loose-residual}
  Let $(t,F)$ be a sub-steady-state of a capacitated splitter network
  $(G,c)$, and let $H$ be its residual graph. If there is a non-loose
  fluid arc $e$ in $E_H$, or a non-loose saturated arc whose reverse is
  in $E_H$, then there is an arc $e' \in F$ such that
  $(t, F \setminus \{e'\})$ is a sub-steady-state.
\end{lemma}

\begin{proof}
  Let $e$ be a fluid edge with $t(e) < c(e)$ such that $e$ is not
  loose. By definition of loose, $e$ is out-coupled to a fluid arc
  $e'$ (possibly $e$ itself) that is upper-tight. If
  $e' \in \delta^--(o)$ for some output $o \in O$ and $t(e') = c(o)$,
  then $(t,F \setminus \{e'\})$ is a sub-steady-state. Otherwise, $e'$
  is in-coupled to a saturated arc $e'' \notin F$ with
  $t(e') = t(e'')$. In that case, $(t,F \setminus \{e'\})$ is a
  sub-steady-state (checking rules~\ref{eqn:incoming-rule}
  and~\ref{eqn:strong-maximization}). Similarly, for a saturated edge
  $e$ with $t(e) > 0$ that is not loose, $e$ is in-coupled to a
  lower-tight arc $e'$. $e'$ is in-coupled to a fluid arc $e''$ with
  $t(e') = t(e'')$. In that case, $(t, F \setminus \{e''\})$ is a
  sub-steady-state.
\end{proof}

Our next result shows that a $\Cequal$-circulation of the residual
graph fills the role of an augmenting flow for the sub-steady-state.

\begin{definition}
  For a sub-steady-state  $(t,F)$, we define $\psi(t,F)$ by
  \begin{equation}
    \psi(t,F) = \left|\{e \in E \setminus F~:~t(e) = 0\}\right| - \left|\{e \in F~:~t(e) < c(e)\}\right|
  \end{equation}
\end{definition}

\begin{lemma}\label{lemma:circulation-augment}
  Let $\splitnetwork$ be a splitter network with capacities
  $c : I \disunion O \to [0,1]$. Let $x$ be a non-zero circulation on
  the residual graph $H$ of a sub-steady-state $(t,F)$, whose support
  contains only loose arcs. Then there is a value $\lambda > 0$ such
  that $(\update{t}, F)$ is a sub-steady-state, where
  $$
  \update{t}(e) \eqdef 
  \left\{
    \begin{array}{ll}
      t(e) + \lambda x_{\rho(e)} & \quad\textrm{if $e \in F$,}\\
      t(e) - \lambda x_{\rho(e)} & \quad\textrm{if $e \in E \setminus F$.}
    \end{array}
  \right.
  $$
  Moreover, either $\psi(\update{t},F) > \psi(t,F)$, or there is an
  arc $e \in F$ such that
  $(\update{t}, \update{F} \eqdef F \setminus \{e\})$ is a
  sub-steady-state with $\psi(\update{t}, \update{F}) > \psi(t,F)$.
\end{lemma}

\begin{proof}
  We compute $\lambda > 0$ to be the maximum value such that
  $(\update{t} \eqdef t + \lambda x, F)$ is a sub-steady-state. A
  positive $\lambda$ exists because the support of $x$ is the set of
  loose arcs of $G$, thus choosing $\lambda$ small enough will not
  break any sub-steady-state rule. Notice that by definition of
  $\update{t}$, the contribution of any arc $e$ to $\psi$ cannot
  decrease, thus we need only find one arc whose contribution
  increases. By the maximality of $\lambda$, there is a loose arc in
  $(t,F)$ that reaches its bound in $(\update{t},F)$, that is:
  \begin{itemize}[label=\bull,nosep]
  \item either there is an arc $e \in F$, $t(e) < \update{t}(e) = c(e)$, hence
    $\psi(\update{t},F) > \psi(t,F)$;
  \item or there is an arc $e \notin F$, $t(e) > \update{t}(e) = 0$, hence
    $\psi(\update{t},F) > \psi(t,F)$;
  \item or there is a vertex $v \in S$ with $\delta^-(v) = \{e,e'\}$,
    $e \in F$, $e' \in E \setminus F$ with $t(e') > t(e)$ and
    $\update{t}(e') = \update{t}(e)$ ($e$ and $e'$ become tight).
    By rule~\ref{eqn:strong-maximization}, arcs in $\delta^+(v)$
    either are saturated or have throughput $1$ in $t$ and in
    $\update{t}$. Then $(\update{t}, F \setminus \{e\})$ is a
    sub-steady-state as rules~\ref{eqn:incoming-rule}
    and~\ref{eqn:strong-maximization} are satisfied, and
    $\psi(\update{t}, F \setminus \{e\}) > \psi(t,F)$, as the
    contribution of $e$ increases.\qedhere
  \end{itemize}
\end{proof}

\begin{lemma}\label{lemma:saturate-an-arc}
  Let $(t,F)$ be a sub-steady-state for a splitter network
  $\splitnetwork$ with capacities $c : I \disunion O \to [0,1]$, and
  let $H$ be its residual graph. If the main connected component of $H$
  contains a sink $v \in S$, then there is an arc $e \in \delta_G(v)$
  such that $(t,F \setminus \{e\})$ is a sub-steady-state.
\end{lemma}

\begin{proof}
  If $H$ contains a non-loose arc, it suffices to apply
  \Cref{lemma:loose-residual}. Assume that all residual arcs are
  loose.

  Consider a sink $v$ in $H$. We may now assume that for any
  $e \in \delta^+_G(v)$, either $t(e) = c(e)$ or $e \notin F$, and for
  any $e \in \delta^-_G(v)$, $t(e) = 0$ or $e \in F$. Let
  $e \in \delta^-_G(v)$ such that $t(e)$ is maximum. If $e \in F$,
  then $(t,F \setminus \{e\})$ is a sub-steady-state, as
  rules~\ref{eqn:incoming-rule} and~\ref{eqn:strong-maximization} are
  satisfied. If $e \notin F$, then $t(e) = 0$, and by
  rule~\ref{eqn:incoming-rule}
  $0 = \sum_{e \in \delta^-_G(v)} t(e) = \sum_{e \in \delta^+_G(v)} t(e)$,
  hence $\delta^+_G(v) \subseteq E \setminus F$. Because $v$ is not
  isolated, the arc $e'$ out-coupled to $e$ must be fluid with
  $t(e') = 0$, and $(t,F \setminus \{e'\})$ is a sub-steady-state, as
  rules~\ref{eqn:incoming-rule} and~\ref{eqn:strong-maximization} are
  satisfied on $v$ and rule~\ref{eqn:outgoing-rule} is satisfied on
  the origin of $e'$.
\end{proof}

As in an augmenting-path algorithm, we will repeatedly build the
residual graph, and deduce either that the current pair $(t,F)$ is a
steady-state, or find a better solution.

\begin{lemma}\label{lemma:sub-steady-state-step}
  There is an algorithm that, given a splitter network
  $\splitnetwork$ with capacities $c : I \disunion O \to [0,1]$, and a
  sub-steady-state $(t,F)$, either checks that $(t,F)$ is a
  steady-state, or find another sub-steady-state
  $(\update{t}, \update{F})$ such that
  $\psi(\update{t},\update{F}) > \psi(t,F)$ and that runs in time
  $O(n + \stationary(G_z))$, where $\stationary(G_z)$ is the
  complexity of finding a stationary distribution on any residual
  graph for $G$.
\end{lemma}

\begin{proof}
  We apply \Cref{lemma:loose-residual} until the residual graph $H$
  contains only loose arcs, then we apply
  \Cref{lemma:circulation-or-sink} to the main component of $H$.

  Case 1: there is a non-zero circulation. By
  \Cref{lemma:circulation-augment}, the sub-steady-state can be
  improved by this circulation into a new sub-steady-state
  $(\update{t}, \update{F})$ with
  $\psi(\update{t},\update{F}) > \psi(t,F)$.

  Case 2: there is a sink $v$ in the main component of $H$. If
  $v \notin S$, $v$ is the special vertex $z$ obtained by identifying
  the vertices of $I \cup O$, then every arc in $\delta^+_G(I)$ is
  saturated or tight, which implies that rule~\ref{eqn:input-capacity} is
  checked, hence $(t,F)$ is a steady-state. If $v \in S$, by
  \Cref{lemma:saturate-an-arc} there is a fluid arc
  $e \in \delta^-_G(v) \cap F$ such that $(t,F \setminus \{e\})$ is a
  sub-steady-state.
\end{proof}

\begin{proof}[Proof of \Cref{thm:equi-algo}]
  The algorithm repeatedly applies \Cref{lemma:sub-steady-state-step}
  from the initial sub-steady-state $(t : e \to 0, F)$. Each iteration
  takes time $O(|E| + \stationary(G))$ and increases $\psi$ by at least one. As
  $\psi$ is initially $-|E|$, and is at most $|E|$, this bounds the number
  of iterations by $2|E|$, from which we get the claimed complexity.
\end{proof}

\begin{figure}
  \centering
  \begin{tikzpicture}[x=2cm,y=2cm,>=latex]
    \node[term] (i) at (0.4,1) {};
    \node[term] (o) at (3.6,1) {};
    \draw (0.4,1) node[anchor=east] {$1$};
    \draw (3.6,1) node[anchor=west] {$1$};
    \foreach \n/\x/\y in {a/1/1,b/2/2,c/2/1,d/2/0,e/3/1} {
      \node[splitter] (s\n) at (\x,\y) {};
    }
    \draw[fluid,->] (i) -- (sa) node[midway,sloped,above=-3pt] {$3/4$};
    \draw[fluid,->] (sa) -- (sb) node[midway,sloped,above=-3pt] {$1$};
    \draw[saturated,->] (sa) -- (sd) node[midway,sloped,below=-3pt] {$3/4$};
    \draw[saturated,->] (sb) -- (sc) node[pos=0.6,right=-3pt] {$1/2$};
    \draw[fluid,->] (sb) -- (se) node[midway,sloped,above=-3pt] {$1/2$};
    \draw[saturated,->] (sc) -- (sa) node[midway,sloped,above=-3pt] {$1$};
    \draw[saturated,->] (sd) -- (sc) node[pos=0.6,left=-3pt] {$1/2$};
    \draw[fluid,->] (se) -- (o) node[midway,sloped,above=-3pt] {$3/4$};
    \draw[fluid,->] (sd) .. controls (2.2,0.4) and (2.6,0.8) ..  (se)
    node[midway,sloped,above=-3pt] {$1$};
    \draw[saturated,->] (se) .. controls (2.8,0.6) and (2.4,0.2) .. (sd)
    node[midway,sloped,below=-3pt] {$3/4$};
    \begin{scope}[xshift=8cm]
      \node[splitter] (i) at (0.4,1) {};
      \node[splitter] (o) at (3.6,1) {};
      \draw (0.4,1) node[anchor=east] {$z$};
      \draw (3.6,1) node[anchor=west] {$z$};
      \foreach \n/\x/\y in {a/1/1,b/2/2,c/2/1,d/2/0,e/3/1} {
        \node[splitter] (s\n) at (\x,\y) {};
      }
      \draw[->] (i) -- (sa) node[midway,sloped,above=-3pt] {$3/18$};
      \draw[<-] (sc) -- (sa) node[midway,sloped,above=-3pt] {$4/18$};
      \draw[<-] (sa) -- (sd) node[midway,sloped,below=-3pt] {$1/18$};
      \draw[<-] (sb) -- (sc) node[pos=0.6,right=-3pt] {$2/18$};
      \draw[->] (sb) -- (se) node[midway,sloped,above=-3pt] {$2/18$};
      \draw[<-] (sd) -- (sc) node[pos=0.6,left=-3pt] {$2/18$};
      \draw[->] (se) -- (o) node[midway,sloped,above=-3pt] {$3/18$};
      \draw[<-] (se) .. controls (2.8,0.6) and (2.4,0.2) .. (sd)
      node[midway,sloped,below=-3pt] {$1/18$};
    \end{scope}  
  \end{tikzpicture}
  \caption{Again the residual graphs contains sinks, that we remove by
    making some arcs saturated. This yields the residual graph on the
    right, free of any sink. We compute a stationary circulation in
    this residual graph, and use it to improve the sub-steady-state
    (taking $\lambda = \frac{9}{14}$, when the incoming arcs of the
    leftmost splitter become tight, see \Cref{fig:algo-round6})}
  \label{fig:algo-round5}
\end{figure}
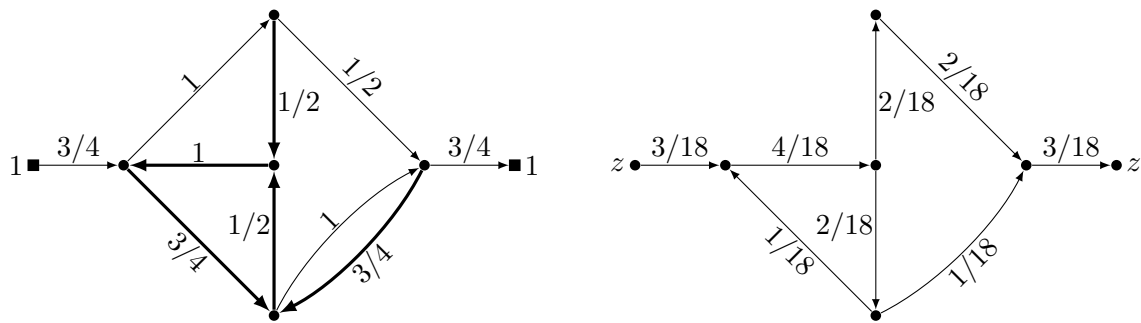

\subsection{Reverse splitter networks and
  consequences}\label{sec:symmetry}

We present some consequences of the algorithms and of the following
fact that splitter networks can be reversed. Because splitters are
fair on their input as well as on their output, splitter networks and
steady-states display a useful symmetry. For an arc set $X$, we denote
$\symmetry{X} := \{vu~:~ uv \in X\}$ the set of reverse arcs of $X$.

\begin{definition}
  Let $\splitnetwork$ be a splitter network, we denote $\symmetry{G}$
  the \emph{reverse} of $G$, defined as the splitter network
  $(O \disunion S \disunion I,\symmetry{E})$, where the inputs and
  outputs are interchanged.
\end{definition}

\begin{lemma}\label{lemma:reversibility}
  Let $\splitnetwork$ be a splitter network with capacities,
  $c : I \disunion O \to [0,1]$, and $(t,F)$ a steady-state for
  $(G,c)$. Then $(t, \symmetry{E} \setminus \symmetry{F})$ is a
  steady-state for $(\symmetry{G},c)$.
\end{lemma}

\begin{proof}
  We check that each steady-state-defining rule is satisfied.
  Rules~\ref{eqn:throughput}, \ref{eqn:fluid-arcs},
  \ref{eqn:conservation} and~\ref{eqn:maximization} can be readily
  checked. Rules~\ref{eqn:input-capacity}
  and~\ref{eqn:output-capacity} translates into each other, as do
  rules~\ref{eqn:incoming-rule} and~\ref{eqn:outgoing-rule}.
\end{proof}

Observe that the strong maximization
rule~\ref{eqn:strong-maximization} does not translate well into
$\symmetry{G}$, hence the reverse of a steady-state does not
necessarily check rule~\ref{eqn:strong-maximization}. But by
\Cref{lemma:equivalence-strong-maximization}, there is a steady-state
$(t,\symmetry{E} \setminus \symmetry{F'})$ of $(\symmetry{G},c)$ with
$F' \subseteq F$ that satisfies rule~\ref{eqn:strong-maximization}.

A careful look at the sub-steady-state algorithm shows that, during
its resolution, each edge $t(e)$ starts in $F$ with $t(e)=0$, then
$t(e)$ increases, before possibly being removed from $F$, then $t(e)$
decreases.

\begin{corollary}\label{coro:monotonicity}
  If $(t,F)$ is a sub-steady-state for $(G,c)$, then there exists an
  steady-state $(\update{t},\update{F})$ with $\update{F} \subseteq F$,
  and for each $e \in E$, if $e \in \update{F}$ then
  $\update{t}(e) \geq t(e)$, and if $e \in E \setminus F$,
  $\update{t}(e) \leq t(e)$.
\end{corollary}

If we increase the input capacities to $\update{c}$, a steady-state
$(t,F)$ for $(G,c)$ becomes a sub-steady-state for $G,\update{C})$. Hence,

\begin{corollary}\label{coro:increase-input}
  If $(t,F)$ is a steady-state for $(G,c)$, and $\update{c}$ be a
  capacity function with $\update{c}(u) \geq c(u)$ for each input or
  output $u \in I \cup O$. There exists a steady-state
  $(\update{t},\update{F})$ for $(G,\update{c})$ such that:
  \begin{enumerate}[label=(\roman*)]
  \item\label{item:case1} if $\update{c}(o) = c(o)$ for each output $o \in O$, then
    $\update{t}(e) \geq t(e)$ for each $e \in \delta^-(O)$;
  \item\label{item:case2} if $\update{c}(i) = c(i)$ for each input $i \in I$, then
    $\update{t}(e) \geq t(e)$ for each $e \in \delta^+(I)$.
  \end{enumerate}
\end{corollary}

\begin{proof}
  $(t,F)$ is a sub-steady-state, which we can improve using the
  algorithm from \Cref{thm:equi-algo}. Let $(\update{t},\update{F})$
  be the resulting steady-state. Notice that if an arc
  $e \in \delta^-(O)$ is saturated, by rule~\ref{eqn:output-capacity},
  $t(e) = c(e)$, hence $e$ is not loose. Thus the throughputs of
  saturated arcs incident to outputs cannot decrease. This
  proves~\ref{item:case1}. Then~\ref{item:case2} follows by taking
  the reverse network and applying~\ref{item:case1} to the reverse
  steady-state, thanks to \Cref{lemma:reversibility}.\qedhere
\end{proof}

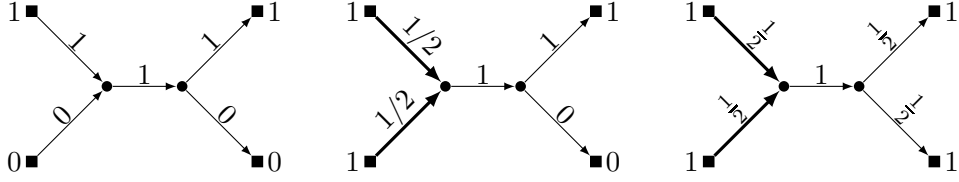
\begin{figure}
  \centering
  \begin{tikzpicture}[x=1cm,y=1cm,>=latex]
    \begin{scope}[xshift=0cm]
      \node[splitter] (s0) at (0,0) {};
      \node[splitter] (s1) at (1,0) {};
      \node[term] (i0) at (-1,1) {}; \draw (-1,1) node[anchor=east] {$1$};
      \node[term] (i1) at (-1,-1) {}; \draw (-1,-1) node[anchor=east] {$0$};
      \node[term] (o0) at (2,1) {}; \draw (2,1) node[anchor=west] {$1$};
      \node[term] (o1) at (2,-1) {}; \draw (2,-1) node[anchor=west] {$0$};
      \draw[->,fluid] (i0) -- (s0) node[midway,above=-3pt,sloped] {$1$};
      \draw[->,fluid] (i1) -- (s0) node[midway,above=-3pt,sloped] {$0$};
      \draw[->,fluid] (s0) -- (s1) node[midway,above=-3pt,sloped] {$1$};
      \draw[->,fluid] (s1) -- (o0) node[midway,above=-3pt,sloped] {$1$};
      \draw[->,fluid] (s1) -- (o1) node[midway,above=-3pt,sloped] {$0$};
    \end{scope}
    \begin{scope}[xshift=4.5cm]
      \node[splitter] (s0) at (0,0) {};
      \node[splitter] (s1) at (1,0) {};
      \node[term] (i0) at (-1,1) {}; \draw (-1,1) node[anchor=east] {$1$};
      \node[term] (i1) at (-1,-1) {}; \draw (-1,-1) node[anchor=east] {$1$};
      \node[term] (o0) at (2,1) {}; \draw (2,1) node[anchor=west] {$1$};
      \node[term] (o1) at (2,-1) {}; \draw (2,-1) node[anchor=west] {$0$};
      \draw[->,saturated] (i0) -- (s0) node[midway,above=-3pt,sloped] {$1/2$};
      \draw[->,saturated] (i1) -- (s0) node[midway,above=-3pt,sloped] {$1/2$};
      \draw[->,fluid] (s0) -- (s1) node[midway,above=-3pt,sloped] {$1$};
      \draw[->,fluid] (s1) -- (o0) node[midway,above=-3pt,sloped] {$1$};
      \draw[->,fluid] (s1) -- (o1) node[midway,above=-3pt,sloped] {$0$};
    \end{scope}
    \begin{scope}[xshift=9cm]
      \node[splitter] (s0) at (0,0) {};
      \node[splitter] (s1) at (1,0) {};
      \node[term] (i0) at (-1,1) {}; \draw (-1,1) node[anchor=east] {$1$};
      \node[term] (i1) at (-1,-1) {}; \draw (-1,-1) node[anchor=east] {$1$};
      \node[term] (o0) at (2,1) {}; \draw (2,1) node[anchor=west] {$1$};
      \node[term] (o1) at (2,-1) {}; \draw (2,-1) node[anchor=west] {$1$};
      \draw[->,saturated] (i0) -- (s0) node[midway,above=-3pt,sloped] {$\frac{1}{2}$};
      \draw[->,saturated] (i1) -- (s0) node[midway,above=-3pt,sloped] {$\frac{1}{2}$};
      \draw[->,fluid] (s0) -- (s1) node[midway,above=-3pt,sloped] {$1$};
      \draw[->,fluid] (s1) -- (o0) node[midway,above=-3pt,sloped] {$\frac{1}{2}$};
      \draw[->,fluid] (s1) -- (o1) node[midway,above=-3pt,sloped] {$\frac{1}{2}$};
    \end{scope}
    
  \end{tikzpicture}
  \caption{Increasing the input capacities may result in reduced
    effective throughputs on some inputs, as if they are in
    competition. Similarly, increasing the output capacities may also
    result in reduced effective throughputs on some outputs. }
  \label{fig:monotonicity-example}
\end{figure}

Thus increasing the input capacities will not decrease the output
throughputs. Notice that increasing the input capacities may lead to
the decrease in the throughput of some of the inputs, as can be seen
in the example of \Cref{fig:monotonicity-example}.

\section{Design of balancer networks}\label{sec:balancers}

We first define formally the simple balancer of order $k$, for
$k \geq 0$, whose recursive definition was suggested in
\Cref{sec:intro-balancer}, then prove that it is a balancer. For
$k = 0$ and $k=1$, the networks without splitter and with a single
splitter respectively are universal balancers. We denote
$\interval{l}{u}$ the set of integers $i$ with $l \leq i \leq u$.

\begin{definition}
  For any integer $k \geq 2$, the \emph{simple balancer of order $k$}
  is the network splitter $\splitnetwork$ where:
  \begin{itemize}[nosep,label=\bull] 
  \item $I \eqdef \left\{i_j : j \in \interval{0}{2^k-1}\right\}$ and
    $O \eqdef \left\{o_j : j \in \interval{0}{2^k-1}\right\}$,
  \item $S \eqdef \left\{s_{(l,j)} : (l,j) \in \interval{0}{k-1} \times \interval{0}{2^{k-1}-1}\right\}$,
  \item $E \eqdef \left\{i_js_{(0,j/2)}, s_{(k-1,j/2)}o_j : j \in \interval{0}{2^{k}-1}\right\} \cup \bigcup_{l \in \interval{0}{k-2}} E_l$,
  \item
    $E_l \eqdef \left\{(s_{(l,j)}s_{(l+1,j)}, s_{(l,j)}s_{(l+1,j \oplus 2^l)} : j \in \interval{0}{2^{k-1}-1}\right\}$.
  \end{itemize}
  where $\oplus$ is the bitwise exclusive or.
\end{definition}

\begin{proposition}\label{prop:simple-balancer}
  For any $k \geq 1$, the simple balancer of order $k$ is a balancer.
\end{proposition}

\begin{proof}
  Let $\splitnetwork$ be the simple balancer of order $k$,
  $c : I \disunion O \to [0,1]$ with $c(o) = 1$ for each output
  $o \in O$. For each arc $e = s_{(l,j)}s_{(l+1,j')}$, we define $J_{(l,j)}$
  to be the set of inputs $i$ such that there is a directed path from
  $i$ to $e$. Clearly, we have that
  $J_{(l,j)} = J_{(l-1,j)} \disunion J_{(l-1,j \oplus 2^{l-1})}$ and
  $J_{(l,j)}$ is the set of integers $j'$ such that the binary
  representations of $j$ and $j'/2$ coincides on the $k-1-l$ heavier
  bits. Then, set
  $$ t(s_{(l,j)}) \eqdef \frac{1}{2^{l+1}} \sum_{\alpha \in J_{(l,j)}} c(i_\alpha) $$
  and $F \eqdef E$. It can be easily checked that $(t,F)$ is a
  steady-state where each output has the same throughput.
\end{proof}

\subsection{Throughput-unlimited balancers}\label{sec:benes}

We now define formally the Beneš network, and prove its property of
being throughput-unlimited.
\begin{definition}
  A \emph{Beneš network} of order $k$ is a splitter network
  $\splitnetwork$ where
  \begin{itemize}[nosep,label=\bull]
  \item $I \eqdef \{i_j : j \in \interval{0}{2^k-1} \}$ and
    $O \eqdef \{ o_j : j \in \interval{0}{2^k-1}\}$,
  \item $S \eqdef \{ s_{(l,j)} : l \in \interval{0}{2k-2}, j \in \interval{0}{
2^{k-1}-1} \}$,
  \item $E \eqdef \{i_js_{(0,j/2)}, s_{(2k-2,j/2)}o_j : j \in \interval{0}{2^{k}-1}\} \cup \bigcup_{l \in \interval{0}{2k-3}} E_l$,
  \item
    $E_l \eqdef \{(s_{(l,j)}s_{(l+1,j)}, s_{(l,j)}s_{(l+1,j \oplus 2^{l'})} : j \in \interval{0}{2^{k-1}-1}\}$
    where $l' = \max \{k - 2 - l, l - k + 1\}$.
  \end{itemize}
\end{definition}

The arcs between levels $l$ and $l+1$ represent swapping a bit of
weight $w$, where $w$ varies from $k-2$ to $0$ then to $k-2$ again.
Alternatively, a recursive definition of a Beneš network of order
$k+1$ consists in placing two Beneš networks of order $k$ in parallel.
Then identify the inputs from both networks, index by index, into a
splitter, and proceed similarly for the outputs (see
\Cref{fig:benes-network}). Because the latter half of a Beneš network
is a simple balancer, it is itself a simple balancer.


\begin{proposition}\label{prop:benes-balancer}
  Beneš networks are balancer.
\end{proposition}

\begin{proof}
  This follows from the fact that the subgraph of splitters $s_{(l,j)}$ for
  $l \in \interval{k-1}{2k-2}$ with their incident arcs is a simple
  balancer.
\end{proof}

\begin{proposition}\label{prop:benes-unlimited}
  Beneš networks are throughput-unlimited.
\end{proposition}

\begin{proof}
  Let $\splitnetwork$ be the Beneš network of order $k$, and
  $c : I \disunion O \to [0,1]$. We may assume that $c(I) \geq c(O)$
  by the reversibility property of splitter
  networks~\Cref{lemma:reversibility}. 
  
  We define a new splitter network $G_L = (V_L,E_L)$, by removing $O$
  and all vertices $s_{(l,j)}$ with $l \geq k$, and adding $2^k$ new
  outputs $O'$, with arcs $s_{(k,j/2)}o'_j$ for each
  $j \in \interval{0}{2^k-1}$. $G_L$ is a simple balancer of order $k$
  between $I$ and $O'$. Symmetrically we define $G_R = (V_R,E_R)$, a
  simple balancer of order $k$, by removing $I$ and all vertices
  $s_{(l,j)}$ with $l < k - 1$, and adding a set $I'$ of new inputs.

  Consider $c'_L : I \cup O' \to [0,1]$ defined by
  $c'_L(I) = c(I)$ and $c'_L(O) = \allone$. By
  \Cref{prop:simple-balancer}, there exists a steady-state
  $(t'_L,F'_L)$ for $G_L,c_L$, such that for each
  $e \in \delta^-(O)$, $t'(e) = c(I) / 2^k$, and
  $F'_L = E_L$.

  Next consider $c_R : I' \cup O \to [0,1]$ defined by $c_R(I') = 1$
  and $c_R(O) = c(O)$. The reverse of $G_R$ is a simple balancer of
  order $k$. Therefore by \Cref{prop:simple-balancer}, there exists a
  steady-state $(\symmetry{t_R},\symmetry{F_R})$ for
  $\symmetry{G_R}, c_R$, such that for each
  $\symmetry{e} \in \delta^-(I')$ of the reverse graph,
  $t(\symmetry{e}) = c(O) / 2^k$ and $\symmetry{e} \in \symmetry{F_R}$. Reversing the
  network and the steady-state, by \Cref{lemma:reversibility},
  $(t_R : e \to \symmetry{t_R}(\symmetry{e}), F_R = E_R \setminus \symmetry{F_R})$
  is a steady-state in $G_R$. Furthermore, for each arc
  $e \in \delta^+(I')$, $t_R(e) = c(O) / 2^k$ and
  $e \in E_R \setminus F_R$ is saturated.

  We plan to build a steady-state using $t'_L$ and $t_R$. However when
  $c(I) > c(O)$, this would lead to positive excess on splitters
  $s_{(k-1,j)}$. To avoid this, we decrease the throughputs on $G_L$
  with the following procedure. Let $c_L : I \disunion O' \to [0,1]$
  denote a capacity function with $c_L(I) = c(I)$ and
  $c_L(o) = c(O) / 2^k$ for each $o \in O'$. Then define a throughput
  function $\update{t}_L$, where $\update{t}_L(e) = c_L(o)$ for each
  $o \in O'$ and $\{e\} = \delta^-(o)$, and
  $\update{t}_L(e) = t'_L(e)$ for any other arc
  $e \notin \delta^-(O')$. Then $(t'_L,E_L)$ is a pre-steady-state for
  $(G_L,c_L)$, with positive excess on splitters $s_{(k-1,j)}$ when
  $c(I) > c(O)$. Using the pre-steady-state algorithm and
  \Cref{lemma:presteady-state-lp}, there exists a steady-state
  $(t_L,F_L)$ for $G_L,c_L$. Furthermore, let $e \in \delta^-(O')$. If
  $e$ is fluid, then by \Cref{coro:mono-pre-steady},
  $c(e) = c(O) / 2^k$. If $e$ is saturated, then by
  rule~\ref{eqn:output-capacity}, $c(e) = c(O) / 2^k$ also.

  Finally, we define $t(e) = t_L(e)$ if $e \in E \cap E_L$, and
  $t(e) = t_R(e)$ if $e \in E \cap E_R$. Let $s = s_{(k,j)}$ for some
  $j \in \interval{0}{2^k-1}$. Then
  $t(\delta^-(s)) = c(O) / 2^{k-1} \geq t'(\delta^+(s)) = c(O) / 2^{k-1}$.
  Let $F = F_L \cup F_R$. Then the arcs in $\delta^+(s)$ are
  saturated. Because rules~\ref{eqn:incoming-rule}, \ref{eqn:outgoing-rule}
  and~\ref{eqn:maximization} hold on $s$ and on each splitter, $(t,F)$
  is a steady-state, with $t$ uniform on $\delta^-(O)$.
\end{proof}


  

\subsection{Universal balancer}\label{sec:universal}

Before giving a design for universal balancer, we first define a
network that behaves like a universal balancer as long as the input
and output capacities are at most $1/2$.

\begin{definition}
  The \emph{half-universal network of order $k$} is a continous
  splitter network $\splitnetwork$ where
  \begin{itemize}[label=\bull,nosep]
  \item $I \eqdef \{ i_j : j \in \interval{0}{2^k-1}\}$ and $O \eqdef \{ o_j : j \in \interval{0}{2^k-1} \}$;
  \item $S \eqdef S_B$;
  \item $E \eqdef \{ uv \in E_B : u,v \in S_B \} \cup F 
    \cup \{i_js_{(0,j)}, s_{(2k,j)}o_j : j \in \interval{0}{2^k-1}\}
    $;
  \item $F \eqdef \{  s_{(2k,j)}s_{(0,j)} : j \in \interval{0}{2^k-1}\}$;
  \item $(I_B \disunion S_B \disunion O_B, E_B)$ is the Beneš network
    of order $k+1$, and $S_B = \{s_{(l,j)} : l \in \interval{0}{2k}, j \in \interval{0}{2^{k+1}-1}\}$.
  \end{itemize}  
\end{definition}

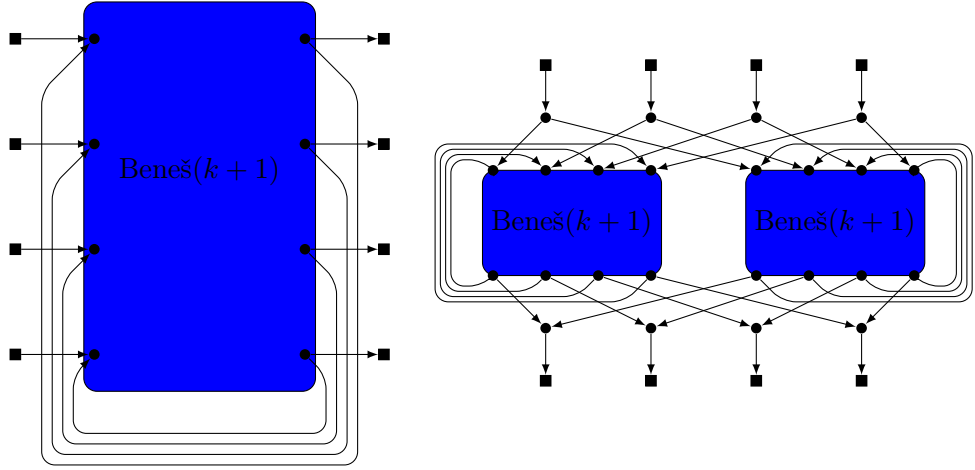
\begin{figure}
  \centering
  \begin{tikzpicture}[x=0.7cm,y=0.7cm,>=latex]
    \fill[rounded corners=5pt,nearly transparent,blue] (0.8,-0.2) rectangle (5.2,7.2);
    \draw[rounded corners=5pt] (0.8,-0.2) rectangle (5.2,7.2);
    \foreach \n/\y in {0/0.5,1/2.5,2/4.5,3/6.5} {
      \node[splitter] (s0\n) at (1,\y) {};
      \node[splitter] (s1\n) at (5,\y) {};
    }
    \draw (3,4) node {Beneš($k+1$)};
    \foreach \y in {0,1,2,3} {
      \node[term] (i\y) at ($(-0.5,6.5) - 2*(0,\y)$) {};
      \node[term] (o\y) at ($(6.5,6.5) - 2*(0,\y)$) {};
    }
    \foreach \u/\v in {i0/s03,i1/s02,i2/s01,i3/s00,s13/o0,s12/o1,s11/o2,s10/o3} {
      \draw[fluid,->] (\u) -- (\v);
    }
    \draw[rounded corners=5pt,fluid,->] (s10) -- ++(0.4,-0.4) -- (5.4,-1) -- (0.6,-1) -- (0.6,0.1) -- (s00);
    \draw[rounded corners=5pt,fluid,->] (s11) -- ++(0.6,-0.6) -- (5.6,-1.2) -- (0.4,-1.2) -- (0.4,1.9) -- (s01);
    \draw[rounded corners=5pt,fluid,->] (s12) -- ++(0.8,-0.8) -- (5.8,-1.4) -- (0.2,-1.4) -- (0.2,3.7) -- (s02);
    \draw[rounded corners=5pt,fluid,->] (s13) -- ++(1,-1) -- (6,-1.6) -- (0,-1.6) -- (0,5.5) -- (s03);
   
    \begin{scope}[xshift=6cm,yshift=0cm]
    \fill[rounded corners=5pt,blue,nearly transparent] (-0.2,2) rectangle (3.2,4);
    \fill[rounded corners=5pt,blue,nearly transparent] (4.8,2) rectangle (8.2,4);
    \draw[rounded corners=5pt] (-0.2,2) rectangle (3.2,4);
    \draw[rounded corners=5pt] (4.8,2) rectangle (8.2,4);
    \foreach \i in {0,1,2,3} {
      \node[term] (i\i) at ($(1,6)+\i*(2,0)$) {};
      \node[splitter] (u\i) at ($(1,5)+\i*(2,0)$) {};
      \foreach \n/\s in {A/0,B/5} {
        \node[splitter] (v\n\i) at ($(\s,4)+\i*(1,0)$) {};
        \node[splitter] (w\n\i) at ($(\s,2)+\i*(1,0)$) {};
      }
      \node[splitter] (x\i) at ($(1,1)+\i*(2,0)$) {};
      \node[term] (o\i) at ($(1,0)+\i*(2,0)$) {};
      \draw[fluid,->] (i\i) -- (u\i);
      \foreach \n in {A,B} {
        \draw[fluid,->] (u\i) -- (v\n\i);
        \draw[fluid,->] (w\n\i) -- (x\i);
      }
      \draw[fluid,->] (x\i) -- (o\i);
    }
    \draw (1.5,3) node {Beneš($k+1$)};
    \draw (6.5,3) node {Beneš($k+1$)};
    \foreach \i in {0,1,2,3} {
      \draw[fluid,->,rounded corners=5pt] 
        (wA\i) -- ($(\i,2)-\i*(0.1,0.1) - (0.2,0.2)$)
        -- ($(-0.8,1.8)-\i*(0.1,0.1)$) -- ($(-0.8,4.2)+\i*(-0.1,0.1)$) 
        -- ($(\i,4)+\i*(-0.1,0.1)+(-0.2,0.2)$) -- (vA\i);
      \draw[fluid,->,rounded corners=5pt] 
        (wB\i) -- ($(5.5,1.5)+\i*(0.9,0.1)$)
        -- ($(9.1,1.5)+\i*(-0.1,0.1)$) -- ($(9.1,4.5)-\i*(0.1,0.1)$) 
        -- ($(5.5,4.5)+\i*(0.9,-0.1)$) -- (vB\i);
    }

    \end{scope}
  \end{tikzpicture}
  \caption{A schematic representation of a half-universal network on
    the left, and of a universal network on the right.}
  \label{fig:half-universal-network}
\end{figure}

See the left side of \Cref{fig:half-universal-network}. The
half-universal network is its own reverse. Its design includes
loopback connections from each output back to each input. When the
throughput of an output reaches its capacity, any excess throughput
can be redirected back through one of these loopback arc, to the entry
of the network. From there, it flows again to different outputs. Hence, the
loopback arcs prevent the occurence of saturated arcs inside the Beneš
network, as long as at least one output has not reached its maximum
capacity. Consequently, the balancing property of the Beneš network is
also preserved. This is done at the cost of reducing by half the
throughput, as we send back half of the flow. This limitation
justifies the name half-universal, and our universal balancer will use
two half-universal networks in parallel.

\begin{proposition}\label{prop:half-universal}
  Let $\splitnetwork$ be the half-universal splitter network, and let
  $c : I \disunion O \to [0,1/2]$. Then there is a steady-state
  $(t,F)$ for $(G,c)$ and $\alpha, \beta \in \mathbb{R}_{\geq 0}$ such
  that
  \begin{enumerate}[label=(\roman*),nosep]
  \item for each input $i$, $t(\delta^+(i)) = \min \{c(i), \alpha\}$,
  \item for each output $o$, $t(\delta^-(o)) = \min \{c(o), \beta\}$.
  \item the total throughput $T \eqdef t(\delta^+(I))$ equals
    $\min \{c(I), c(O)\}$.
  \end{enumerate}  
\end{proposition}

\begin{proof}
  We may assume that $c(I) \leq c(O)$ by \Cref{lemma:reversibility},
  because the half-universal network is its own reverse. We build a
  steady-state with the stated properties. Let $\beta \geq 0$ be such
  that $\sum_{o \in O} \min \{ c(o), \beta \} = c(I)$. Observe that
  $\beta \leq \max_{o \in O} c(o) \leq \frac{1}{2}$. Let $f_j$ be the
  loopback arc $s_{(2k,j)}s_{(0,j)}$ and
  $J' \eqdef \{ j \in \interval{0}{2^k-1} : c(o_j) < \beta\}$.

  For $j \in \interval{0}{2^k-1}$, set 
  \begin{align*}
    t(\delta^-(o_j)) &\eqdef \min \{ c(o_j), \beta \},\\
    t(\delta^+(i_j)) &\eqdef c(i_j),\\
    t(f_j) &\eqdef 2\beta - \min \{c(o_j), \beta\} \leq 1.
  \end{align*}
  Let $(\optt,E_B)$ be the steady-state of the Beněs network of order
  $k$, when the input capacities are given by the values of
  $t(\delta^+(i_j))$ and $t(f_j)$ (for $j \in \interval{0}{2^k-1}$)
  and the output capacities are uniformly $1$. For any arc $e$ in the
  Beneš subnetwork, set $t(e) \eqdef \optt(e)$.

  We claim that $(t,E \setminus \{\delta^-(o_j) : j \in J'\})$ is a
  steady-state. Indeed, by \Cref{prop:benes-unlimited}, the
  flow going through the Beneš subnetwork is
  $$t(I) + t(F) = c(I) + \sum_{o \in O} 2\beta - \min \{c(o), \beta\} = \sum_{o \in O} 2\beta.$$
  By \Cref{prop:benes-balancer}, the flow entering any node
  $s_{(2k,j)}$ must be $2\beta$, which equals the leaving flow.
  Moreover, Rule~\ref{eqn:outgoing-rule} is clearly satisfied on node
  $s_{(2k,j)}$ as $t(f_j) \geq t(\delta^-(o_j))$, concluding the
  proof.
\end{proof}

\begin{definition}
  The \emph{universal network of order $k$} is the splitter
  network built from two disjoint copies of the half universal network
  of order $k$, by pairwise identifying the inputs and outputs from both copies,
  and adding $2^k$ input nodes and $2^k$ output nodes, each with one
  arc linking it to one of the identified vertices.
\end{definition}

The right side of \Cref{fig:half-universal-network} illustrates the
construction of a half-universal network. Again, universal networks
are their own reverse. 

\begin{theorem}\label{prop:universal}
  The universal network of order $k$ is universally balancing.
\end{theorem}

\begin{proof}
  Let $c : I \disunion O \to [0,1]$ be a capacity function on the
  universal balancer $G$ of order $k$. Consider a half-universal
  balancer $H = (I_H \disunion S_H \disunion O_H,E_H)$ of order $k-1$,
  let $c_H : I_H \disunion O_H \to \interval{0}{\frac{1}{2}}$ be a
  capacity function defined by $c_H(i_j) = \frac{1}{2} c(i_j)$ and
  $c_H(o_j) = \frac{1}{2}c(o_j)$. Let $(t_H,F_H)$ a steady-state for
  $(H,c_H)$ given by \Cref{prop:half-universal}. Then one can define
  a steady-state $(t,F)$ for $(G,c)$ by using $t_H$ on both sides of
  $G$ and completing $t$ to the arcs incident to output and input
  nodes, and easily check that it has the expected properties.
\end{proof}

\subsection{Balancers of all sizes}

We extend the construction of balancers to general
$(n_i,n_o)$-balancers. Given $n_i$ and $n_o$, two positive integers
representing respectively the number of inputs and outputs, let
$k \in \mathbb{N}$ such that $m \eqdef \max \{n_i, n_o\} \leq 2^k$.
Let $\splitnetwork$ be the universal network with $2^k$ inputs and
$2^k$ outputs. Remove any surplus inputs or
outputs to achieve a network with $n_i$ inputs and $n_o$ outputs. This
can be accomplished by setting their capacities to $0$. We obtain a
universally balancing network with $n_i$ inputs and $n_0$ outputs. 

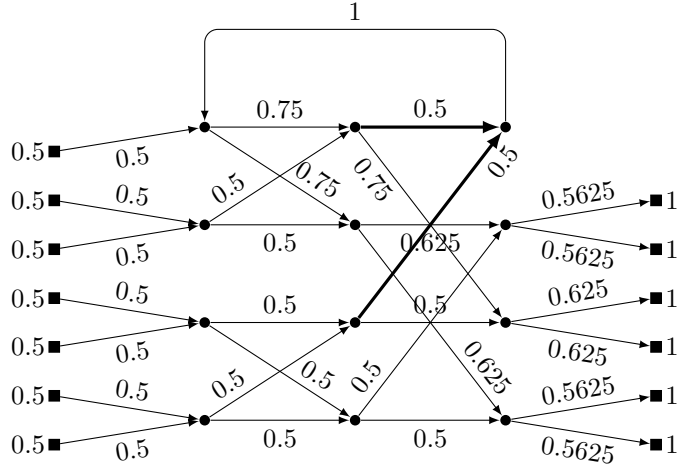
\begin{figure}
  \centering
  \small
  \begin{tikzpicture}[x=2cm,y=1.3cm,>=latex]
    \foreach \n/\y/\c in {0/-0.25/1,1/0.25/1,2/0.75/1,3/1.25/1,4/1.75/1,5/2.25/1} {
      \node[term] (i\n) at (0,\y) {};
      \node[term] (o\n) at (4,\y) {};
      \draw (0,\y) node[anchor=east] {$0.5$};
      \draw (4,\y) node[anchor=west] {$\c$};
    }     
    \node[term] (i6) at (0,2.75) {};
    \draw (0,2.75) node[anchor=east] {$0.5$};
    \foreach \y in {0,1,2,3} {
      \foreach \x in {1,2,3} {
        \node[splitter] (s\x\y) at (\x,\y) {};
      }
    }
    \draw[fluid,->,rounded corners=5pt] (s33) -- (3,4) -- node[midway,above] {$1$} (1,4) -- (s13);

    \draw[fluid,->] (i0) -- (s10) node[sloped,midway,below] {$0.5$};
    \draw[fluid,->] (i1) -- (s10) node[sloped,midway,above] {$0.5$};
    \draw[fluid,->] (i2) -- (s11) node[sloped,midway,below] {$0.5$};
    \draw[fluid,->] (i3) -- (s11) node[sloped,midway,above] {$0.5$};
    \draw[fluid,->] (i4) -- (s12) node[sloped,midway,below] {$0.5$};
    \draw[fluid,->] (i5) -- (s12) node[sloped,midway,above] {$0.5$};
    \draw[fluid,->] (i6) -- (s13) node[sloped,midway,below] {$0.5$};

    \draw[fluid,->] (s10) -- (s20) node[midway,below] {$0.5$};
    \draw[fluid,->] (s11) -- (s21) node[midway,above] {$0.5$};
    \draw[fluid,->] (s12) -- (s22) node[midway,below] {$0.5$};
    \draw[fluid,->] (s13) -- (s23) node[midway,above] {$0.75$};
    \draw[fluid,->] (s10) -- (s21) node[sloped,pos=0.2,above] {$0.5$};
    \draw[fluid,->] (s11) -- (s20) node[sloped,pos=0.7,above] {$0.5$};
    \draw[fluid,->] (s12) -- (s23) node[sloped,pos=0.2,above] {$0.5$};
    \draw[fluid,->] (s13) -- (s22) node[sloped,pos=0.7,above] {$0.75$};

    \draw[fluid,->] (s20) -- (s30) node[midway,below] {$0.5$};
    \draw[fluid,->] (s21) -- (s31) node[midway,above] {$0.5$};
    \draw[fluid,->] (s22) -- (s32) node[midway,below] {$0.625$};
    \draw[saturated,->] (s23) -- (s33) node[midway,above] {$0.5$};
    \draw[fluid,->] (s20) -- (s32) node[sloped,pos=0.15,above] {$0.5$};
    \draw[saturated,->] (s21) -- (s33) node[sloped,pos=0.9,below] {$0.5$};
    \draw[fluid,->] (s22) -- (s30) node[sloped,pos=0.8,above] {$0.625$};
    \draw[fluid,->] (s23) -- (s31) node[sloped,pos=0.2,below] {$0.75$};

    \draw[fluid,->] (s30) -- (o0) node[sloped,midway,below] {$0.5625$};
    \draw[fluid,->] (s30) -- (o1) node[sloped,midway,above] {$0.5625$};
    \draw[fluid,->] (s31) -- (o2) node[sloped,midway,below] {$0.625$};
    \draw[fluid,->] (s31) -- (o3) node[sloped,midway,above] {$0.625$};
    \draw[fluid,->] (s32) -- (o4) node[sloped,midway,below] {$0.5625$};
    \draw[fluid,->] (s32) -- (o5) node[sloped,midway,above] {$0.5625$};

  \end{tikzpicture}
  \caption{Adding loops on a simple balancer to reduce the number of
    inputs and outputs does not necessarily preserve the balancing
    property.}
  \label{fig:not-a-simple-balancer}
\end{figure}

Another method to decrease the number of outputs and inputs would be
to add loopback arcs from the surplus output to the surplus input.
This would avoid the usage of dummy terminals. Loopback arcs are also
useful to reduce the number of inputs and outputs in non-universal
balancer. However, preserving the balancer property when adding
loopback arcs requires some care. Indeed, in a simple balancer,
balancing is achieved when the capacities of the outputs are uniformly
$1$. Therefore it is necessary to ensure that the loopback arcs remain
fluid. A careless implementation of this process may result in a
network failing to be a balancer, as illustrated by
\Cref{fig:not-a-simple-balancer}.

\section{Lower bounds for the size of
  balancers}\label{sec:lowerbounds}

We begin this section by establishing \Cref{prop:count-splitters},
followed by the derivation of a lower bound on the number of splitters
in a balancer.

\begin{proof}[Proof of \Cref{prop:count-splitters}]
  This follows directly from the definitions for $S(k)$ and $B(k)$. A
  universal network of order $k$ consists in its two Beneš subnetworks
  of order $k+1$, plus $2 \cdot 2^k$ additional splitters to connect
  the inputs and outputs to each half-universal network. In total, we
  get
  \[U(k) = 2 \cdot (2k+1) 2^k + 2 \cdot 2^k = (k+1) \cdot 2^{k+2}\qedhere\]
\end{proof}

As a consequence, the number of splitters of any balancer discussed in
\Cref{sec:balancers} is $O(n \log n)$ where $n = \max\{|I|,|O|\}$. We
proceed to establish a corresponding lower bound.

\begin{proof}[Proof of \Cref{thm:lower-bound}] 

  Let $\splitnetwork$ be a balancer network. Let
  $c : I \disunion O \to [0,1]$ be a capacity function. We initialize
  a capacity $c'$ with $c'(i) = 0$ for each $i \in I$ and
  $c'(o) = c(o)$ for each $o \in O$. We then incrementally increase
  the capacity of each input $i$, from $0$ to $c(i)$, recalculating a
  new steady-state at each step. Increasing an input capacity
  transforms a steady-state into a sub-steady-state, therefore each
  iteration performs a series of augmentations. Denote by
  $f_1,\ldots,f_m$ the sequence of augmenting circulations computed
  until reaching the final steady-state.

  As $G$ is a balancer, and each intermediary sub-steady-state is a
  steady-state for some choice of input capacities, each augmenting
  circulation $f_k$ is uniform on $\delta^-(O)$. By construction $f_k$
  is non-zero on at most one input arc $e_k \in \delta^+(I)$. The
  support of $f_k$ is defined as the subgraph of all arcs $e$ such
  that $f_k(e) > 0$. We will assume that the support of each
  circulation intersects $\delta^+(I)$ (and thus increases the global
  throughput), as other circulations will not contribute to the proof
  and thus can be ignored.

  Consider the support of $f_k$ on $G$. Reverse all saturated arcs, to
  ensure that $f_k$ is a flow on this graph. Then contract each arc
  $uv \neq e_k$ with $|\delta^+(u)| = 1$, by identifying
  $u$ and $v$ and removing $uv$. This action eliminates all the
  vertices with out-degree one, except for one input $i_k$. The
  resulting graph is denoted $H_k$, and the flow $g_k$, the
  restriction of $f_k$ on $H_k$, represents a flow from $i_k$ to $O$.
  Let $\delta^+_k$ and $\delta^-_k$ denote the incidence functions of
  $H_k$. The following properties hold:

\begin{claim}
  \begin{enumerate}[nosep,label=(\roman*)]
  \item\label{item:outdeg2} each remaining splitter $s$ in $H_k$ has out-degree 2;
  \item\label{item:splitconst} for each vertex $v$ in $H_k$, $g_k$ is constant on
    $\delta^+_k(v)$;
  \item\label{item:outconst} $g_k$ is constant on $\delta^-_k(O) \cap F$.
  \end{enumerate}
\end{claim}

\begin{proof}
  \ref{item:outdeg2} and \ref{item:splitconst} follow from the fact
  that each vertex other than $i_k$ in the residual graph has
  out-degree at most 2 and its outgoing arcs are coupled. This last
  fact follows by a simple case analysis on which incident arcs are
  saturated.

  By construction, $f_k$ is an augmenting circulation between two
  steady-states of a balancer. Consequently, $f_k$ is the difference
  of two throughput functions that are both uniform on $\delta^-(O)$.
  Therefore $f_k$ is itself uniform on fluid arcs of $\delta^-(O)$,
  and so is $g_k$, proving~\ref{item:outconst}.
\end{proof}

We proceed by constructing a directed arborescence rooted at $i_k$,
which may be infinite. This is achieved by establishing a parent-child
relation among the walks originating from $i_k$ in $H$. A walk $w$ is a
parent of a walk $w'$ if the length of $w'$ is one plus the length of
$w$, and $w$ is a prefix of $w'$. This defines an arborescence $(T,E_T)$
whose root $r$ is the empty walk and each node is a walk.
Additionally, there is a natural morphism $\phi$ from each walk
$w \in T$ to its end vertex, mapping each parent-child pair to a
directed path in $H$ from $\phi(\textrm{parent})$ to
$\phi(\textrm{child})$. Observe that nodes in the arborescence, whose
end vertices are splitters, possess two children. In contrast, nodes
whose end vertices are outputs have no children. Let us introduce the
function $p : E_T \to [0,1]$, where $p(w,w')$ denotes the probability of
a random walk originating from $i_k$ having $w'$ as a prefix.
Specifically, because the tree is binary,
$p(w,w') = 2^{-\textrm{depth}(w)}$. For a splitter $s$, let $W_s$ be
the set of all walks whose end vertex is $s$. The expected number of
occurences of $s$ in a random walk starting from $i_k$, is
$\Lambda(s) = \sum_{w' \in W_s} p(ww')$.

Using $p$, we construct a flow on the graph $H_k$. For each arc $a$,
let $\phi^{-1}(a) \subseteq E_T$ be the set of edges $e$ in the
arborescence such that $a \in \phi(e)$. Then then flow value $f'_k(a)$
on an arc $a$ is defined by $f'_k(a) = \sum_{e \in \phi^-1(e)} p(e)$.
Due to its construction, $f'_k$ satisfies the conservation rules on
each splitter, hence is a flow from $i_k$ to $O$. Observe that the
augmenting flows are defined by a linear system. For each node
$v \in S \cup \{i_k\}$ of $H_k$, this linear system contains one
variable, representing the amount of flow on each outgoing arc, and
one constraint, representing the conservation rule. The constraints
are linearly dependant, as their sum is zero, but have rank one less
than the number of variables. Hence the solution space has dimension
1. Therefore $f_k$ and $f'_k$ are identical up to a scaling factor:
$f_k = f_k(e_k) \cdot f'_k$. Because $f_k$ is uniform on
$\delta^-(O)$, so is $f'_k$. This implies that $(T,E_T)$ is the binary
decision tree of a sampling process over the uniform distribution on
$O$. By \Cref{thm:knuth-yao}, in the residual graph,
\[
  \sum_{s \in S} \sum_{e \in \delta_k^-(s)} f_k(e)
  = f_k(e_k) \sum_{s \in S} \Lambda(s) 
  \geq f_k(e_k) |O| \sum_{k \in \mathbb{N}} \frac{k}{2^k} \binary[k]{|O|}.
\]
Summing over all augmenting circulation $f_1,\ldots,f_m$, we obtain:
\[ \sum_{k=1}^m \sum_{s \in S} \sum_{e \in \delta_k^-(s)} f_k(e) 
   \geq |I| |O| \sum_{k \in \mathbb{N}} \frac{k}{2^k} \binary[k]{|O|}.
\]

We conclude the proof by analysing the contribution of each arc to the
left-hand side of the inequality. Because of \Cref{coro:monotonicity},
a fluid arc entering a splitter contributes up to a value of $1$,
until it reaches a throughput of $1$. Then it may contribute as a
saturated arc to the splitter from which it as a leaving arc, again by
a total amount of at most $1$, until it reaches a throughput of $0$.
Each leaving arc of a splitter contributes at most 1 when
saturated, and each entering arc of a splitter contributes at most 1
when fluid. Therefore, the total contribution is at most $4|S|$, yielding:
\[|S| \geq \frac{1}{4} |I| |O| \sum_{k \in \mathbb{N}} \frac{k}{2^k} \binary[k]{|O|}.\]

In the case when the final steady-state contains only fluid arcs, we
do not count the contribution of saturated arcs. This leads to
\Cref{thm:weak-lower-bound}:
\[|S| \geq \frac{1}{2} |I| |O| \sum_{k \in \mathbb{N}} \frac{k}{2^k} \binary[k]{|O|}.\qedhere\]
\end{proof}




\section{Simulating an arbitrary capacity}\label{sec:capacity}

We present a way to build, for any rational
$r \in [0,1] \cap \mathbb{Q}$, a continuous splitter network with one
input vertex $i$ and one output vertex $o$, such that, when
$c(i) = 1$, then the throughput of any equilibrium of this network is
$\min \{r, c(o)\}$. Those networks can be used as gadgets to simulate
arcs with rational capacities in larger networks, thanks to the
additional property that the arc leaving $i$ may be set as saturated
if and only if its throughput is $r$ or $c(o)$.

Let $r = \frac{p}{q} \in \mathbb{Q}$ be a positive
rational with $0 < p < q$. Let $k \in \mathbb{N}$ be such that
$2^{k-1} < q \leq 2^k$, $k >= 1$.

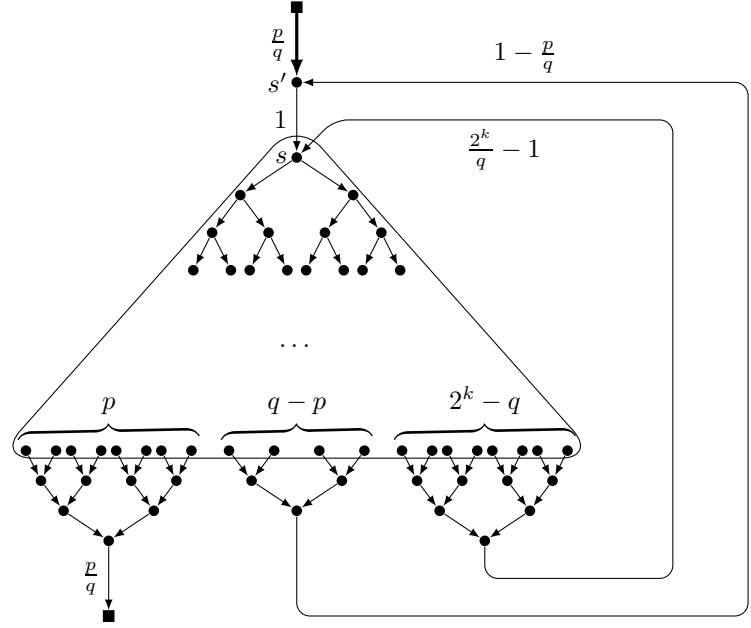
\begin{figure}
  \begin{center}
    \small
    \begin{tikzpicture}[x=1cm,y=1cm,>=latex]
      \begin{scope}[->,level distance=0.5cm, level/.style={sibling distance=1.5cm/#1}]
        \node[splitter] (root) at (3,3) {}
          child foreach \x in {0,1} { node[splitter] {}
            child foreach \y in {0,1} { node[splitter] {}
              child foreach \z in {0,1} { node[splitter] {}
              }
            }
          };
      \end{scope}
      \node[term] (input) at (3,5) {};
      \node[splitter] (mix-input) at (3,4) {};
      \draw (root) node[anchor=east] {$s$};
      \draw (mix-input) node[anchor=east] {$s'$};
      \draw[->,saturated] (input) -- (mix-input) node[midway,left] {$\frac{p}{q}$};
      \draw[->,fluid] (mix-input) -- (root) node[midway,left] {$1$}; 
      \draw[rounded corners=14pt,thin] (-1,-1) -- (3,3.5) -- (7,-1) -- cycle;
      
      \begin{scope}[<-,fluid,grow=up,level distance=0.4cm, level/.style={sibling distance=1.2cm/#1}]
        \node[splitter] (outn) at (0.5,-2.1) {}
          child foreach \x in {0,1} { node[splitter] {}
            child foreach \y in {0,1} { node[splitter] {}
              child foreach \z in {0,1} { node[splitter] (n\x\y\z) {}
              }
            }
          };
      \end{scope}     
      \begin{scope}[<-,fluid,grow=up,level distance=0.4cm, level/.style={sibling distance=1.2cm/#1}]
        \node[splitter] (outp) at (3,-1.7) {}
          child foreach \x in {0,1} { node[splitter] {}
            child foreach \y in {0,1} { node[splitter] (p\x\y) {}
            }
          };
      \end{scope}     
      \begin{scope}[<-,fluid,grow=up,level distance=0.4cm, level/.style={sibling distance=1.2cm/#1}]
        \node[splitter] (outa) at (5.5,-2.1) {}
          child foreach \x in {0,1} { node[splitter] {}
            child foreach \y in {0,1} { node[splitter] {}
              child foreach \z in {0,1} { node[splitter] (a\x\y\z) {}
              }
            }
          };
      \end{scope} 
      \node[term] (output) at (0.5,-3.1) {};
      \draw[fluid,->] (outn) -- (output) node[midway,left] {$\frac{p}{q}$};
      \draw[fluid,rounded corners=5pt,->] (outp) -- (3,-3.1) -- (9,-3.1) -- (9,4) -- (mix-input)
        node[pos=0.5,above] {$1 - \frac{p}{q}$};
      \draw[fluid,rounded corners=5pt,->] (outa) -- (5.5,-2.6) -- (8,-2.6) -- (8,3.5) -- (3.5,3.5)
        node[pos=0.5,below] {$\frac{2^k}{q} - 1$} --(root);
      \draw (3,0.5) node {\ldots};
      \draw[decoration={calligraphic brace,amplitude=5pt}, decorate, line width=1pt]
        ($(n111)+(-0.1,0.15)$) -- ($(n000) +(0.1,0.2)$) node[midway,above=5pt] {$p$};
      \draw[decoration={calligraphic brace,amplitude=5pt}, decorate, line width=1pt]
        ($(p11)+(-0.1,0.15)$) -- ($(p00) +(0.1,0.2)$) node[midway,above=5pt] {$q-p$};
      \draw[decoration={calligraphic brace,amplitude=5pt}, decorate, line width=1pt]
        ($(a111)+(-0.1,0.15)$) -- ($(a000) +(0.1,0.2)$) node[midway,above=5pt] {$2^k - q$};
    \end{tikzpicture}
    \end{center}
    \caption{An inefficient network to simulate a rational capacity of
      $\frac{p}{q}$, where $2^{k-1} < q \leq 2^k$. This only works
      when $\frac{p}{q} \geq 1 - \frac{p}{q}$, because we expect the
      arc leaving the input node to saturate as soon as the throughput
      reaches $\frac{p}{q}$. }
  \label{fig:trees-network}
\end{figure}

We construct a splitter network from a complete binary
out-arborescence of depth $k$ from a source splitter $s$. The
throughput entering $s$ is fairly split among the $2^k$ leaves. Then
we partition the leaves into three sets $P, Q, R$, with $|P| = p$,
$|Q| = q - p$ and $|R| = 2^k - q$. Route all the flow from $R$ back to
the root of the tree, by adding an in-arborescence from these leaves
to $s$. Then a random walk from $s$ ends in $P$ with probability
$\frac{p}{q}$, and in $Q$ with probability $1 - \frac{p}{q}$. If we
send all the flow from $P$ to an output, and all the flow from $Q$
back to the input, as illustrated in Figure~\ref{fig:trees-network},
we get a splitter network with maximal throughput $r$.

\begin{figure}
  \begin{center}
    \small
    \begin{tikzpicture}[>=latex,x=1.2cm,y=1.2cm]
      \node[term] (i) at (9,1) {};
      \node[term] (o) at (9,-1) {};
      \foreach \x in {0,1,...,8} {
        \node[splitter] (u\x) at (\x,0) {};
      }
      \foreach \x/\p/\q  in {0/1/2,2/1/8,4/2/63,6/1/126} {
        \node[splitter] (v\x) at (\x,1) {};
        \draw[fluid,->] (u\x) -- (v\x) node[right=-3pt,midway] {$\frac{\p}{\q}$};
      }
      \foreach \x/\p/\q in {1/1/4,3/4/63,5/1/53,7/1/252,8/1/504} {
        \node[splitter] (w\x) at (\x,-1) {};
        \draw[fluid,->] (u\x) -- (w\x) node[right=-3pt,midway] {$\frac{\p}{\q}$};
      }
      \foreach \i\j\p\q in {6/4/173/504,4/2/3/8,2/0/1/2} {
        \draw[saturated,->] (v\i) -- (v\j) node[midway,above=-2pt] {$\frac{\p}{\q}$};
      }
      \foreach \i\j\p\q in {0/1/1/2,1/2/1/4,3/4/4/63,5/6/1/63,7/8/1/252} {
        \draw[fluid,->] (u\i) -- (u\j) node[midway,above=-2pt] {$\frac{\p}{\q}$};
      }
      \foreach \i\j\p\q in {2/3/1/8,4/5/2/63,6/7/1/126} {
        \draw[fluid,->] (u\i) -- (u\j) node[midway,below=-2pt] {$\frac{\p}{\q}$};
      }
      \foreach \i\j\p\q in {1/3/1/4,3/5/79/252,5/7/83/252,7/8/84/252} {
        \draw[fluid,->] (w\i) -- (w\j) node[midway,below=-2pt] {$\frac{\p}{\q}$};
      }
      \draw[saturated,->] (i) -- (v6) node[midway,above=-2pt] {$\frac{169}{504}$};
      \draw[fluid,->] (w8) -- (o) node[midway,below=-2pt] {$\frac{169}{504}$};
      \draw[fluid,->] (v0) .. controls (-1,1) and (-1,0) .. (u0) node[midway,left=-3pt] {$1$};
      \draw[fluid,->,rounded corners=5pt] (u8) -- (10,0) -- (10,-2)
        node[midway,right=-3pt] {$\frac{1}{504}$}  -- (2.75,-2) -- (2.75,-0.25) -- (u3);

      \node (u) at (-1.5,0) {$u$:};
      \node (v) at (-1.5,1) {$v$:};
      \node (w) at (-1.5,-1) {$w$:};
      \foreach \x/\b in {0/0,1/1,2/0,3/1,4/0,5/1,6/0,7/1,8/1} {
        \node (b\x) at (\x,2) {$\b$};
      }
    \end{tikzpicture}
  \end{center}
  \caption{A network with maximal throughput $r = \frac{169}{504} = \frac{169}{8 \cdot 63}$,
    built from the binary representation of $r$: $0.010(101011)^\omega$.}
  \label{fig:smart-binary-capacity-network}
\end{figure}
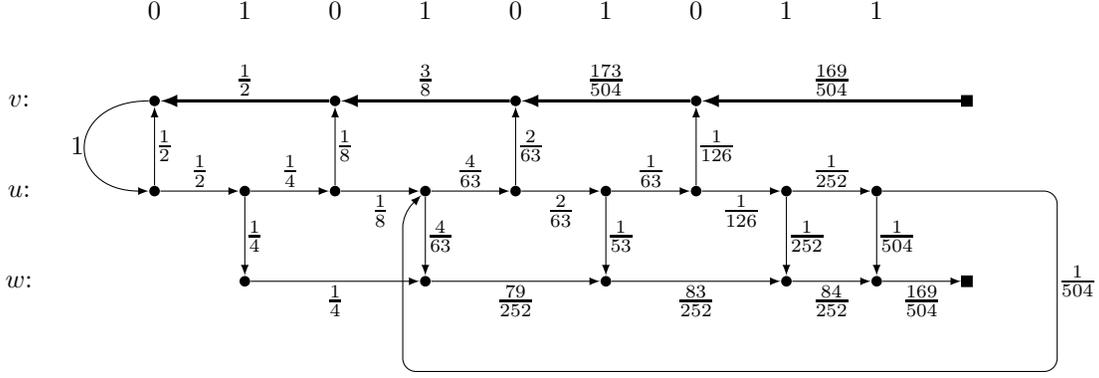

Informally, the arc $s's$ entering the root of the main arborsecence
serves as the bottleneck of that network. Its throughput, when capped
at $1$, equals the sum of the input throughput and the throughputs
from $q-p$ leaves. By the rule of conservation~\ref{eqn:conservation},
the input throughput equals the output throughput, which is the sum of
throughputs of $p$ leaves. Therefore the throughput $x$ at each leaf
satisfies $px + (q-p) x = t(ss')$, that is $x = \frac{t(ss')}{q}$. It
follows that, when $t(ss')$ is capped at $1$, the input throughput is
$\frac{p}{q} = r$. We do not prove precisely this result because this
construction has two main shortcomings:
\begin{itemize}[label=\bull,nosep]
\item it only holds as long as $\frac{p}{q} \geq \frac{1}{2}$.
  Otherwise, the tree collecting the flow back from the $p-q$ leaves
  to the input would become saturated, while we would rather have that
  the arc $is'$ is saturated.
\item the size of this construction is linear in $q$ (hence
exponential in the size of the encoding of the capacities).
\end{itemize}
Optimizing over that solution would overcome these shortcomings, but
we will rather directly define a correct network with an optimal
number of splitters (up to a multiplicative constant). Before
introducing an efficient design, we consider another construction
based on the binary expansion of $r$. Because $r = \frac{p}{q}$ is
rational, its binary expansion is ultimately periodic: there exist
binary words $x, y \in \{0,1\}^{\star}$ with
$\textrm{binary}(r) = 0.xy^\omega$. Let $l = |x| + |y|$. We define the
splitter network
$G'_r \eqdef (V \eqdef I \disunion S \disunion O, E \eqdef E_0 \cup E_r \cup E_{1-r})$
with
\begin{itemize}[label=\bull,nosep]
\item $I \eqdef \{ i \}$, $O \eqdef \{ o \}$,
\item $S \eqdef \bigcup_{j=1}^{l} \{u_j, v_j, w_j\}$,
\item
  $E_0 \eqdef \{ iv_l, v_1u_1, w_lo, u_lu_{|y|+1} \} \cup \bigcup_{j=1}^{l-1} \{u_ju_{j+1}, v_{j+1}v_j, w_jw_{j+1}\}$,
\item $E_r \eqdef \{u_iv_i~:~\textrm{for each
    $i \in \interval{1}{l}$ with $(xy)_i = 0$}\}$,
\item $E_{1-r} \eqdef \{u_iw_i~:~\textrm{for each $i \in \interval{1}{l}$ with $(xy)_i = 1$}\}$.  
\end{itemize}
Then $G_r$ is obtained by contracting any splitter $s$ with
$d^+(s) = d^-(s) = 1$ into a single arc.
Figure~\ref{fig:smart-binary-capacity-network} shows an example of
network $G_r$ with $r = \frac{169}{504}$. One way to understand this design
is to unfold the loop made by $u_{l-1}u_{|y|+1}$, making the $u$, $v$
and $w$ lines extend infinitely to the right; this gives a network
that simply split the flow following the binary expansion of $r$, each
vertex $u_i$ distributing $2^{-i}$ unit of flow. This construction
seems more compact than the construction based on trees, but can still
have a large number of splitters, as $l$ can be as large as $q-1$.
Also, it still needs to be adapted for some values of $r$ to make sure
that the only saturated arcs are the arcs $v_{i+1}v_i$. Namely, it is
sufficient to choose a periodic representation where the period starts
with a $1$.

Our best construction relies on combining the main ideas of the two
previous constructions: 
\begin{itemize}[label=-,nosep]
\item split the flow into $2^k$ equal chunks, and make three groups of
  size $p$, $q - p$, $2^k - q$, so that we get three flows of size
  $\lambda p 2^{-k}$, $\lambda (q-p) 2^{-k}$ and $1 - q 2^{-k}$. One
  flow is output, one loops back to the head of the bottleneck arc,
  and one loops back to the tail of the bottleneck arc.
\item use the binary representations of each group size to create the
  three flows.
\end{itemize}
We first give the construction for $r = \frac{p}{q} \geq \frac{1}{2}$,
and we will later show how to extend it to any rational value.

For an integer $n \in \mathbb{N}$, denote $\binary{n}$ the binary
representation of $n$, and $\binary[i]{n}$ the value of the bit of weight
$2^i$ in that representation. 

\begin{definition}
  Let $p_1, p_2,\ldots,p_l,k \in \mathbb{N}_{>0}$ be integers with
  $\sum_{i=1}^l p_i = 2^k$. Let $T_{p_1,p_2,\ldots,p_l,k}$ be a binary
  out-arborescence where each leaf is labeled by an integer in
  $\interval{1}{l}$, such that for any label $i \in \interval{1}{l}$
  and any depth $d \in \interval{1}{k}$, the number of leaves with
  label $i$ at depth $d$ is $\binary[k-d]{p_i}$.
\end{definition}

Notice that $T_{p_1,p_2,\ldots,p_l,k}$ always exists, but is not
necessarily unique. We prove its existence by showing that
the number $n_d$ of inner nodes at depth $d \in \interval{0}{k}$ is
$$ n_d = 2^d - \sum_{i=1}^l \left\lfloor \frac{p_i}{2^{k-d}} \right\rfloor.$$

The proof is by induction on $d$, where $n_0 = 1$. Let
$d \in \interval{1}{k}$, and assume that
$n_{d-1} = 2^{d-1} - \sum_{i=1}^l \left\lfloor \frac{p_i}{2^{k-d+1}} \right\rfloor$.
Then the number of nodes at depth $d$ is twice $n_{d-1}$, to which we
subtract the leaves at depth $d$. This yields
\begin{align*}
  n_d 
  &= 2n^{d-1} - \sum_{i=1}^l \binary[k-d]{p_i} \\
  &= 2 \cdot \left(2^{d-1} - \sum_{i=1}^l \left\lfloor \frac{p_i}{2^{n-d+1}}\right\rfloor\right) - \sum_{i=1}^l \binary[k-d]{p_i}\\
  &= 2^d - \sum_{i=1}^l 2\left\lfloor \frac{p_i}{2^{n-d+1}}\right\rfloor + \binary[k-d]{p_i}\\
  &= 2^d - \sum_{i=1}^l \left\lfloor \frac{p_i}{2^{k-d}} \right\rfloor.
\end{align*}
As $\sum_{i=1}^l p_i \leq 2^k$, $n_d$ is non-negative, hence
$T_{p_1,p_2,\ldots,p_l,k}$ exists.

\begin{definition}
  Given $r = \frac{p}{q} \in \mathbb{Q}$, with
  $\frac{1}{2} \leq \frac{p}{q} < 1$, $\textrm{gcd}(p,q) = 1$, and
  $k \in \mathbb{N}_{>0}$ such that $2^{k-1} < q \leq 2^k$. Let
  $T = T_{p,q-p,2^k-q,k}$ with root $r$. We define the splitter network
  $G_{p/q} \eqdef (V \eqdef I \disunion S \disunion O, E)$ where
\begin{itemize}[label=\bull,nosep]
\item $I \eqdef \{i\}$, $O = \{o\}$ and $S = \{r' \} \disunion V(T)$,
\item
  $E \eqdef E(T) \disunion E(P_1) \disunion E(P_2) \disunion E(P_3) 
     \disunion \{ir', r'r,t_1o,t_2r',t_3r\}$,
\item for each label $\gamma \in \{1,2,3\}$, $P_\gamma$ is a
  path connecting all the leaves with label $\gamma$ in increasing
  depth order, and $t_\gamma$ is the end vertex of $P_\gamma$.
\end{itemize}
\end{definition}

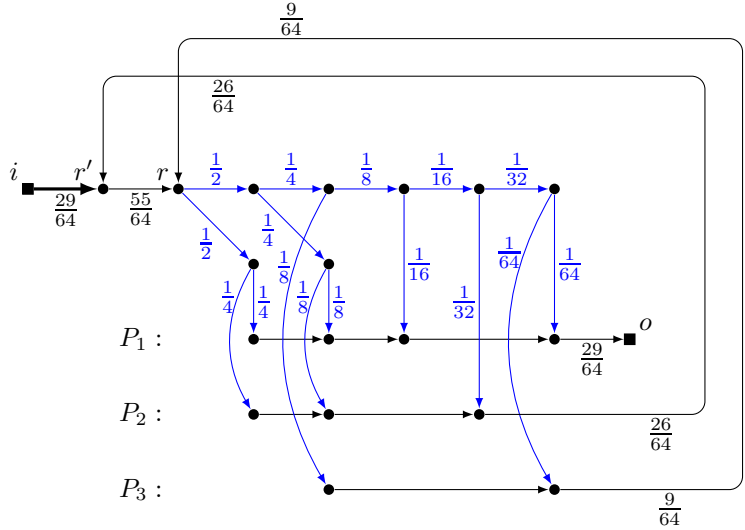
\begin{figure}
  \centering
  \small
  \begin{tikzpicture}[x=1cm,y=1cm,>=latex]
    \foreach \x in {0,1,...,5} {
      \node[splitter] (u\x) at (\x,4) {};
    }
    \foreach \x in {1,2} {
      \node[splitter] (v\x) at (\x,3) {};
    }
    \foreach \x in {1,2,3,5} {
      \node[splitter] (w\x) at (\x,2) {};
    }
    \foreach \x in {1,2,4} {
      \node[splitter] (y\x) at (\x,1) {};
    }
    \foreach \x in {2,5} {
      \node[splitter] (z\x) at (\x,0) {};
    }
    \node[term] (i) at (-2,4) {};
    \node[splitter] (mix) at (-1,4) {};
    \node[term] (out) at (6,2) {};
    \foreach \i/\j/\p/\q in {0/1/1/2,1/2/1/4,2/3/1/8,3/4/1/16,4/5/1/32} {
      \draw[->,fluid,blue] (u\i) -- (u\j) node[midway,above=-2pt] {$\frac{\p}{\q}$};
    }
    \foreach \l/\i/\j in {w/1/2,w/2/3,w/3/5,y/1/2,y/2/4,z/2/5} {
      \draw[->,fluid] (\l\i) -- (\l\j);
    }
    \draw[->,fluid,blue] (u0) -- (v1) node[midway,below left=-3pt] {$\frac{1}{2}$};
    \draw[->,fluid,blue] (u1) -- (v2) node[pos=0.3,below left=-3pt] {$\frac{1}{4}$};
    \draw[->,fluid,blue] (u3) -- (w3) node[midway,right=-3pt] {$\frac{1}{16}$};
    \draw[->,fluid,blue] (u4) -- (y4) node[midway,left=-3pt] {$\frac{1}{32}$};
    \draw[->,fluid,blue] (u5) -- (w5) node[midway,right=-3pt] {$\frac{1}{64}$};
    \draw[->,fluid,blue] (v1) -- (w1) node[midway,right=-3pt] {$\frac{1}{4}$};
    \draw[->,fluid,blue] (v2) -- (w2) node[midway,right=-3pt] {$\frac{1}{8}$};
    \draw[->,fluid,blue] (v1) to[bend right] node[pos=0.2,left=-3pt] {$\frac{1}{4}$} (y1);
    \draw[->,fluid,blue] (v2) to[bend right] node[pos=0.2,left=-3pt] {$\frac{1}{8}$} (y2);
    \draw[->,fluid,blue] (u2) to[bend right] node[pos=0.25,left=-3pt] {$\frac{1}{8}$}  (z2);
    \draw[->,fluid,blue] (u5) to[bend right] node[pos=0.2,left=-3pt] {$\frac{1}{64}$} (z5);
    \draw[->,saturated] (i) -- (mix) node[midway,below=-2pt] {$\frac{29}{64}$};
    \draw[->,fluid] (mix) -- (u0) node[midway,below=-2pt] {$\frac{55}{64}$};
    \draw[->,fluid] (w5) -- (out) node[midway,below=-2pt] {$\frac{29}{64}$};
    \draw[->,fluid,rounded corners=5pt] (y4) -- (7,1) node[pos=0.8,below=-2pt] {$\frac{26}{64}$}
      -- (7,5.5) -- (-1,5.5) node[pos=0.8,below=-2pt] {$\frac{26}{64}$} -- (mix);
    \draw[->,fluid,rounded corners=5pt] (z5) -- (7.5,0) node[pos=0.6,below=-2pt] {$\frac{9}{64}$}
      -- (7.5,6) -- (0,6) node[pos=0.8,above=-2pt] {$\frac{9}{64}$} -- (u0);
    \node (u) at (-0.5,2) {$P_1:$};
    \node (v) at (-0.5,1) {$P_2:$};
    \node (w) at (-0.5,0) {$P_3:$};
    \path (i) node[anchor=south east] {$i$};
    \path (mix) node[anchor=south east] {$r'$};
    \path (u0) node[anchor=south east] {$r$};
    \path (out) node[anchor=south west] {$o$};

  \end{tikzpicture}
  \caption{A network of size $O(\log q)$ with maximum throughput
    $\frac{29}{55}$. The throughput values shown on the figure should
    be scaled by a factor $\frac{64}{55}$ to get the value at
    equilibrium. The lines $P_1$, $P_2$, $P_3$ are given by the binary
    encoding of $p = 29 = 011101$, $q = 26 = 011010$ and
    $2^k - q = 9 = 001001$.
    \Cref{fig:reversed-optimal-capacity-network} $(a)$ gives another
    representation of the same network.}
  \label{fig:optimal-capacity-network}
\end{figure}

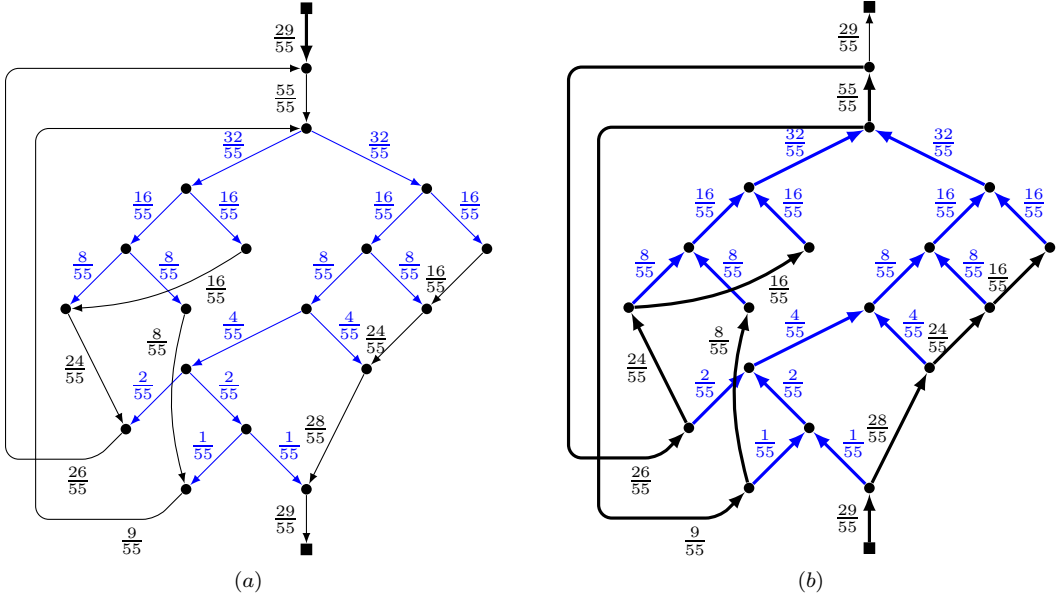
\begin{figure}
  \centering\scriptsize
  \begin{tabular}{cp{0.15cm}c}
  \begin{tikzpicture}[x=0.8cm,y=0.8cm,>=latex]
    \node[term] (out) at (4,9) {};
    \node[term] (in) at (4,0) {};
    \node[splitter] (mix) at (4,8) {};
    \node[splitter] (r) at (4,7) {};
    \node[splitter] (a) at (2,6) {};
    \node[splitter] (b) at (6,6) {};
    \node[splitter] (c) at (1,5) {};
    \node[splitter] (d) at (3,5) {};
    \node[splitter] (e) at (5,5) {};
    \node[splitter] (f) at (7,5) {};
    \node[splitter] (g) at (0,4) {};
    \node[splitter] (h) at (2,4) {};
    \node[splitter] (i) at (4,4) {};
    \node[splitter] (j) at (6,4) {};
    \node[splitter] (k) at (2,3) {};
    \node[splitter] (l) at (5,3) {};
    \node[splitter] (m) at (1,2) {};
    \node[splitter] (n) at (3,2) {};
    \node[splitter] (o) at (2,1) {};
    \node[splitter] (p) at (4,1) {};
    \foreach \u/\v/\p/\q in {a/r/32/55,c/a/16/55,e/b/16/55,g/c/8/55,i/e/8/55,k/i/4/55,m/k/2/55,o/n/1/55} {
      \draw[fluid,blue,<-] (\u) -- (\v) node[midway,above left=-3pt] {$\frac{\p}{\q}$};
    }
    \foreach \u/\v/\p/\q in {b/r/32/55,d/a/16/55,f/b/16/55,h/c/8/55,j/e/8/55,l/i/4/55,n/k/2/55,p/n/1/55} {
      \draw[fluid,blue,<-] (\u) -- (\v) node[midway,above right=-3pt] {$\frac{\p}{\q}$};
    }
    \foreach \u/\v/\p/\q in {m/g/24/55,in/p/29/55,p/l/28/55,l/j/24/55,j/f/16/55,r/mix/55/55} {
      \draw[fluid,<-] (\u) -- (\v) node[midway,left] {$\frac{\p}{\q}$};
    }
    \draw[fluid,<-] (g) to[bend right=15] node[pos=0.85,below] {$\frac{16}{55}$} (d);
    \draw[saturated,<-] (mix) -- (out) node[left,midway] {$\frac{29}{55}$};
    \draw[fluid,<-] (o) to[bend left=15] node[left,pos=0.85] {$\frac{8}{55}$}  (h);
    \draw[fluid,<-,rounded corners=5pt] (mix) -- (-1,8) -- (-1,1.5) 
    -- node[pos=0.8,below] {$\frac{26}{55}$} (0.5,1.5) -- (m);
    \draw[fluid,<-,rounded corners=5pt] (r) -- (-0.5,7) -- (-0.5,0.5) 
    -- node[pos=0.8,below] {$\frac{9}{55}$} (1.5,0.5) -- (o);
  \end{tikzpicture}
    & &
  \begin{tikzpicture}[x=0.8cm,y=0.8cm,>=latex]
    \node[term] (out) at (4,9) {};
    \node[term] (in) at (4,0) {};
    \node[splitter] (mix) at (4,8) {};
    \node[splitter] (r) at (4,7) {};
    \node[splitter] (a) at (2,6) {};
    \node[splitter] (b) at (6,6) {};
    \node[splitter] (c) at (1,5) {};
    \node[splitter] (d) at (3,5) {};
    \node[splitter] (e) at (5,5) {};
    \node[splitter] (f) at (7,5) {};
    \node[splitter] (g) at (0,4) {};
    \node[splitter] (h) at (2,4) {};
    \node[splitter] (i) at (4,4) {};
    \node[splitter] (j) at (6,4) {};
    \node[splitter] (k) at (2,3) {};
    \node[splitter] (l) at (5,3) {};
    \node[splitter] (m) at (1,2) {};
    \node[splitter] (n) at (3,2) {};
    \node[splitter] (o) at (2,1) {};
    \node[splitter] (p) at (4,1) {};
    \foreach \u/\v/\p/\q in {a/r/32/55,c/a/16/55,e/b/16/55,g/c/8/55,i/e/8/55,k/i/4/55,m/k/2/55,o/n/1/55} {
      \draw[saturated,blue,->] (\u) -- (\v) node[midway,above left=-3pt] {$\frac{\p}{\q}$};
    }
    \foreach \u/\v/\p/\q in {b/r/32/55,d/a/16/55,f/b/16/55,h/c/8/55,j/e/8/55,l/i/4/55,n/k/2/55,p/n/1/55} {
      \draw[saturated,blue,->] (\u) -- (\v) node[midway,above right=-3pt] {$\frac{\p}{\q}$};
    }
    \foreach \u/\v/\p/\q in {m/g/24/55,in/p/29/55,p/l/28/55,l/j/24/55,j/f/16/55} {
      \draw[saturated,->] (\u) -- (\v) node[midway,left] {$\frac{\p}{\q}$};
    }
    \draw[saturated,->] (r) -- (mix) node[midway,left] {$\frac{55}{55}$};
    \draw[saturated,->] (g) to[bend right=15] node[pos=0.85,below] {$\frac{16}{55}$} (d);
    \draw[fluid,->] (mix) -- (out) node[left,midway] {$\frac{29}{55}$};
    \draw[saturated,->] (o) to[bend left=15] node[left,pos=0.85] {$\frac{8}{55}$}  (h);
    \draw[saturated,->,rounded corners=5pt] (mix) -- (-1,8) -- (-1,1.5) 
    -- node[pos=0.8,below] {$\frac{26}{55}$} (0.5,1.5) -- (m);
    \draw[saturated,->,rounded corners=5pt] (r) -- (-0.5,7) -- (-0.5,0.5) 
    -- node[pos=0.8,below] {$\frac{9}{55}$} (1.5,0.5) -- (o);
  \end{tikzpicture} \\
    $(a)$ & & $(b)$
  \end{tabular}
  \caption{Two networks with maximum throughput $\frac{29}{55}$,
    reverse from each other.}
  \label{fig:reversed-optimal-capacity-network}
\end{figure}

The construction is illustrated in \Cref{fig:optimal-capacity-network}
and~\Cref{fig:reversed-optimal-capacity-network} $(a)$.

\begin{lemma}\label{lemma:capacitated-arc-equilibrium}
  $G_{p/q}$ has a steady-state $(t,F)$ with throughput
  $t^\star = \min \left\{c(i), c(o), \frac{p}{q}\right\}$. Moreover
  \begin{enumerate}[nosep,label=(\roman*)]
  \item if $t^\star = \frac{p}{q}$, we may take $F = E(G_{p/q}) \setminus \{ii'\}$;
  \item if $t^\star = c(i)$, we may take $F = E(G_{p/q})$;
  \item if $t^\star = c(o)$, we may take $\{ii',o'o\} \subseteq F$.
  \end{enumerate}
\end{lemma}

\begin{proof}
  Let $t$ be the following throughput function:
  \begin{equation*}
    t(e) \eqdef \left\{
      \begin{array}{ll}
        \frac{2^{k-d}}{q} & \textrm{if $e \in T$ is from depth $d-1$ to depth $d$,} \\
        \frac{2^{k+d}}{q} \left\lfloor \frac{v(\gamma)}{2^d}\right\rfloor & \textrm{if $e \in P_{\gamma}$, $\gamma \in \{1,2,3\}$, is from depth $d$ to depth $d' > d$,}\\
        \frac{p}{q} & \textrm{if $e \in \{ir', t_1o\}$,} \\
        1 - \frac{p}{q} & \textrm{if $e = t_2r'$} \\
        \frac{2^k}{q} - 1 & \textrm{if $e = t_3r$} \\
        1 & \textrm{if $e = r'r$}
      \end{array}
    \right.
  \end{equation*}
  where $v(1) = p, v(2) = q-p, v(3) = 2^k - q$. We check that
  $(t,E \setminus \{ir'\})$ is a steady-state when $c(i) = c(o) = 1$.
  As $2^{k-1} < q \leq 2^k$, the throughputs are between $0$ and $1$.
  The conservation rule~\ref{eqn:conservation} is easily checked on
  every vertex, except vertices of $P_\gamma$. Let $u \in V(P_\gamma)$
  at depth $d$. let $wu$ the incoming arc on $P_\gamma$, with $w$ at
  depth $d'$. Then
  $\binary[d']{v(\gamma)} = \binary[d]{v(\gamma)} = 1$, and
  $\binary[\delta]{v(\gamma)} = 0$ for each
  $\delta \in \interval{d'+1}{d-1}$. This implies
  \[2^{d'-d} \cdot 2^{d'} \left\lfloor\frac{v(\gamma)}{2^{d'}}\right\rfloor + 1 = \frac{1}{2^d} \left\lfloor \frac{v(\gamma)}{2^d} \right\rfloor.\]
  Scaling this equation by a factor $\frac{2^k}{q}$ yields the
  conservation rule for $u$. Rule~\ref{eqn:incoming-rule} at $r'$
  requires that $\frac{p}{q} \geq 1 - \frac{p}{q}$, which holds by the
  assumption that $\frac{p}{q} \geq \frac{1}{2}$. The other rules can
  be readily checked. This steady-state remains valid as long as
  $\min\{c(i),c(o)\} \geq \frac{p}{q}$.

  Suppose now that $\min\{c(i),c(o)\} \leq \frac{p}{q}$. When
  $c(i) \geq c(o)$, we build a pre-steady-state by decreasing
  $t(t_1o)$ to $c(o)$, generating some excess on $t_1$, and removing
  $t_1o$ from the fluid arcs. Then using the pre-steady-state
  algorithm, the flow will be push back to $i$ through $ir'$, when
  $ir'$ is saturated. Therefore there is a steady-state with total
  throughput $c(o)$ and $ir', t_1o$ are saturated. When $c(i) < c(o)$,
  we start from $(t,E \setminus \{ir'\})$, take the reverse
  steady-state in the reverse network, then decrease its output
  capacity on output $i$ to $c(i)$, saturating $r'i$ and inducing some
  excess at $r'$. Then we apply the pre-steady-state algorithm to find
  a steady-state. The first step of the algorithm will be to make
  $rr'$ saturated, in order to push back the excess at $r'$. From
  there all arcs are saturated. Running the algorithm to its
  termination, we obtain a steady-state for the reverse network, that
  we reverse into a steady-state for $G_{p/q}$ with only fluid arcs.
\end{proof}

When $\frac{p}{q} < \frac{1}{2}$, we construct $G_{p/q}$ from
$G_{2p/q}$. This adds at most $2\log q$ additional splitters, thus the
number of splitters of $G_{p/q}$ remains $O(\log q)$.

\begin{definition}
  Let $\frac{p}{q} \in \mathbb{Q}_{>0}$ with
  $\frac{p}{q} < \frac{1}{2}$, we define the splitter network
  $G_{p/q} \eqdef (V \eqdef I \disunion S \disunion O, E)$ where
  \begin{itemize}[label=-,nosep]
  \item $I = \{i\}$, $O = \{o\}$, $S = V(H)$,
  \item $E = \{ii', o'o, o'i'\} \disunion E(H)$,
  \end{itemize}
  where $H$ is a copy of $G_{2p/q}$ with input $i'$ and output $o'$.
\end{definition}

We extend the proof of Lemma~\ref{lemma:capacitated-arc-equilibrium}
to every rational $\frac{p}{q}$.

\begin{proof}
  We prove it by induction on the number of recursive steps in the
  definition of $G_{p/q}$. The base case is given by the first proof
  of the lemma, when $\frac{p}{q} \geq \frac{1}{2}$. Let
  $\frac{p}{q} < \frac{1}{2}$, and suppose that $G_{2p/q}$ behaves as
  expected.

  Case 1: $\frac{p}{q} \leq \min\{c(i),c(o)\}$. Let $(t',F')$ be a
  steady-state on $G_{2p/q}$ when $c'(i') = 1 = c'(o')$, obtained by
  induction hypothesis. Then $\delta^+(i') \cap F' = \emptyset$. Let
  $t(ii') = t(o'o)= t(o'i') = \frac{p}{q}$ and $t(e) = t'(e)$ for any
  other arc $e$. Then, checking all the rules on $i$, $o$, $i'$, and
  $o'$, we get that $(t, F' \cup \{ii',oi'\}$ is a steady-state, as
  expected.

  Case 2: $c(i) \leq c(o)$ and $c(i) < \frac{p}{q}$. Let $(t',F')$
  be a steady-state for $G_{2p/q}$ and
  $c'(i') = c'(o') = 2c(i) < \frac{1p}{q}$, obtained by induction
  hypothesis. Then $F' = E(G_{2p/q})$, all arcs are fluid. We extend
  this steady-state, by setting $t(ii') = t(o'o) = t(o'i') = c(i)$ and
  $t(e) = t'(e)$ for any other arcs. Then $(t,E(G_{2p/q}))$ is a
  steady-state, with only fluid arcs.

  Case 3: $c(o) < \min\{c(i), \frac{p}{q}\}$. Let $(t',F')$ be a
  steady-state on $G_{2p/q}$ when
  $c'(i') = c'(o') = 2c(o) \leq \frac{2p}{q}$, obtained by induction
  hypothesis. Then $\delta^-(o')$ and $\delta^+(i')$ are saturated in
  this steady-state. We extend it by setting
  $t(ii') = t(o'o) = t(o'i') = c(o)$, $t(e) = t'(e)$ for any other arc
  $e$, and $F = F'$. Then $(t,F)$ is a steady-state for $(G,c)$, with
  $ii'$ and $o'o$ saturated, as expected.
\end{proof}

The next theorem summarizes the previous results.

\begin{theorem}
  For any rational $r = \frac{p}{q}\in \mathbb{Q}$ with $r \in [0,1]$,
  there is a splitter network with maximum throughput $\frac{p}{q}$.
  This network may be used as a gadget to simulate an arc with
  capacity $r$. The size of this network is linear in the size of $r$
  when $r$ is encoded either as a ratio or by its binary expansion
  (prefix and period).
\end{theorem}


The network $G_{p/q}$ exhibits a curious asymmetry. It operates by
splitting the input flow carefully until reaching the exact throughput
$\frac{p}{q}$. Intuitively $G_{p/q}$ acts on the input. The reversed
network $\symmetry{G_{p/q}}$ is also a capacity-simulating network,
with the same maximum throughput. Intuitively $\symmetry{G_{p/q}}$
acts on the output, by carefully grabbing amount of flows until
reaching throughput $\frac{p}{q}$. $G_{p/q}$ and $\symmetry{G_{p/q}}$
share the same general behaviour, with two distinct strategies.
\Cref{fig:reversed-optimal-capacity-network} shows an example of
$G_{p/q}$ together with its reverse networks and the throughputs at
maximum capacity.

We cannot extend these constructions to build network with an
irrational maximum throughput.

\begin{lemma}
  For any irrational $x \in [0,1] \setminus \mathbb{Q}$, there is no
  splitter network with a single input $i$ and a single output $o$,
  such that $c(i) = c(o) = 1$, the total throughput of a steady-state
  is $x$.
\end{lemma}

\begin{proof}
  The maximal throughput $t$ of a network under some capacities is an
  optimal solution to the linear program~\eqref{eqn:presteady-state-lp}
  for some set $F$ of fluid arcs, hence $t$ must be a convex
  combination of rational vectors. By maximality of $t$, all those
  rational vectors have the same total throughput, which is a
  rational.
\end{proof}

\begin{figure}
  \centering\scriptsize
  \begin{tikzpicture}[x=0.7cm,y=0.7cm,>=latex]
    \foreach \name/\x/\y in {i1/6/13,i2/9/13,i3/12/13,o1/11/0,o2/13/0,o3/15/0,o4/17/0} {
      \node[term] (\name) at (\x,\y) {};
    }
    \foreach \x in {6,9,12,15} {
      \foreach \y in {8,9} {
        \node[splitter] (x\x-\y) at (\x,\y) {};
      }
    }
    \foreach \x in {6,9,12} {
      \foreach \y in {10,12} {
        \node[splitter] (x\x-\y) at (\x,\y) {};
      }
    }
    \foreach \x/\y in {
      5/7,7/7,8/7,10/7,11/7,13/7,14/7,16/7,
      6/6,8/6,9/6,11/6,15/6,17/6,
      7/5,9/5,10/5,12/5,14/5,16/5,
      5/4,8/4,11/4,13/4,15/4,17/4,
      4/3,6/3,7/3,9/3,10/3,12/3,14/3,16/3,
      5/2,7/2,8/2,10/2,11/2,13/2,15/2,17/2} {
      \node[splitter] (x\x-\y) at (\x,\y) {};
    }
    \foreach \u/\v in {
      x6-10/x6-9,x6-10/x9-9,x9-10/x6-9,x9-10/x9-9,x12-10/x12-9,x12-10/x15-9,
      x6-9/x6-8,x6-9/x12-8,x9-9/x9-8,x9-9/x15-8,x12-9/x12-8,x12-9/x6-8,x15-9/x9-8,x15-9/x15-8,
      x6-8/x5-7,x6-8/x7-7,x9-8/x8-7,x9-8/x10-7,x12-8/x11-7,x12-8/x13-7,x15-8/x14-7,x15-8/x16-7,
      x7-7/x6-6,x7-7/x8-6,x10-7/x9-6,x10-7/x11-6,x16-7/x15-6,x16-7/x17-6,
      x8-6/x7-5,x8-6/x9-5,x11-6/x10-5,x11-6/x12-5,x15-6/x14-5,x15-6/x16-5,
      x7-5/x5-4,x7-5/x8-4,x12-5/x11-4,x12-5/x13-4,x16-5/x15-4,x16-5/x17-4,
      x5-4/x4-3,x5-4/x6-3,x8-4/x7-3,x8-4/x9-3,x11-4/x10-3,x11-4/x12-3,x15-4/x14-3,x15-4/x16-3,
      x6-3/x5-2,x6-3/x7-2,x9-3/x8-2,x9-3/x10-2,x12-3/x11-2,x12-3/x13-2,x16-3/x15-2,x16-3/x17-2} {
      \draw[blue,->] (\u) -- (\v);
    }
    \foreach \u/\v/\p/\q in {x11-2/o1/27/49,x13-2/o2/33/49,x15-2/o3/11/49,x17-2/o4/37/49} {
      \draw[fluid,->] (\u) -- node[midway,right] {$\frac{\p}{\q}$} (\v);
    }
    \foreach \x in {6,9,12} {
      \draw[fluid,->] (x\x-12) -- node[pos=0.25,right] {$1$} (x\x-10);
    }
    \draw[fluid,->] (i1) -- node[midway,right] {$\frac{10}{49}$} (x6-12);
    \draw[fluid,->] (i2) -- node[midway,right] {$1$} (x9-12);
    \draw[fluid,->] (i3) -- node[midway,right] {$1$} (x12-12);
    \draw[fluid,->,rounded corners=5pt] (x5-2) -- node[pos=0.9,right] {$\frac{11}{49}$} (5,1) 
    -- (1.5,1) -- (1.5,11) -- (11,11) -- (x12-10);
    \draw[fluid,->,rounded corners=5pt] (x7-2) -- node[pos=0.6,right] {$\frac{49}{49}$} (7,0.5) 
    -- (1,0.5) -- (1,10.5) -- (8.5,10.5) -- (x9-10);
    \draw[fluid,->,rounded corners=5pt] (x8-2) -- node[pos=0.45,right] {$\frac{49}{49}$} (8,0) 
    -- (0.5,0) -- (0.5,10) -- (x6-10);
    \draw[fluid,->,rounded corners=5pt] (x10-2) -- node[pos=0.3,right] {$\frac{39}{49}$} (10,-1) 
    -- (-0.5,-1) -- (-0.5,12) -- (x6-12);
    \foreach \u/\v/\p in {
      x4-3/x5-2/10, 
      x5-7/x6-6/32,
      x8-7/x9-6/32,
      x10-5/x10-3/24,x10-3/x11-2/26,
      x14-5/x14-3/8,x14-3/x15-2/10,
      x17-4/x17-2/36} {
      \draw[fluid,->] (\u) -- node[midway,above right=-3pt] {$\frac{\p}{49}$} (\v);
    }
    \draw[fluid,->] (x11-7) -- node[pos=0.3,above right=-3pt] {$\frac{32}{49}$} (x13-4);
    \draw[fluid,->] (x7-3) -- node[pos=0.7,below=-1pt] {$\frac{38}{49}$} (x10-2);
    \draw[fluid,->] (x9-5) -- node[pos=0.1,above left =-3pt] {$\frac{8}{49}$} (x4-3);
    \draw[fluid,->] (x13-4) -- node[pos=0.1,below right =-3pt] {$\frac{36}{49}$} (x7-3);
    \draw[fluid,->] (x17-6) -- node[pos=0.05,below right=-3pt] {$\frac{16}{49}$} (x10-5);
    \draw[fluid,->] (x6-6) -- node[pos=0.15, left=-3pt] {$\frac{48}{49}$} (x7-2);
    \draw[fluid,->] (x9-6) -- node[pos=0.45,right =-3pt] {$\frac{48}{49}$} (x8-2);
    \draw[fluid,->] (x13-7) to[bend left=15] node[pos=0.15,above right =-3pt] {$\frac{32}{49}$} (x13-2);
    \draw[fluid,->] (x14-7) to[bend left=15] node[pos=0.15,above right =-3pt] {$\frac{32}{49}$} (x17-4);
    \path (18,9.5) node[text=blue,anchor=west] {$\frac{128}{49}$ in average};
    \path (18,8.5) node[text=blue,anchor=west] {$\frac{64}{49}$ in average};
    \foreach \y/\p in {7.5/32,6.5/16,5.5/8,4.5/4,3.5/2,2.5/1} {
      \path (18,\y) node[text=blue,anchor=west] {$\frac{\p}{49}$};
    }
  \end{tikzpicture}
  \caption{An splitter network defined from a rational vector
    $(\frac{27}{49},\frac{33}{49},\frac{11}{49},\frac{37}{49})$, with
    $(r_1,r_2,r_3,r_4) = (49,49,11)$. The acyclic subgraph $H$,
    colored in blue, is obtained by replicating the highest layers of
    a tree. All blue arcs (in $H$) between two consecutive levels have
    equal throughput indicated on the right, except for the first two
    levels that appeared by replicating the top of the tree and form a
    simple balancer.}
  \label{fig:unlimited-rational-network}
\end{figure}
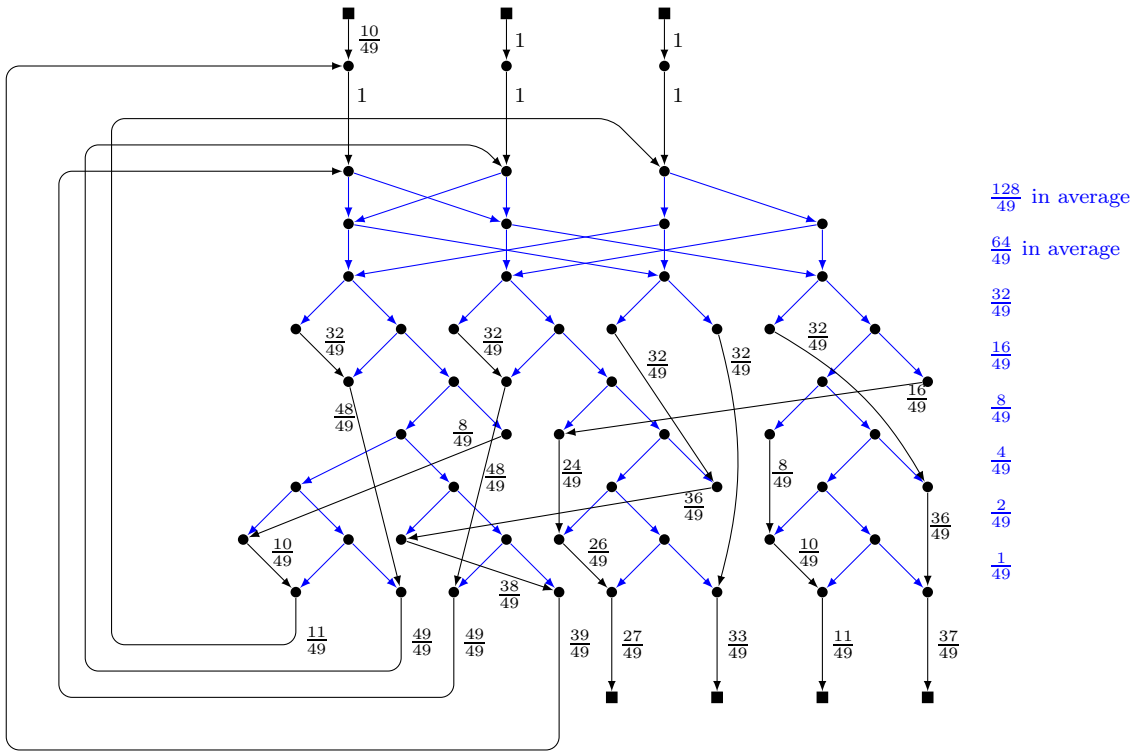

We conclude this section, by mentioning that the construction of a
capacity-simulating network can be extended to design networks that
distribute their incoming flow over their outputs following an
arbitrary rational distribution. Considering a rational vector
$(p_1,\ldots,p_l)$, and $q \in \mathbb{N}$, where $l$ is the required
number of outputs and $\frac{p_o}{q}$ is the expected throughput for
output $o$. Choose $k$ with $2^{k-1} < q \leq 2^k$. We replace the
single edge bottleneck by a bottleneck cut of size
$s = \left\lceil\frac{p_1 + \ldots + p_l}{q}\right\rceil$. Then we
define $\frac{r_i}{q}$ for $i \in \interval{1}{s}$, the amount of flow
that loop back after the bottleneck, and $\frac{q'}{q}$ the amount of
flow that loop back before the bottleneck. We start from a binary
out-arborescence $T = T_{p_1,\ldots,p_l,q',r_1,\ldots,r_s}$. As the
maximum throughput may be larger than $1$, we replicate the highest
nodes of $T$, until each layer contains at least $s$ nodes. For each
of the $s$ replicated roots, we add to incoming arc, one is in the
bottleneck cut, the other is a loopback arc with throughput
$\frac{r_i}{q}$. \Cref{fig:unlimited-rational-network} illustrates an
example of distribution network. From there, it is possible to glue a
simple balancer over the inputs to make the network throughput
unlimited, and then to add even more loopback arcs, to make a
\emph{half-universal distribution network}, and duplicate it to define
ultimately a \emph{universal distribution network}. In such a network,
the throughputs on fluid outputs arcs stay proportional to each other,
following the required distribution.

\section{Structure of the steady-states of a splitter network}\label{sec:uniformSSS}

\subsection{Uniqueness of steady-states}\label{sec:unique}

In this section, we investigate the structure of the set of
sub-steady-states, and prove that, in a splitter network with input
and output capacities, all the steady-states induces the same
throughput of its inputs and outputs. Our strategy relies on
restricting sub-steady-states to uniform sub-steady-states, where all
inputs contribute with an equal throughput as much as possible. Then
we study the two operations: making some arc saturated, and improving
the throughput along a circulation. We show that they are confluent,
therefore for any order of operations processed by the
sub-steady-state algorithm, we obtain the same final throughput. The
proof will be concluded by showing that any steady-state is reachable
by the sub-steady-state algorithm.

We will impose that all the inputs with a fluid outgoing arc must be
augmented at the same rate. This reduces the non-determinism in the
sub-steady-state algorithm and will make the proof of confluence
easier. To this end, we introduce the following rule:

\begin{enumerate}[label=R\arabic*]
\setcounter{enumi}{8}
\item \label{eqn:input-fair} there exists $\gamma \geq 0$ such that for
  any input $i \in I$ with $\delta^+(i) = \{e\}$, if $e \in F$, then
  $t(e) = \max \{c(i), \gamma\}$.
\end{enumerate}

A sub-steady-state that satisfies rule~\ref{eqn:input-fair} will be
called \emph{uniform sub-steady-state}. To simplify the proof
further, we will assume in this section that the capacities of inputs
and outputs are always $1$. There is no loss of generality in this
assumption: if an arbitrary network $(G,c)$ has several steady-states,
then its derived network $(G',\allone)$, obtained by replacing each
input and output by a rate-limiting network simulating its capacity
will also have several steady-states. Therefore, in this context of
all-one capacities, rule~\ref{eqn:input-fair} becomes: there exists
$\gamma \geq 0$ such that for each $e \in \delta^+(I) \cap F$,
$t(e) = \gamma$.

Because of the restriction to uniform sub-steady-state, we make two
modifications to the residual graph of a uniform sub-steady-state.
First, $\delta^+(I) \cap F$ is a single class in $\Cequal$. Second, we
remove from $\delta^+(z)$ the saturated arcs
$\delta^-(O) \setminus F$. Therefore, $\delta^+(z)$ corresponds to
$\delta^+(I)$, and consequently any $\Cequal$-circulation will be a
stationary circulation. In particular it will be uniform on
$\delta^+(I) \cap F$, and be able to augment the current uniform
sub-steady-state to another uniform sub-steady-state in
\Cref{lemma:circulation-augment}. The removal of
$\delta^-(O) \setminus F$ has no consequence as the throughput on these
arcs cannot change anymore.



The sub-steady-state algorithm computes a steady state by a sequence
of three \emph{elementary operations}: increasing along a stationary
circulation of the residual graph whose support intersects $\delta(z)$
(an \emph{augmentation}), removing an arc from $F$ (a
\emph{saturation}), or increasing along a stationary circulation of
the residual graph whose support does not intersect $\delta(z)$ (a
\emph{move}). By increasing along a circulation, we mean augmenting
the throughput value of fluid arcs and decreasing those of saturated
arcs. The amplitude of the change is always chosen maximum subject to
the constraint of obtaining a uniform sub-steady-state.

We recall some basic facts that will be necessary in the following
proofs. First, It can be readily checked that all saturations done in
the proof of the sub-steady-state algorithm are of one of the three
following cases.

\begin{claim}\label{claim:when-saturation}
  During the uniform sub-steady-state algorithm, an arc $e=uv$ becomes
  saturated when one of these conditions occur:
  \begin{enumerate}[label=(\roman*),nosep]
  \item $v \in O$ and $t(e) = c(e)$. Because we modified the residual
    graph in this section, when we remove $e$ from $F$, we also remove
    it from the residual graph instead of reversing it;
  \item $e$ is non-loose: it is in-coupled to an arc $e' \notin F$, and $t(e) = t(e')$;
  \item $v$ is a sink in the residual graph: arcs in $\delta^+(v)$
    either are saturated or have throughput 1, and the arc
    $e' \in \delta^-(v) \setminus \{e\}$ is fluid with
    $t(e) \geq t(e')$. $e'$ may also be saturated with
    $t(e) = t(e') = 0$, in which case $e$ is also non-loose.
  \end{enumerate}
\end{claim}

We also need to understand when an augmentation or a move happens.
Both come from $\Cequal$-circulations of the residual graph. The
support of $\Cequal$-circulations must be strongly connected
components (by virtue of being a circulation) of the residual graphs,
with no leaving arcs (because the $\Cequal$-classes correspond to the
out-incidencies of the vertices). Therefore, each such
$\Cequal$-circulation is (up to a multiplicative factor) a stationary
circulation over a sink strongly connected component of the residual
graph. This implies:

\begin{claim}\label{claim:disjoint-circulations}
  The supports of $\Cequal$-circulations that can be used in a
  \emph{move} or \emph{augment} operation are edge-disjoint and
  vertex-disjoint. Moreover, we can increase along a
  $\Cequal$-circulation if and only if its support contains only loose
  arcs. Therefore, not all $\Cequal$-circulation can be used.
\end{claim}

These two claims allow us to prove:

\begin{lemma}[Local confluence]\label{lem:confluence}
  Let $(t,F)$, $(t_1,F_1)$ and $(t_2,F_2)$ be three uniform
  sub-steady-state for a splitter network $(G,\allone)$, such that
  $(t_1,F_1)$ and $(t_2,F_2)$ are each derived from $(t,F)$ by a
  single elementary operation, respectively $o_1$ and $o_2$. Then
  $o_2$ can be applied to $(t_1,F_1)$ and $o_1$ can be applied to
  $(t_2,F_2)$, resulting in the same uniform sub-steady-state $(t',F')$.
\end{lemma}

\begin{proof}
  The proof is by a case analysis, each case corresponding to the
  choice of which operations where used to derive $(t_1,F_1)$ and
  $(t_2,F_2)$. Notice that from $(t,F)$, there can be at most one
  possible augmentation operation, using the stationary circulation
  from $z$ in the residual graph. Let $Z, Z_1, Z_2$ be the strongly
  connected sink components of the residual graphs of $(t,F)$,
  $(t_1,F_1)$ and $(t_2,F_2)$ containing $z$, respectively.

  \begin{enumerate}[label=Case \arabic*:,wide]
  \item Saturation of $e_1$ and saturation of $e_2$. By
    \Cref{claim:when-saturation}, the conditions that allow the
    saturation of an arc are local to the head of that arc, hence the
    saturations of the two arcs are independant as long as the two
    arcs have no common extremity. Specifically, there are two
    subcases: $e_1 = uv$ and $e_2 = vw$, or $e_1= uv$ and $e_2 = wv$.

    Consider the former case. Using again the locality of the
    necessary conditions to saturate $e_2$, $e_2$ can be saturated
    after $e_1$, leading from $(t,F \setminus \{e_1\})$ to
    $(t,F \setminus \{e_1,e_2\})$. If $e_1$ is non-loose (respectively
    a sink) in $(t,F)$, it is still non-loose (respectively a sink) in
    $(t,F_2)$, and hence can be removed from $F_2$ to get
    $(t,F \setminus \{e_1,e_2\})$.

    Suppose now that $e_1 = uv$ and $e_2 = wv$. As the saturation of
    each arc $e_1$, $e_2$, is possible, they are both fluid and loose,
    $v$ is a sink and $t(e_1) = t(e_2)$. Therefore, $e_2$ is non-loose
    in $(t,F_1)$ and therefore can be saturated to get
    $(t,F \setminus \{e_1,e_2\})$. Symmetrically, $e_1$ can be
    saturated in $(t,F_2)$.

  \item Move over $h_1$ and move over $h_2$. As the supports of $h_1$
    and $h_2$ are vertex-disjoint by
    \Cref{claim:disjoint-circulations}, the residual graph does not
    change from $(t,F)$ to $(t_1,F)$ on the support of $h_2$ and all
    its incident arcs. Therefore the support of $h_2$ is still a
    strongly connected sink component in the residual graph of
    $(t_1,F)$, with only loose arcs. Hence a move operation with $h_2$
    is possible, and its amplitude is also not changed as the
    throughput value have also not changed on the support of $h_2$ and
    its incident arcs. Similarly, we can apply a move with $h_1$ from
    $(t_2, F)$. In both case, we finally obtain $(t_1 \lor t_2, F)$.

  \item Augmentation over $g$ and saturation of $e$. Let $(t_1,F)$ be
    the uniform sub-steady-state obtained by augmenting over $g$. We
    consider the possible cases for $e$ from
    \Cref{claim:when-saturation}
    
    Suppose that $e$ is a non-loose arc from $u$ to an output $o$. If
    $u \in V(Z)$, then $e$ is a non-loose arc in the support of $g$,
    and augmenting over $g$ would not be possible. Therefore,
    $u \notin Z$, and the saturation of $e$ is still possible from
    $(t_1,F)$. Moreover, $Z_2 = Z$, as only $e$ is removed from the
    residual graph when it is saturated. Thus augmenting over $g$ from
    $(t, F \setminus e)$ also gives $(t_1,F \setminus \{e\})$.

    Suppose that $e$ is a non-loose arc from $u$ to $v \in S$, with
    in-coupled saturated arc $e' = wv$. Here again $u$ is not in $Z$
    (otherwise $e$ is in the support of $g$), and $v \notin Z$
    (otherwise $\symmetry{e}$ is in the support of $g$ but is
    non-loose). Thus we may saturate $e$ from $(t_1,F)$ to obtain
    $(t_1, F \setminus \{e\})$. Moreover saturating $e$ reverses the
    orientation of $e$, and both extremities of $e$ being outside $Z$,
    this implies that $Z_2 = Z$, therefore we can augment over $g$
    from $(t,F \setminus \{e\})$ to obtain $(t_1, F \setminus \{e\})$.

    The last case occurs when $e = uv$ is loose but $v$ is a sink in
    the residual graph. Hence $u, v \notin Z$, because $Z$ is a sink
    component. This implies again that $Z = Z_2$, and that the
    saturation and augmentation can be done in any order.

  \item Augmentation over $g$ and move over $h$. The support of $g$
    and $h$ are two distinct strongly connected sink component of the
    residual graph by \Cref{claim:disjoint-circulations}. Hence the
    two operations are independant and commute.

  \item Move over $h$ and saturation of $e$. The proof is similar to
    the case augmention and saturation. If $e$ is a non-loose arc, it
    must be disjoint for the support of $h$, making the two operations
    independant. If $e$ is a loose arc, then $v$ is a sink. As the
    support of $h$ is a non-trivial sink component, $e$ and the
    support are vertex-disjoint. Therefore the two operations are
    independant and commute. \qedhere
  \end{enumerate}
\end{proof}

By Newman's lemma on locally confluent binary relations, we get:

\begin{corollary}\label{coro:confluence}
  Let $(t_1,F_1)$ and $(t_2,F_2)$ be two uniform sub-steady-states
  obtained by two sequences of operations from a common uniform
  sub-steady-state $(t,F)$. Then there are two sequences of
  operations, one from $(t_1,F_1)$, the other from $(t_2,F_2)$, ending
  at a common uniform sub-steady-state $(t',F')$.
\end{corollary}

It remains to prove that any uniform-sub-steady-state is obtainable
from $(0,E)$.

\begin{lemma}\label{lemma:sss-reachable}
  For any uniform sub-steady-state $(t,F)$ distinct from $(0,E)$,
  either there is $e \in E \setminus F$ such that $(t,F \cup \{e\})$
  is a uniform sub-steady-state, or there exists a uniform
  sub-steady-state $(t',F)$ and a non-zero circulation of the residual graph
  for $(t',F)$, such that $(t,F)$ is obtained by a move or augment
  operation with that circulation from $(t',F)$.
\end{lemma}

\begin{proof}
  First we may assume that for any arc $e$ with $t(e) = 1$, $e$ is
  fluid. Otherwise, $(t,F \cup \{e\})$ is a uniform
  sub-steady-state, as required. Consider the graph $H$ with arcs
  $\{e \in F : t(e) > 0\} \cup (E \setminus F)$. If $H$ contains a
  non-zero $\Cequal$-circulation, where $\Cequal$ are the
  out-incidencies of the vertices, then we can decrease $t$ along that
  circulation, decreasing $t$ on fluid arcs and increasing $t$ on
  backward saturated arcs. This yields a throughput $t'$ with $(t',F)$
  a uniform sub-steady-state, whose residual graph admits the same
  $\Cequal$-circulation (because none of the arcs whose throughput
  differs between $t$ and $t'$ is tight). Hence $(t,f)$ is obtained
  from $(t',F)$ by a single \emph{augment} or \emph{move} operation.

  Therefore we can assume that $H$ as no non-zero
  $\Cequal$-circulation, which implies that it contains a sink vertex
  $u$, and because $t$ is non-zero, we may assume that
  $t(\delta^-(u)) \neq 0$. Because $u$ is a sink, and by the
  assumption that saturated arcs have throughput less than $1$, its
  incoming arcs must be fluid. Up to reindexing, we assume that
  $t(e_1) \geq t(e_2)$. Then
  $t(e_1) \geq \frac{1}{2}t(\delta^+(u)) = \frac{1}{2}t(\delta^-(u)) > 0$,
  therefore $e_1$ must be saturated. Then $(t,F \cup \{e_1\})$ is a
  uniform sub-steady-state: rule~\ref{eqn:outgoing-rule} holds because
  $t(e_1) \geq t(e_2)$, and rule~\ref{eqn:strong-maximization} holds
  because $\delta^-(u) \in F$.
\end{proof}

\begin{lemma}\label{lemma:once-sss}
  If $(t,F)$ is a steady-state, no \emph{augment} operation can be
  applied. Any other operations result into another steady-state with
  the same throughput on $\delta^+(I) \cup \delta^-(O)$.
\end{lemma}

\begin{proof}
  Because $(t,F)$ is a steady-state, rule~\ref{eqn:input-capacity}
  holds. Therefore for any input $i \in I$ with outgoing belt $e$,
  either $t(e) = c(i) = 1$ (recall that we assumed $c = \allone$), or
  $e \notin F$. Thus $z$ is a sink in the residual graph, proving that
  no \emph{augment} operation can happen.

  Also the throughput on $\delta^+(I) \cup \delta^-(O)$ cannot be
  changed by a \emph{move} operation, therefore
  rule~\ref{eqn:input-capacity} will still hold after any valid
  operation.
\end{proof}

\begin{theorem}
  For any splitter network $\splitnetwork$ and any capacity function
  $c: I \cup O \to [0,1]$, all steady-states have the same restriction
  on $\delta^+(I) \cup \delta^-(O)$.
\end{theorem}

\begin{proof}
  First assume $c = \allone$. Then any steady-state is reachable from
  $(0,E)$ by \Cref{lemma:sss-reachable}, and can be modified into a
  unique steady-state $(t^\uparrow_F,F)$ by \Cref{coro:confluence}
  (this notation will be justified later). By \Cref{lemma:once-sss},
  $t$ and $t^\uparrow_F$ coincides on $\delta^+(I) \cup \delta^-(O)$.

  When $c$ is not uniformly $1$, we replace each arc in
  $\delta^+(I) \cup \delta^-(O)$ by a rate-limiting network whose rate
  is the capacity of the incident input or output. Then there is an
  obvious morphism between the steady-state of the original splitter
  network and this extended splitter networks, where the restriction
  of the throughputs to $\delta^+(I) \cup \delta^-(O)$ coincides.
  Therefore the result extends to splitter networks with arbitrary
  capacities.
\end{proof}

\subsection{The lattice of uniform sub-steady-states}

Another consequence of the local confluence \Cref{lem:confluence} is
that the sequences of admissible operations starting from $(0,E)$ are
an antimatroid (see Lemma 1.2 in~\cite{greedoids}). An
\emph{antimatroid} is a family $\mathcal{A}$ of finite sequences of
symbols, satisfying the following properties
\begin{itemize}
\item (normal) each sequence in $\mathcal{A}$ contains each symbol at
  most once;
\item (hereditary) each prefix of a sequence in $\mathcal{A}$ is in
  $\mathcal{A}$ ($\mathcal{A}$ is closed by taking prefixes);
\item (anti-exchange) for each two sequences distinct $R$ and $S$ of
  $\mathcal{A}$, such that $R$ contains a symbol not in $S$, there is
  a symbol $x$ in $R$ such that the sequence $S,x$ is in
  $\mathcal{A}$.
\end{itemize}
By repeatedly applying the anti-exchange property, we can make a
sequence $R'$ from all the elements of $R$ there are not in $S$, such
that $S,R' \in \mathcal{A}$. For uniform sub-steady-states, the
symbols are the \emph{move}, \emph{augment} and \emph{saturate}
operations, and a sequence of operation can be associated to the
uniform sub-steady-state obtained by applying these operations from
$(0,E)$.

It is known that antimatroids are semimodular lattices, therefore the
set of uniform sub-steady-states is a semimodular lattice, and admits
meet and join operations. In this section, we define these two
operations explicitely, focusing at first on the restriction to a
fixed set $F$ of fluid arcs. We consider a splitter network
$\splitnetwork$, with a capacity function on $I \cup O$ and $t_1, t_2$
be two throughput functions such that $(t_1,F)$ and $(t_2,F)$ are
uniform sub-steady-states over the same set $F$ of fluid arcs.

Consider the graph $H = (z \cup S, E^+ \cup E^-)$, where
$z$ is the identification of all inputs and outputs, and
\begin{align*}
  E^+ &= \{e \in F : t_2(e) > t_1(e)\} \cup \{\symmetry{e} : e \in E \setminus F, t_2(e) < t_1(e) \},\\
  E^- &= \{\symmetry{e} : e \in F, t_2(e) < t_1(e)\} \cup \{e \in E \setminus F : t_2(e) > t_1(e), \}
\end{align*}
(some arcs are reversed, but for simplicity we will abusively identify
$e$ and $\symmetry{e}$). Let $E^= = \{ e \in E : t_1(e) = t_2(e) \}$,
such that $E = E^= \disunion E^+ \disunion E^-$.

\begin{lemma}\label{lemma:diff-sc}
  $|t_2 - t_1|$ is a circulation in $H$. Therefore the
  connected components of $H$ are strongly connected.
\end{lemma} 

\begin{proof}
  Indeed, it follows from the
  conservation rule~\ref{eqn:conservation} applied to $t_1$ and $t_2$
  at $u$ that
  \[ (t_2 - t_1)(\delta^+_G(u)) = (t_2 - t_1)(\delta^-_G(u)),\]
  which, by the choice of reversing arcs $e$ with $t_2(e) < t_1(e)$, is equivalent to 
  \[ |t_2 - t_1|(\delta^+_H(u)) = |t_2 - t_1|(\delta^-_H(u)).\qedhere\]
\end{proof}

\begin{lemma}\label{lemma:diff-sign}
  For each connected component $C$ of $H$, either
  $E(C) \subseteq E^+$ or $E(C) \subseteq E^-$. 
\end{lemma}

\begin{proof}
  Let $u \in V(H)$ and $e_1 \in \delta^-_H(u)$ and
  $e_2 \in \delta^+_H(u)$. We prove that if $e_1 \in E^-$ then
  $e_2 \in E^-$. Because we consider uniform sub-steady-states,
  $u \neq z$. For the sake of contradiction, assume that $e_1 \in E^-$
  and $e_2 \in E^+$. We consider four cases:
  \begin{enumerate}[label=Case \arabic*:,wide]
  \item $e_1, e_2 \in F$. Then
    $\delta^+_G(u) = \{\symmetry{e_3},e_4\}$, and applying
    rule~\ref{eqn:outgoing-rule} we get a contradiction:
    \[t_1(e_4) > t_2(e_4) = t_2(\symmetry{e_3}) > t_1(\symmetry{e_3}) = t_1(e_4).\]
  \item $e_3, e_4 \in E \setminus F$. Then
    $\delta^-_G(u) = \{\symmetry{e_3},e_4\}$, and
    applying~\ref{eqn:incoming-rule} we get a contradiction:
    \[t_1(e_4) < t_2(e_4) = t_2(\symmetry{e_3}) < t_1(\symmetry{e_3}) = t_1(e_4).\]
  \item $e_3 \in F$, $e_4 \in E \setminus F$. Then
    $e_4 \in \delta^-_G(u)$ and $\symmetry{e_3} \in \delta^+_G(u)$.
    But $t_1(e_4) < t_2(e_4) \leq 1$ and
    $t_1(\symmetry{e_3}) < t_2(\symmetry{e_3}) \leq 1$, therefore
    $t_1$ contradicts rule~\ref{eqn:strong-maximization}.
  \item $e_3 \in E \setminus F$, $e_4 \in F$. Then
    $\symmetry{e_3} \in \delta^-_G(u)$ and $e_4 \in \delta^+_G(u)$.
    Then $t_2(\symmetry{e_3}) < t_1(\symmetry{e_3}) \leq 1$ and
    $t_2(e_4) < t_2(e_3) \leq 1$, therefore $t_2$ contradicts
    rule~\ref{eqn:strong-maximization}.
  \end{enumerate}

  Next we prove that if $e_1 \in E^+$ then $e_2 \in E^+$. Otherwise,
  $e_1 \in E^+$ and $e_2 \in E^-$ appears consecutively in a cycle of $H$,
  as the connected components of $H$ are strongly connected by
  \Cref{lemma:diff-sc}. This cycle must contain a pair of
  consecutive arcs $e_3 \in E^-$, $e_4 \in E^+$, which cannot happen
  as we have already proved.

  As each connected component of $H$ is strongly connected, this
  implies that each component of $H$ is either in $E^+$ or in $E^-$.
\end{proof}

We define two new throughput functions from $t_1$ and $t_2$.

\begin{definition}
  Let $(t_1,F)$, $(t_2,F)$ be two uniform sub-steady-states over a
  capacitated splitter network $(G,c)$. Let $t_1 \land t_2$ and
  $t_1 \lor t_2$ be the throughput functions defined by 
  \[ (t_1 \land t_2)(e) = 
    \left\{
      \begin{array}{ll} 
        \min \{ t_1(e), t_2(e) \} & \quad\textrm{ if $e \in F$;}\\
        \max \{ t_1(e), t_2(e) \} & \quad\textrm{ if $e \notin F$.}
      \end{array}
    \right.
  \]
  and 
  \[ (t_1 \lor t_2)(e) = 
    \left\{
      \begin{array}{ll} 
        \max \{ t_1(e), t_2(e) \} & \quad\textrm{ if $e \in F$;}\\
        \min \{ t_1(e), t_2(e) \} & \quad\textrm{ if $e \notin F$.}
      \end{array}
    \right..
  \]
\end{definition}

\begin{proposition}\label{prop:sss-lattice}
  $(t_1 \land t_2,F)$ and $(t_1 \lor t_2,F)$ are uniform
  sub-steady-states.
\end{proposition}

\begin{proof}
  For each vertex $u \in V(H)$, by \Cref{lemma:diff-sign}, all arcs
  incident to $u$ are in either $E^- \cup E^=$ or $E^+ \cup E^0$. In
  the former case, the throughput of its incident arcs in
  $t_1 \land t_2$ are equals to those in $t_2$, in the latter case to
  those in $t_1$. Therefore, all the rules defining a uniform
  sub-steady-state hold for $t_1 \land t_2$. The same reasoning
  applies for $t_1 \lor t_2$, because $t_1 \lor t_2$ agrees with $t_1$
  on vertices incident to $E^-$, and with $t_2$ on vertices incident
  to $E^+$.
\end{proof}

Consequently, for any set $F \subset E$ for which a uniform
sub-steady-state exists, there are two special uniform
sub-steady-states $(t_F^\down,F)$ and $(t_F^\up,F)$, defined by
$t_F^\down = \bigwedge \{t : (t,F) \textrm{ uniform
  sub-steady-state}\}$ and
$t_F^\up = \bigvee \{t : (t,F) \textrm{ uniform sub-steady-state}\}$.
When all the input capacities are equal, these corresponds to the two
extremal solutions (minimum and maximum) corresponding to optimizing
the linear form
$\sum_{e \in F} t(e) - \sum_{e \in E \setminus F} t(e)$ over the
polyhedron of uniform sub-steady-states. Observe that with non-uniform
input capacities, the set of uniform sub-steady-states, for a given
$F$, is not necessarily a polyhedron. When $F$ is not fixed, let
$c_1 < c_2 < \ldots < c_l$ be the values of input capacities, the
uniform sub-steady-states can be described as the union of several
polyhedra, each defined by adding a variable $\gamma$ whose value is
the common throughput to all non-limited inputs, and adding
constraints $c_k \leq \gamma \leq c_{k+1}$, and for each fluid arc $e$
leaving an input $i \in I$, either $t(e) = c(i)$ (if $c(i) < \gamma$),
or $t(e) = \gamma$ (if $c(i) \geq \gamma$). Then $t_F^\up$ and
$t_F^\down$ are the extremal points for the same linear form, in this
union of polyhedra. Therefore, $t_F^\up$ and $t_F^\down$ can also be
computed using linear programming.

We extend the definition of $\lor$ and $\land$ to arbitrary uniform
sub-steady-states. Let $(t_1,F_1)$ and $(t_2,F_2)$ be two uniform
sub-steady states. We define $(t_1,F_1) \land (t_2,F_2)$ as
$(t_0, F_1 \cup F_2)$ where for each $e \in E$,
\[
  t_0(e) = \left\{
    \begin{array}{ll}
      \max \{t_1(e), t_2(e)\} & \quad\textrm{if $e \in F_1 \cap F_2$}\\
      t_1(e) & \quad\textrm{if $e \in F_1 \setminus F_2$}\\
      t_2(e) & \quad\textrm{if $e \in F_2 \setminus F_1$}\\
      \min \{t_1(e), t_2(e)\} & \quad\textrm{if $e \notin F_1 \cup F_2$}\\
    \end{array}\right.
\]
and we define $(t_1,F_1) \lor (t_2,F_2)$ as $(t',F_1 \cap F_2)$ where
for each $e \in E$,
\[
  t'(e) = \left\{
    \begin{array}{ll}
      \min \{t_1(e), t_2(e)\} & \quad\textrm{if $e \in F_1 \cap F_2$}\\
      t_2(e) & \quad\textrm{if $e \in F_1 \setminus F_2$}\\
      t_1(e) & \quad\textrm{if $e \in F_2 \setminus F_1$}\\
      \max \{t_1(e), t_2(e)\} & \quad\textrm{if $e \notin F_1 \cup F_2$}\\
    \end{array}\right.
\]
These two operations coincides with the previous definitions when $F_1 = F_2$.

By the anti-exchange property of antimatroids, if $R$ and $S$ are two
sequences of operations from $(0,E)$ (or any uniform sub-steady-state
$(t_0,F_0)$, by removing a common prefix to both sequences), and $r$
is an operation in $R$ that does not appear in $S$, then $r$ appears
in a sequence $S,R'$, therefore appears after any element in $S$. We
use this property several times in the proof of the next theorem, and
say that an operation $\tau$ is \emph{in the future} of a sequence $R$
if there exists $S$ containing $\tau$ such that $R,S$ is a valid
sequence.

\begin{theorem}
  The operations $\land$ and $\lor$ are the meet and join operations
  of the semimodular lattice of uniform sub-steady states.
\end{theorem}

\begin{proof}
  Let $(t_0, F_0)$ be the meet of $(t_1,F_1)$ and $(t_2,F_2)$, and
  let's prove that it satisfies the formula given above. There are two
  sequences of operations $R$ and $S$, such that $(t_1,F_1)$
  arises from $(t_0,F_0)$ by the sequence $R$ and $(t_2,F_2)$ arises
  from $(t_0,F_0)$ by the sequence $S$.

  We claim that $R$ and $S$ contain distinct operations. Suppose, for
  the sake of contradiction, that there is a common operation $\rho$.
  We may assume that $\rho$ appears as soon as possible in both
  $R = R',\rho,R''$ and $S = S',\rho,S''$, and no common operation
  appears earlier in both sequences. Hence $\rho$ commutes with
  neither $\tau_1 := \textrm{last}(R')$ nor
  $\tau_2 := \textrm{last}(S')$ (and $R'$ and $S'$ cannot be empty).
  If $\rho$ is a \emph{move} or \emph{augment} operation, then as
  $\tau_1$ does not commute with $\rho$, there must be an arc $uv$,
  with $u$ in the support of $\rho$, such that
  \begin{itemize}
  \item either $\tau_1$ is the saturation of $uv$. As $\tau_1$ is not
    in $S'$, $uv$ is not saturated after applying $S'$, therefore $uv$
    is not in the support of $\rho$. Thus to apply $\rho$ after
    $S'$, we must have $t(uv) = c(uv)$. As an \emph{augment} or \emph{move}
    operation whose support contain $uv$ cannot be in the future of
    $R'$, and by our choice of the earliest common operation $\rho$,
    no such operation occurs in $S'$. This implies that $t(uv)$ is not
    increased by $S'$, and $t_0(uv) = c(uv)$. Hence $\tau_1$ and $\rho$
    commutes in $R$, contradiction;
  \item or $\tau_1$ is an \emph{augment} or \emph{move} that increases
    the throughput value of $uv$ to $c(uv)$. As $\tau_1$ is in the future
    of $S'$, the throughput value of $uv$ after $S'$ is less than $c(uv)$,
    and $\rho$ is not applicable, contradiction.
  \end{itemize}

  Else $\rho$ is the saturation of some arc $uv$. Then $\tau_1$ is 
  \begin{itemize}
  \item either an \emph{augment} or a \emph{move} whose support
    contains $uv$. But then $\tau_1$ cannot be in the future of
    $S',\rho$ (where $uv$ is saturated), therefore $\tau_1$ must be in
    $S'$, contradicting the choice of the earliest common operation
    $\rho$;r
  \item or the saturation of an arc $vw$. Then after $R'$, $t(vw) < 1$
    as otherwise $\tau_1$ and $\rho$ commutes. But then after $S'$,
    $vw$ is not saturated as $\tau_1$ is in the future of $S'$, thus
    by Rule~\ref{eqn:maximization} $t(vw) = 1$, implying that
    $S' \setminus R'$ contains an \emph{augment} of \emph{move}
    operation whose support contains $vw$. But such an operation
    cannot be in the future of $R'$, as $vw$ is saturated after $R'$,
    contradiction;
  \item or an \emph{augment} or \emph{move} operation whose support
    contains an arc $vw$, such that after $R'$, $t(vw) = 1$. As
    $\tau_1$ is in the future of $S'$, after $S'$ $vw$ is neither
    saturated nor at throughput $1$, therefore $\rho$ is not
    applicable, contradiction.
  \end{itemize}
  Therefore, $R$ and $S$ contain distinct operations.

  Next we claim that for any arc $e$, the first operations $\rho$ in
  $R$, $\tau$ in $S$, that increase the throughput of $e$, must be
  equal. To prove this, we decompose $R = R',\rho,R''$ and
  $S = S',\tau,S''$. By the anti-exchange property, there is a
  permutation $\tilde{S}'$ of $S'$ such that $R',\tilde{S}'$ is
  applicable, and moreover both $\rho$ and $\tau$ may be applied
  immediately after $R',\tilde{S}'$. Therefore as the support of
  $\rho$ and $\tau$ intersects at $e$, they must be equal. Because $R$
  and $S$ have distinct operations, this implies that only one of $R$
  and $S$ can modify the throughput of any given arc.
  
  It follows from the fact that $R$ and $S$ have distinct operations,
  hence distinct saturations, that $F_0 = F_1 \cup F_2$. Now consider
  some operation $\rho$ in $R$ that increase the throughput of an arc
  $e$. Then $\rho$ is in the future of $S$, therefore $e \in F_2$. By
  contraposition, if $e \notin F_2$, the throughput on $e$ does not
  increase when applying $R$, hence $t_1(e) = t_0(e)$, proving that
  $t_0(e) = t_1(e)$ when $e \in F_1 \setminus F_2$ (and its symmetric
  case). If $e \in F_1 \cap F_2$, $e$ does not increase in either $R$
  or $S$, hence $t_0(e) = \min \{t_1(e), t_2(e)\}$. The case
  $e \notin F_1 \cup F_2$ is symmetric, either $R$ or $S$ does not
  decrease $e$, therefore $t_0(e) = \max \{t_1(e), t_2(e)\}$, proving
  the formula for $\land$.

  Finally, $t_1 \lor t_2$ is obtained by applying the sequences
  $R,\tilde{S}$ or $S,\tilde{R}$, for some permutations $\tilde{S}$
  and $\tilde{R}$ of $S$ and $R$ respectively. The effects of
  $\tilde{R}$ (respectively $\tilde{S}$) on a uniform sub-steady state
  are the same as applying $R$ (respectively $S$). The formula for
  $\lor$ immediately follows.
\end{proof}

\section{Priority splitters}\label{sec:priority}

\subsection{Definition and algorithms}

Factorio allows players to assign priorities to splitters, altering
their behavior. A splitter that prioritizes some output belt will send
as much throughput to that belt, until saturating or reaching its
capacity. Any excess flow will then be sent on the other output belt.
Similarly, a splitter will pull its flow from the prioritized input
belt, and use the other output belt only to complete its throughput.

We modify the definition of steady-states to
accomodate priority splitters.

\begin{definition}
  Let $\splitnetwork$ be a splitter network, $c : I \cup O \to [0,1]$
  a capacity function, $p^+, p^- : S \to E \cup \{\bot\}$ the
  so-called the output-priority and input-priority functions, such
  that for each $s \in S$, $p^+(s) \in \delta^+(s) \cup \{\bot\}$ and
  $p^-(s) \in \delta^-(s) \cup \{\bot\}$. A \emph{steady-state} for
  $(G,c,p^+,p^-)$ is a pair $(t,F)$ where

  \begin{enumerate}[label=P\arabic*,nosep]
  \item \label{p:throughput} $t : E \to [0,1]$ is the throughput function;
  \item \label{p:fluid-arcs} $F \subseteq E$ is the set of \emph{fluid
      arcs}, $E \setminus F$ is the set of \emph{saturated} arcs;
  \item \label{p:input-capacity} for each $i \in I$ with $\delta^+(i) = \{e\}$,
    $t(e) \leq c(i)$ and moreover if $e \in F$ then $t(e) = c(i)$;
  \item \label{p:output-capacity} for each $o \in O$ with $\delta^-(o) = \{e\}$,
    $t(e) \leq c(o)$ and moreover if $e \notin F$ then $t(e) = c(o)$;
  \item \label{p:conservation} for each $s \in S$, with
    $\delta^-(s) = \{ e_1, e_2 \}$ and $\delta^+(s) = \{e_3, e_4\}$,
    $t(e_1) + t(e_2) = t(e_3) + t(e_4)$;
  \item \label{p:incoming-rule} for any $e_1,e_2 \in E$ with
    $\{e_1, e_2\} = \delta^-(s)$ and $e_1 \notin F$, if $p^-(s) = e_1$
    then $t(e_1) = 1$ or $t(e_2) = 0$, and if $p^-(s) \neq e_2$ then
    $ t(e_1) \geq t(e_2)$;
  \item \label{p:outgoing-rule} for any $e_1, e_2 \in E$ with
    $\{e_1, e_2\} = \delta^+(s)$ and $e_1 \in F$, if $p^+(s) = e_1$
    then $t(e_2) = 0$ or $t(e_1) = 1$, and if $p^+(s) \neq e_2$ then
    $t(e_1) \geq t(e_2)$;
  \item \label{p:maximization} for any $uv \in E \setminus F$ and
    $vw \in F$, $t(uv) = 1$ or $t(vw) = 1$.
  \end{enumerate}
\end{definition}

Compared to the definition of steady-states for $(G,c)$, all rules are
identical except rules~\ref{p:incoming-rule} and~\ref{p:outgoing-rule}
which generalizes rules~\ref{eqn:incoming-rule}
and~\ref{eqn:outgoing-rule}. Therefore when $p^-$ and $p^+$ are
uniformly $\bot$, the two notions of steady-states coincide.

The extended definition of steady-state preserves the symmetry already
observed, therefore if $(t,F)$ is a steady-state for $(G,c,p^+,p^-)$,
then $(t,\symmetry{E \setminus F})$ is a steady-state for
$(\symmetry{G},c,\symmetry{p^-},\symmetry{p^+})$, where
$\symmetry{p}(e) = \symmetry(p(e))$.

In the rule~\ref{p:outgoing-rule}, for $\delta^+(s) = \{e_1,e_2\}$,
when $e_1 \in F$ and $p^+(s) = e_1$, the constraint $t(e_1) = 1$ or
$t(e_2) = 0$ is not immediately representable as a linear constraint.
However, in the context of iteratively solving the linear system, and
removing arcs from $F$ at each step, at the start of an iteration, if
$e_1 \in F$ and $t(e_1) < 1$, we may force the constraint
$t(e_2) = 0$. Hence $t(e_2)$ will stay null until $e_1$ becomes tight
or saturated. Then we can remove the constraint $t(e_2) = 0$ as it is
no longer induced by the definition of steady-state. The same
principle applies to rule~\ref{p:incoming-rule}. Therefore, the
LP-based algorithms can be adapted to compute the steady-state of a
splitter network with priorities. Likewise, the definition of residual
graph can be adapted, to accomodate priorities.

Priorities allow to change the feasible steady-states. Therefore, the
question of finding a steady-state optimizing the total throughput
becomes relevant in the context of splitter network with priorities.
Formally, the input of a throughput maximization problem consists in a
splitter network $(G,c)$. Its output is a quadruple $(p^+, p^-, t, F)$
such that $(t,F)$ is a steady-state for $(G,c,p^+,p^-)$. The goal is
to maximize the global throughput $t(\delta^+(I))$.

We study several possible scenarii for this optimization problem: in
\Cref{sec:in-out-prio} when we may choose all the priorities, in and
out of each splitter; in \Cref{sec:out-prio} we impose the in
priorities and ask for maximizing out priorities; and finally in
\Cref{sec:np-hard} we impose some priorities arbitrarily, and must
choose the other priorities.

\subsection{Choose input and output priorities for each splitter}\label{sec:in-out-prio}

\begin{proposition}\label{prop:complexity-choose-all}
  The throughput maximization problem is polynomial-time solvable when
  $c = \allone$. Let $\splitnetwork$ be a splitter network. The
  maximum throughput equals the value of a maximum flow from $I$ to
  $O$ in $G$ with unit capacities.
\end{proposition}

\begin{proof}
  Consider an integral maximum flow $t$ from $I$ to $O$. Because there is a
  minimum cut having the same value, the maximum throughput cannot
  exceed $t(\delta^+(I))$. We show that we can choose $F$, $p^+$ and
  $p^-$ such that $(t,F)$ is a steady-state in $(G,c,p^+,p^-)$.

  For each splitter $s \in S$, with $\delta^+(s) = \{e_1,e_2\}$, we define
  \[ p^+(s) = 
    \left\{
      \begin{array}{ll}
        \bot & \textrm{if $t(e_1) = t(e_2)$;} \\
        e_1 & \textrm{if $t(e_1) = 0$ and $t(e_2) = 0$;}\\
        e_2 & \textrm{if $t(e_2) = 1$ and $t(e_1) = 1$.}
      \end{array}
    \right.
  \]
  Define $p^-(s)$ similarly. Finally, we define $F$ by setting
  saturated arcs to be precisely the arcs $e$ with $t(e) = 0$ that are
  reachable from $I$ in the residual graph of the maximal flow $t$.
  This choice guarantees that rule~\ref{p:maximization} is satisfied.
  All the other rules are readily checked.
\end{proof}

\subsection{Choose output priority for each splitter}\label{sec:out-prio}

\begin{proposition}\label{prop:complexity-choose-output}
  Given $(G,c = \allone)$ and input priorities $p^- : S \to E$, the problem of
  finding a steady-state $(t,F,p^+,p^-)$ with maximum throughput is
  polynomial-time solvable. The maximal throughput equals the value of
  a maximum flow from $I$ to $O$ in $G$ with unit capacities.
\end{proposition}

\begin{proof}
  For an integral maximum flow $t$ from $I$ to $O$, and setting
  $p^+(s)$ as in the proof of \Cref{prop:complexity-choose-all},
  $(t,E)$ is a sub-steady-state, as can be readily checked. We
  initiate the sub-steady-state algorithm with $(t,E)$, yielding a
  steady-state $(t',F)$. Because the sub-steady-state cannot reduce
  the global throughput, and by the existence of a minimum cut of size
  $t(\delta^+(I))$, the throughput of $(t',F)$ is precisely the value
  of the maximum flow.
\end{proof}

\begin{corollary}\label{prop:complexity-choose-input}
  Given $(G,c=\allone)$ and output priorities $p^+ : S \to E$, the problem of
  finding a steady-state $(t,F,p^+,p^-)$ with maximum throughput is
  polynomial-time solvable. The maximal throughput equals the value of
  a maximum flow from $I$ to $O$ in $G$ with unit capacities.
\end{corollary}

\begin{proof}
  Apply \Cref{prop:complexity-choose-output} to the reverse splitter
  network.
\end{proof}

In contrast, when we force each splitter to have an output priority
distinct from $\bot$ and the input capacities are not uniform, finding
a maximal throughput is hard.

\begin{proposition}
  Let $\splitnetwork$ be a splitter network, capacities
  $c : I \cup O \to [0,1]$ and input priorities $p^-$, with $c(O) = 1$
  for each $o \in o$. The problem of finding output priorities
  $p^+ : S \to E$ and a steady-state $(t,F,p^-,p^+)$ of maximum
  throughput is NP-hard, even when $p^-(s) = \bot$ for each splitter
  $s \in S$.
\end{proposition}

\begin{figure}
  \begin{center}
    \begin{tikzpicture}[x=1cm,y=1cm,>=latex]
      \foreach \n/\x/\y/\pos/\capa in {
        sn/0/0/left/s_n,
        sm/0/1/left/{},
        s2/0/3/left/s_2,
        s1/0/4/left/s_1,
        t1/7/4/right/t_1,
        t2/7/0/right/t_2,
        t3/7/3/right/s_0,
        t4/7/1/right/t_0}
      {
        \node[term,label=\pos:$\capa$] (\n) at (\x,\y) {};
      }
      \foreach \n\x/\y in {
        a1/1/4,a2/1/3,am/1/1,an/1/0,
        u2/2/4,u3/3/4,um/4/4,un/5/4,
        w2/2/5,w3/3/5,wm/4/5,wn/5/5,
        v2/2/0,v3/3/0,vm/4/0,vn/5/0,
        x2/2/-1,x3/3/-1,xm/4/-1,xn/5/-1,
        r1/6/4,r2/6/0,d/6/2,e/7/2} 
      {
        \node[chooser] (\n) at (\x,\y) {};
      }
      \foreach\u/\v in {
        s1/a1,s2/a2,sm/am,sn/an,
        a1/u2,a2/u2,am/um,an/un,
        a1/vn,a2/vm,am/v2,an/v2,
        um/un,vm/vn,un/r1,vn/r2,
        u2/u3,v2/v3,w2/w3,x2/x3,u3/w3,v3/x3,
        u2/w2,w2/w3,um/wm,un/wn,wn/r1,wm/wn,xm/xn,
        v2/x2,x2/x3,vm/xm,vn/xn,xn/r2,
        r1/t1,r2/t2,r1/d,r2/d,d/e,t3/e,e/t4}
      {
        \draw[->] (\u) -- (\v);
      }
      \foreach \u/\v in {u3/um,v3/vm,x3/xm,w3/wm} {
        \draw[->,dotted] (\u) -- (\v);
      }
      \draw (0,2) node {\vdots};
      \draw (d) node[left] {$u$};
      \draw (e) node[right] {$v$};
    \end{tikzpicture}
  \end{center}
  \caption{An illustration of the reduction from the partition problem
    to finding output priorities with maximal throughput.}
  \label{fig:partition-reduction}
\end{figure}
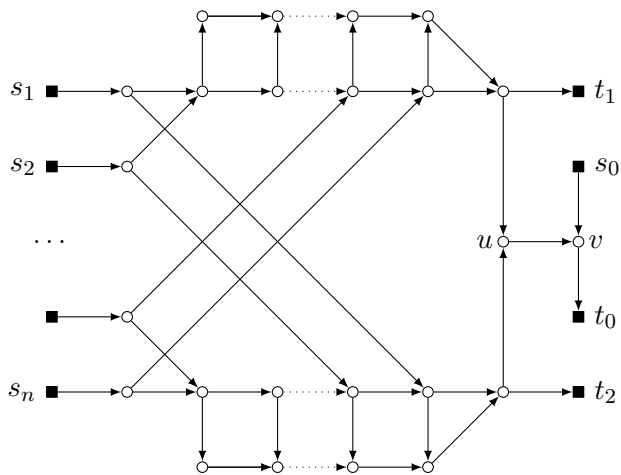

\begin{proof}
  We reduce the partition problem: given $s_1,\ldots,s_n$ with
  $\sum_{i=1}^n s_i = 2$, is there a subset
  $I \subseteq \interval{1}{n}$ with $\sum_{i \in I} s_i = 1$. The
  reduction is straightforward, and illustrated in
  \Cref{fig:partition-reduction}. The $n$ leftmost terminal have
  output capacities $s_1$, $s_2$,\ldots, $s_n$ respectively. $s_0$ and
  the outputs have capacity $1$. The maximal throughput is $3$ if and
  only if the partition instance has a solution. Indeed, when a
  partition exists, routing all the flows from terminals $s_i$ with
  $i \in I$ up, and all the other flows down, provides a routing where
  no flow goes through $u$. Inversely, if any flow goes through $u$,
  it must go through $v$ too. That flow will concur with the one unit
  flow between $s_0$ and $t_0$, therefore forbidding the total
  throughput out of $s_0$ to be one. Howerever, in order to avoid $u$
  to receive any flow, the output flow of each input $s_i$ must be
  fairly split between the bottom and top parts of the splitter
  network, hereby defining a partition.
\end{proof}

\subsection{Choose output priorities for some splitters}\label{sec:np-hard}

When an arbitrary subset of priorities are imposed, the maximum
throughput problem becomes intractable, stays so even when we only
choose some of the output priorities and all other splitters act
fairly.

\begin{theorem}\label{thm:np-hard}
  Let $\splitnetwork$ be a splitter network, $c : I \cup O \to [0,1]$
  a capacity function, $p^- : S \to E \cup \{\bot\}$ input priorities
for each splitter, and $p^+_0 : S' \to E \cup \{\bot\}$ be a partial output-priority
function, where $S' \subseteq S$. The problem of finding a
steady-state $(t,F,p^-,p^+)$ of maximum throughput with $p^+_0$ being
the restriction of $p^+$ to $S'$ is NP-hard, even when $p^-$ and
$p^+_0$ are uniformly $\bot$ and $c=\allone$.
\end{theorem}

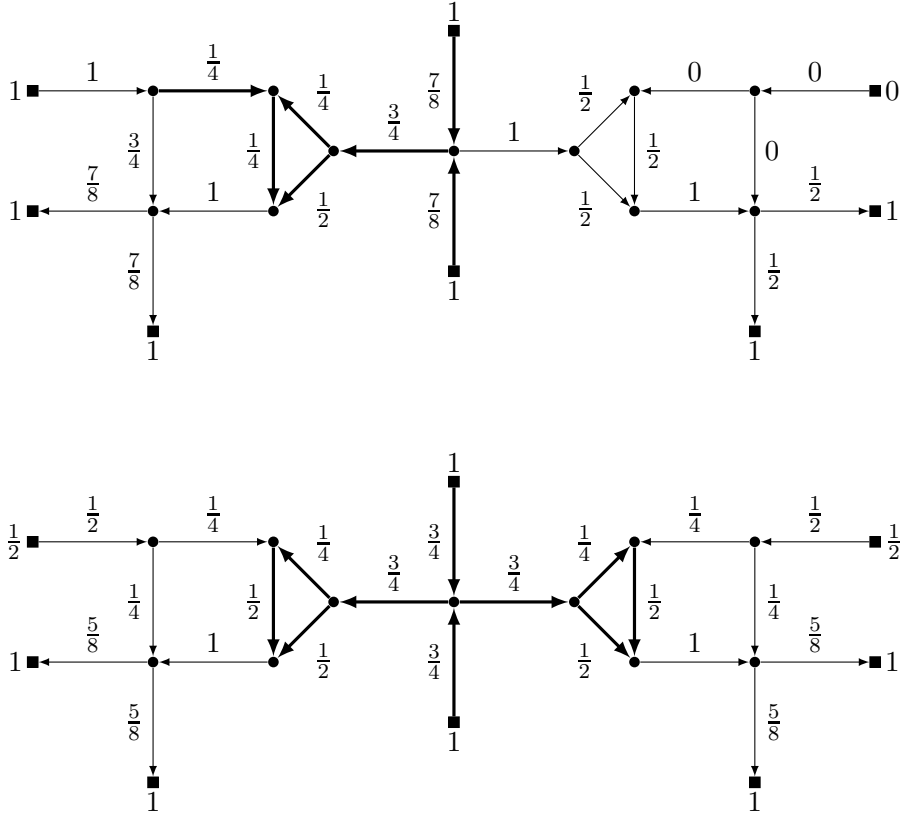
\begin{figure}
  \begin{center}
    \begin{tikzpicture}[x=0.8cm,y=0.8cm,>=latex]
      
      \foreach \u/\x/\y/\pos/\c in {
        s1/0/4/left/1,
        s2/7/5/above/1,
        s3/7/1/below/1,
        s4/14/4/right/0,
        t1/0/2/left/1,
        t2/2/0/below/1,
        t3/12/0/below/1,
        t4/14/2/right/1}
      {
        \node[term] (\u) at (\x,\y) {};
        \draw (\u) node[\pos] {$\c$};
      }
      \foreach \u/\x/\y in {
        a/2/4,b/2/2,c/4/4,d/4/2,e/5/3,f/7/3,
        g/9/3,h/10/4,i/10/2,j/12/4,k/12/2
      } {
        \node[splitter] (\u) at (\x,\y) {};
      }
      \foreach \u/\v/\t/\pos in {
        b/t1/\frac{7}{8}/above,
        d/b/1/above,
        j/h/0/above,
        f/g/1/above,
        i/k/1/above,
        k/t4/\frac12/above,
        s1/a/1/above,
        a/b/\frac34/left,
        b/t2/\frac78/left,
        s4/j/0/above,
        h/i/\frac12/right,
        g/i/\frac12/below left,
        g/h/\frac12/above left,
        j/k/0/right,
        k/t3/\frac12/right}
      {
        \draw[->,fluid] (\u) -- (\v) node[midway,\pos] {$\t$};
      }
      \foreach \u/\v/\t/\pos in {
        a/c/\frac14/above,
        c/d/\frac14/left,
        e/c/\frac14/above right,
        e/d/\frac12/below right,
        f/e/\frac34/above,
        s2/f/\frac78/left,
        s3/f/\frac78/left}
      {
        \draw[->,saturated] (\u) -- (\v) node[midway,\pos] {$\t$};
      }

      \begin{scope}[yshift=-6cm]
        \foreach \u/\x/\y/\pos/\c in {
          s1/0/4/left/\frac{1}{2},
          s2/7/5/above/1,
          s3/7/1/below/1,
          s4/14/4/right/\frac{1}{2},
          t1/0/2/left/1,
          t2/2/0/below/1,
          t3/12/0/below/1,
          t4/14/2/right/1}
        {
          \node[term] (\u) at (\x,\y) {};
          \draw (\u) node[\pos] {$\c$};
        }
        \foreach \u/\x/\y in {
          a/2/4,b/2/2,c/4/4,d/4/2,e/5/3,f/7/3,
          g/9/3,h/10/4,i/10/2,j/12/4,k/12/2
        } {
          \node[splitter] (\u) at (\x,\y) {};
        }
        \foreach \u/\v/\t/\pos in {
          a/c/\frac14/above,
          b/t1/\frac58/above,
          d/b/1/above,
          j/h/\frac14/above,
          i/k/1/above,
          k/t4/\frac58/above,
          s1/a/\frac12/above,
          a/b/\frac14/left,
          b/t2/\frac58/left,
          s4/j/\frac12/above,
          j/k/\frac14/right,
          k/t3/\frac58/right}
        {
          \draw[->,fluid] (\u) -- (\v) node[midway,\pos] {$\t$};
        }
        \foreach \u/\v/\t/\pos in {
          c/d/\frac12/left,
          e/c/\frac14/above right,
          e/d/\frac12/below right,
          f/e/\frac34/above,
          f/g/\frac34/above,
          s2/f/\frac34/left,
          s3/f/\frac34/left,
          h/i/\frac12/right,
          g/i/\frac12/below left,
          g/h/\frac14/above left}
        {
          \draw[->,saturated] (\u) -- (\v) node[midway,\pos] {$\t$};
        }
      \end{scope}

    \end{tikzpicture}
  \end{center}
  \caption{A network splitter with two capacity functions and their
    steady-states.}
  \label{fig:non-convex-network}
\end{figure}

\begin{figure}
  \begin{center}
    \begin{tikzpicture}[x=0.8cm,y=0.8cm,>=latex]
      \foreach \u/\x/\y/\pos/\name in {
        s1/4/5/right/s_1,
        s2/4/2/left/s_2,
        s3/4/0/left/s_3}
      {
        \node[term] (\u) at (\x,\y) {};
        \draw (\u) node[\pos] {$\name$};
      }
      \foreach \u/\x/\y/\pos/\name in {
        t1/-1/0/left/x,
        t2/0/-1/below/x,
        t3/8/-1/below/\update{x},
        t4/9/0/right/\update{x}} 
      {
        \coordinate (\u) at (\x,\y) {};
        \draw (\u) node[\pos] {$\name$};
      }
      \foreach \u/\x/\y in {
        a/0/2,b/0/0,c/2/2,d/2/0,e/3/1,f/4/1,g/5/1,h/6/2,i/6/0,j/8/2,k/8/0} 
      {
        \node[splitter] (\u) at (\x,\y) {};
      }
      \node[chooser] (x) at (4,4) {};
      \draw (x) node[below] {$x?$};
      \foreach \u/\v in {
        s1/x,a/b,a/c,c/d,d/b,b/t1,b/t2,e/c,e/d,f/e,
        f/g,g/h,g/i,h/i,j/h,j/k,i/k,k/t3,k/t4,s2/f,s3/f} 
      {
        \draw[->] (\u) -- (\v);
      }
      \draw[->] (x) to[out=180,in=90] (a);
      \draw[->] (x) to[out=0,in=90] (j);
      \draw[rounded corners=4pt] (-0.5,-0.5) rectangle (8.5,5.5); 

    \end{tikzpicture}
  \end{center}
  \caption{The variable gadget. The node $x?$ represents a splitter
    whose output priority is not predetermined. All other splitters
    $s$ are fair: $p_0^+(s) = p^-(s) = \bot$. The leaving arcs indexed
    by $x$ are connected to gadgets of clauses containing $x$
    positively, those indexed by $\overline{x}$ are connected to
    gadgets of clauses containing $x$ negatively. If there are not
    enough clauses, each superfluous arc has a new terminal as
    destination.}
  \label{fig:variable-gadget}
\end{figure}
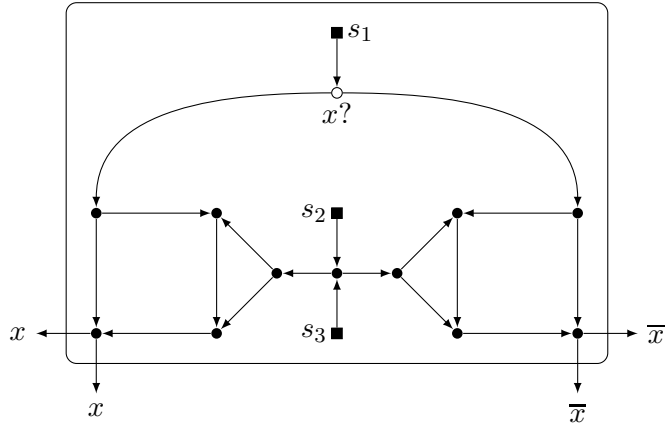

The reduction is based on two gadgets, representing respectively the
variables and the clauses of a 3-SAT formula. Before establishing the
reduction, we study the two properties of the two gadgets. The
variable gadget is based on the splitter network described in
\Cref{fig:non-convex-network}. We will exploit the fact that its
global throughput is not a convex function of the input capacities.
Indeed, as depicted in \Cref{fig:non-convex-network}, for the input
capacities $(1,1,0)$, $(\frac{1}{2},1,\frac{1}{2})$ and $(0,0,1)$, the
global throughput are respectively $\frac{11}{4}$, $\frac{10}{4}$ and
$\frac{11}{4}$. Therefore, by adding a splitter with undetermined
priority, to provide the flow to the two opposite inputs, we
incentivize a choice of priority different from $\bot$ for that
additional splitter. Implementing this modification yields the
variable gadget, defined in \Cref{fig:variable-gadget}.

\begin{proposition}\label{prop:variable-gadget}
  Depending on the choice of $p^+(x?)$, and assuming that each arc
  leaving the gadget is fluid, the throughput on the leaving arcs of
  the splitter network depicted in \Cref{fig:variable-gadget} are
  either $(\frac78,\frac78,\frac12,\frac12)$, or
  $(\frac58,\frac58,\frac58,\frac58)$, or
  $(\frac12,\frac12,\frac78,\frac78)$.
\end{proposition}

\begin{proof}
  This follows immediately from the steady-states depicted in
  \Cref{fig:non-convex-network}, considering the throughput on the
  leaving arc of $x?$ will be either $1,0$, or $\frac12,\frac12$, or
  $0,1$, depending on the choice of out-priority.
\end{proof}

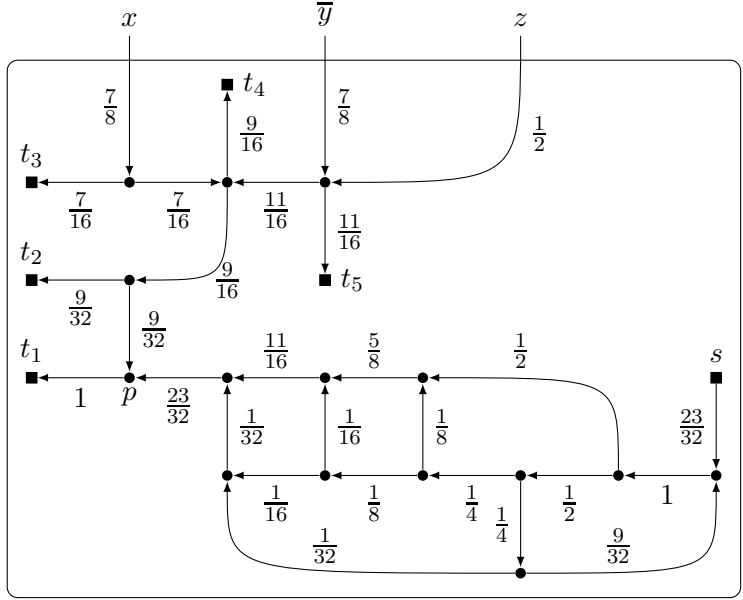
\begin{figure}
  \begin{center}
    \begin{tikzpicture}[x=1.3cm,y=1.3cm,>=latex]
      \foreach \u/\x/\y/\pos/\name in {
      t1/0/2/above/t_1,
      t2/0/3/above/t_2,
      t3/0/4/above/t_3,
      t4/2/5/right/t_4,
      s1/3/3/right/t_5,
      s2/7/2/above/s} 
      {
        \node[term,label=\pos:$\name$] (\u) at (\x,\y) {};
      }
      \foreach \u/\x/\y/\pos/\name in {
        x/1/5.5/above/x,
        y/3/5.5/above/\overline{y},
        z/5/5.5/above/z} 
      {
        \coordinate[label=\pos:$\name$] (\u) at (\x,\y) {};
      }
      \draw[rounded corners=4pt] (-0.25,-0.25) rectangle (7.25,5.25);
      \foreach \u/\x/\y in {
        a/1/2,
        b/1/3,
        c/1/4,
        d/2/1,
        e/2/2,
        f/2/4,
        g/3/1,
        h/3/2,
        i/3/4,
        j/4/1,
        k/4/2,
        m/5/0,
        n/5/1,
        o/6/1,
        p/7/1}
      {
        \node[splitter] (\u) at (\x,\y) {};
      }
      \foreach \u/\v/\pos/\text in {
        a/t1/below/1,
        b/t2/below/\frac{9}{32},
        c/t3/below/\frac{7}{16},
        b/a/right/\frac{9}{32},
        x/c/left/\frac78,
        e/a/below/\frac{23}{32},
        c/f/below/\frac{7}{16},
        f/t4/right/\frac{9}{16},
        d/e/right/\frac{1}{32},
        g/d/below/\frac{1}{16},
        h/e/above/\frac{11}{16},
        i/f/below/\frac{11}{16},
        g/h/right/\frac{1}{16},
        i/s1/right/\frac{11}{16},
        y/i/right/\frac78,
        k/h/above/\frac58,
        j/g/below/\frac18,
        j/k/right/\frac18,
        n/j/below/\frac14,
        n/m/left/\frac14,
        o/n/below/\frac12,
        p/o/below/1,
        s2/p/left/\frac{23}{32}} 
      {
        \draw[->] (\u) -- (\v) node[midway,\pos] {$\text$};
      }
      \draw[->] (f) .. controls (2,3) .. (b);
      \path (2,3) node {$\frac{9}{16}$};
      \draw[->] (z) .. controls (5,4) .. (i);
      \path (5,4.5) node[right] {$\frac{1}{2}$};
      \draw[->] (m) .. controls (2,0) .. (d);
      \path (3,0) node[above] {$\frac{1}{32}$};
      \draw[->] (m) .. controls (7,0) .. (p);
      \path (6,0) node[above] {$\frac{9}{32}$}; 
      \draw[->] (o) .. controls (6,2) .. (k);
      \path (5,2) node[above] {$\frac{1}{2}$};
      \draw (a) node[below] {$p$};

    \end{tikzpicture}
  \end{center}
  \caption{The clause gadget, for a clause
    $x \lor \overline{y} \lor z$.}
  \label{fig:clause-gadget}
\end{figure}

Hence, each variable gadget induces a choice, that translates into
some arcs carrying a throughput of $\frac78$ and others $\frac12$. We
propose a clause gadget that can accept a throughput of up to
$\frac{9}{4} = \frac{1}{2} + \frac78 + \frac78$, forbidding the three
inputs from having all a throughput of $\frac78$. The splitter network
presented in \Cref{fig:clause-gadget} features two parts: the upper
part receives the flow from the variable gadget, and reduces it to an
eightth of its original value. Hopefully the remaining throughput
should be at most $\frac{9}{32} < \frac12$. The bottom part consists
in a capacity simulating network, with capacity
$1 - \frac{9}{32} = \frac{23}{32} > \frac12$. Then we join this two
flows toward the output $t_1$ in the splitter $p$. Increasing the
input flow from the variable gadgets will increase the flow entering
$p$ from above. As $\frac{9}{32} < \frac{23}{32}$, by respecting
rule~\ref{p:incoming-rule}, the throughput on the arc entering $p$
from the right would then decrease, and thus $s_2$ would provide
strictly less than $\frac{23}{32}$.

\begin{proposition}\label{prop:clause-gadget}
  For any throughput values $f_1,f_2,f_3$ on the three entering arcs
  of the clause gadget, depicted in \Cref{fig:clause-gadget}, there is
  a steady-state $(t,F)$ for which the three entering arcs
  $e_1,e_2,e_3$ are fluid, with $t(e_i) = f_i$ for each
  $i \in \{1,2,3\}$. Moreover, $f_1 + f_2 + f_3 \leq \frac{9}{4}$ if
  and only if $t(\delta^+(s)) = \frac{23}{32}$.
\end{proposition}

\begin{proof}
  In order to show that the arcs entering the gadget are always fluid,
  we consider the case when the throughput on the three entering arcs
  $e_1$, $e_2$ and $e_3$ are uniformly $1$. Then the throughput on
  each arc of the upper part of the gadget can be easily computed, and
  the throughput on the arc $e$ entering $p$ from above is
  $\frac{3}{8}$. This arc is fluid, and the other arc $e'$ entering
  $p$ from the right is saturated with throughput $\frac{5}{8}$.
  Therefore $e_1$, $e_2$ and $e_3$ are fluid in this steady-state.

  More precisely, the throughput on $e$ is
  $t(e) = \frac{f_1 + f_2 + f_3}{8}$, and the throughput on $e'$ is
  $t(e') = \max \{\frac{23}{32}, 1 - t(e)\}$, with $e'$ saturated when
  $t(e) > \frac{9}{32}$. The steady-state at the critical point, when
  $f_1 + f_2 + f_3 = \frac{9}{4}$, is illustrated in
  \Cref{fig:clause-gadget}.
\end{proof}

Using the two gadgets, we prove that the throughput maximization
problem is NP-hard.

\begin{proof}[Proof of \Cref{thm:np-hard}]
  We reduce 3SAT, where each variable appears at most twice positively
  and twice negatively. Let $\phi$ be a set of 3-clauses on variables
  $x_1,\ldots,x_n$. We define a splitter network, with one copy of the
  variable gadget for each variable in the formula, and one copy of
  the clause gadget for each of its clauses. A leaving arc labeled $x$
  (respectively $\overline{x}$) is identified with an entering arc of
  a clause gadget containing the literal $x$ (respectively
  $\overline{x}$). Any superfluous leaving arc is redirected to a new
  distinct output node. The priorities of every splitter are set to
  $\bot$, except for the output priority of the splitter $x?$ from the
  variable gadget, for each variable $x$. The output priority of that
  splitter is free. Formally
  $S' = S \setminus \{x?~:~x \in \{x_1,\ldots,x_n\}\}$, $p^-$ is
  uniformly $\bot$ on $S$, and $p^+_0$ is uniformly $\bot$ on $S'$.
  The size of the splitter network thus constructed is a polynomial in
  $|\phi|$ and $n$.

  Suppose that $\phi$ is satisfiable, and consider a satisfying
  assignment of the variables. For each variable $x$, set the output
  priority $p^+(x?)$ to the arc toward the right if $x$ is assigned to
  \emph{true}, and to the arc toward the left if $x$ is assigned to
  \emph{false}. As a consequence, in the steady-state, by
  \Cref{prop:variable-gadget}, the throughput on the arcs labeled $x$
  is $\frac{1}{2}$ if $x$ is \emph{true}, and $\frac{3}{8}$ otherwise,
  and symmetrically for the arcs labeled $\overline{x}$. Then, because
  each clause is satisfied, at least one arc entering each clause has
  throughput $\frac{1}{2}$, and the two others have throughput at most
  $\frac{7}{8}$. Hence each clause receives at most $\frac{9}{4}$, and
  by \Cref{prop:clause-gadget}, the throughput of its input $s$ is
  $\frac{23}{32}$. Thus the total throughput is
  $\frac{11}{4} n + \frac{23}{32}|\phi|$.

  Conversely, we claim that if their is a choice of priority with a
  steady-state whose global throughput is at least
  $\frac{11}{4}n + \frac{23}{32}|\phi|$, then $\phi$ is satisfiable.
  By \Cref{prop:variable-gadget}, a variable gadget can produce at
  most $\frac{11}{4}$ unit of flow from its input, and a clause gadget
  at most $\frac{23}{32}$ by \Cref{prop:clause-gadget}. Therefore such
  a throughput can only happen if every variable gadget produces
  $\frac{11}{4}$, which means that the priority $p^+(x?)$ is not set
  to $\bot$. It is also necessary that each clause receives at most
  $\frac{9}{4}$ unit of flows from the variable gadgets. The
  priorities of the splitters $x?$ induce an assignment of truth
  values to the variable (\emph{true} if the priority is to the arc to
  the right, and \emph{false} to the left). The $\frac{9}{4}$-bound
  imposes that for each clause, there is a arc from a variable gadget
  entering that clause's gadget with throughput $\frac{1}{2}$. By
  construction, that arc corresponds to a literal that
  is satisfied by the assignment. Therefore $\phi$ is satisfiable. The
  splitter network admits a choice of priorities with a steady-state
  whose global throughput is at least
  $\frac{11}{4}n + \frac{23}{32}|\phi|$ if and only if $\phi$ is
  satisfiable, therefore the throughput maximization problem is
  NP-hard.
\end{proof}

As the gadgets are planarly embedded, reducing from planar-3SAT is
also possible, and thus the maximization throughput problem stays
NP-hard with the constraint that the splitter network is planar.

\section{A balancer with saturated arcs}\label{sec:saturating-balancer}

In \Cref{thm:lower-bound}, we proved that a $(2^k,2^k)$-balancer must
have at least $k 2^{k-2}$ fair splitters. We also define the simple
balancer of order $k$, a $(2^k,2^k)$-balancer with
$S(k) = (k-1) 2^{k-1}$ splitters. In this section we exhibit a
balancer with priorities whose number of fair splitters is
only $(k+1)2^{k-2}$. Hence this balancer is much closer to the lower bound in
regards to the number of fair splitters. Unfortunately, it also
contains many splitters with non-fair output priorities, making it
asymptotically larger than the simple balancer. The construction is
nonetheless interesting, not only for its smaller number of fair
splitters, but also in demonstrating that saturation may also be used
in an effective way to balance the outputs.

\begin{figure}
  \begin{center}
    \begin{tikzpicture}[x=0.5cm,y=0.5cm,>=latex]
      \foreach \i in {0,1,...,15} {
        \node[term] (i\i) at (\i,0) {};
      }
      \foreach \x/\y/\name in {1/1/x1y1,2/1/x2y1,2/2/x2y2,3/1/x3y1,3/2/x3y2,3/3/x3y3,4/1/x4y1,4/2/x4y2,4/3/x4y3,4/4/x4y4,5/1/x5y1,5/2/x5y2,5/3/x5y3,5/4/x5y4,5/5/x5y5,6/1/x6y1,6/2/x6y2,6/3/x6y3,6/4/x6y4,6/5/x6y5,6/6/x6y6,7/1/x7y1,7/2/x7y2,7/3/x7y3,7/4/x7y4,7/5/x7y5,7/6/x7y6,7/7/x7y7,8/1/x8y1,8/2/x8y2,8/3/x8y3,8/4/x8y4,8/5/x8y5,8/6/x8y6,8/7/x8y7,8/8/x8y8,9/1/x9y1,9/2/x9y2,9/3/x9y3,9/4/x9y4,9/5/x9y5,9/6/x9y6,9/7/x9y7,9/8/x9y8,9/9/x9y9,10/1/x10y1,10/2/x10y2,10/3/x10y3,10/4/x10y4,10/5/x10y5,10/6/x10y6,10/7/x10y7,10/8/x10y8,10/9/x10y9,10/10/x10y10,11/1/x11y1,11/2/x11y2,11/3/x11y3,11/4/x11y4,11/5/x11y5,11/6/x11y6,11/7/x11y7,11/8/x11y8,11/9/x11y9,11/10/x11y10,11/11/x11y11,12/1/x12y1,12/2/x12y2,12/3/x12y3,12/4/x12y4,12/5/x12y5,12/6/x12y6,12/7/x12y7,12/8/x12y8,12/9/x12y9,12/10/x12y10,12/11/x12y11,12/12/x12y12,13/1/x13y1,13/2/x13y2,13/3/x13y3,13/4/x13y4,13/5/x13y5,13/6/x13y6,13/7/x13y7,13/8/x13y8,13/9/x13y9,13/10/x13y10,13/11/x13y11,13/12/x13y12,13/13/x13y13,14/1/x14y1,14/2/x14y2,14/3/x14y3,14/4/x14y4,14/5/x14y5,14/6/x14y6,14/7/x14y7,14/8/x14y8,14/9/x14y9,14/10/x14y10,14/11/x14y11,14/12/x14y12,14/13/x14y13,14/14/x14y14,15/1/x15y1,15/2/x15y2,15/3/x15y3,15/4/x15y4,15/5/x15y5,15/6/x15y6,15/7/x15y7,15/8/x15y8,15/9/x15y9,15/10/x15y10,15/11/x15y11,15/12/x15y12,15/13/x15y13,15/14/x15y14,15/15/x15y15}
      {
          \node[chooser] (\name) at (\x,\y) {};
          \draw[thin,-{Straight Barb[length=1pt]}] ($(\name.center) + (-0pt,-2pt)$) -- ($(\name.center) + (0pt,2pt)$);
      }
      \foreach \y in {10,12,14,16} {
        \node[chooser] (a\y) at (16,\y) {};
        \draw[thin,-{Straight Barb[length=1pt]}] 
          ($(a\y.center) + (180:2pt)$) .. controls (a\y.center) .. ($(a\y.center) + (315:2pt)$);
      }
      \foreach \y in {9,11,13,15} {
        \node[chooser] (a\y) at (16,\y) {};
        \draw[thin,-{Straight Barb[length=1pt]}] 
          ($(a\y.center) + (180:2pt)$) .. controls (a\y.center) .. ($(a\y.center) + (329:2pt)$);
      }
      \foreach \y in {1,3,5,7} {
        \node[chooser] (b\y) at (26,\y) {};
        \draw[thin,-{Straight Barb[length=1pt]}] 
          ($(b\y.center) + (180:2pt)$) .. controls (b\y.center) .. ($(b\y.center) + (0:2pt)$);

      }
      \foreach \y in {2,4,6,8} {
        \node[chooser] (b\y) at (26,\y) {};
        \draw[thin,-{Straight Barb[length=1pt]}] 
          ($(b\y.center) + (180:2pt)$) .. controls (b\y.center) .. ($(b\y.center) + (0:2pt)$);

      }
      \foreach \y in {1,3,...,15} {
        \node[splitter] (c\y) at ($(30,\y) + (0,0.5)$) {};
      }

      \foreach \u/\v in {x2y1/x2y2,x3y1/x3y2,x3y2/x3y3,x4y1/x4y2,x4y2/x4y3,x4y3/x4y4,x5y1/x5y2,x5y2/x5y3,x5y3/x5y4,x5y4/x5y5,x6y1/x6y2,x6y2/x6y3,x6y3/x6y4,x6y4/x6y5,x6y5/x6y6,x7y1/x7y2,x7y2/x7y3,x7y3/x7y4,x7y4/x7y5,x7y5/x7y6,x7y6/x7y7,x8y1/x8y2,x8y2/x8y3,x8y3/x8y4,x8y4/x8y5,x8y5/x8y6,x8y6/x8y7,x8y7/x8y8,x9y1/x9y2,x9y2/x9y3,x9y3/x9y4,x9y4/x9y5,x9y5/x9y6,x9y6/x9y7,x9y7/x9y8,x9y8/x9y9,x10y1/x10y2,x10y2/x10y3,x10y3/x10y4,x10y4/x10y5,x10y5/x10y6,x10y6/x10y7,x10y7/x10y8,x10y8/x10y9,x10y9/x10y10,x11y1/x11y2,x11y2/x11y3,x11y3/x11y4,x11y4/x11y5,x11y5/x11y6,x11y6/x11y7,x11y7/x11y8,x11y8/x11y9,x11y9/x11y10,x11y10/x11y11,x12y1/x12y2,x12y2/x12y3,x12y3/x12y4,x12y4/x12y5,x12y5/x12y6,x12y6/x12y7,x12y7/x12y8,x12y8/x12y9,x12y9/x12y10,x12y10/x12y11,x12y11/x12y12,x13y1/x13y2,x13y2/x13y3,x13y3/x13y4,x13y4/x13y5,x13y5/x13y6,x13y6/x13y7,x13y7/x13y8,x13y8/x13y9,x13y9/x13y10,x13y10/x13y11,x13y11/x13y12,x13y12/x13y13,x14y1/x14y2,x14y2/x14y3,x14y3/x14y4,x14y4/x14y5,x14y5/x14y6,x14y6/x14y7,x14y7/x14y8,x14y8/x14y9,x14y9/x14y10,x14y10/x14y11,x14y11/x14y12,x14y12/x14y13,x14y13/x14y14,x15y1/x15y2,x15y2/x15y3,x15y3/x15y4,x15y4/x15y5,x15y5/x15y6,x15y6/x15y7,x15y7/x15y8,x15y8/x15y9,x15y9/x15y10,x15y10/x15y11,x15y11/x15y12,x15y12/x15y13,x15y13/x15y14,x15y14/x15y15} 
      {
        \draw[->] (\u) -- (\v);
      }
      \foreach \u/\v in {x1y1/x2y1,x2y1/x3y1,x2y2/x3y2,x3y1/x4y1,x3y2/x4y2,x3y3/x4y3,x4y1/x5y1,x4y2/x5y2,x4y3/x5y3,x4y4/x5y4,x5y1/x6y1,x5y2/x6y2,x5y3/x6y3,x5y4/x6y4,x5y5/x6y5,x6y1/x7y1,x6y2/x7y2,x6y3/x7y3,x6y4/x7y4,x6y5/x7y5,x6y6/x7y6,x7y1/x8y1,x7y2/x8y2,x7y3/x8y3,x7y4/x8y4,x7y5/x8y5,x7y6/x8y6,x7y7/x8y7,x8y1/x9y1,x8y2/x9y2,x8y3/x9y3,x8y4/x9y4,x8y5/x9y5,x8y6/x9y6,x8y7/x9y7,x8y8/x9y8,x9y1/x10y1,x9y2/x10y2,x9y3/x10y3,x9y4/x10y4,x9y5/x10y5,x9y6/x10y6,x9y7/x10y7,x9y8/x10y8,x9y9/x10y9,x10y1/x11y1,x10y2/x11y2,x10y3/x11y3,x10y4/x11y4,x10y5/x11y5,x10y6/x11y6,x10y7/x11y7,x10y8/x11y8,x10y9/x11y9,x10y10/x11y10,x11y1/x12y1,x11y2/x12y2,x11y3/x12y3,x11y4/x12y4,x11y5/x12y5,x11y6/x12y6,x11y7/x12y7,x11y8/x12y8,x11y9/x12y9,x11y10/x12y10,x11y11/x12y11,x12y1/x13y1,x12y2/x13y2,x12y3/x13y3,x12y4/x13y4,x12y5/x13y5,x12y6/x13y6,x12y7/x13y7,x12y8/x13y8,x12y9/x13y9,x12y10/x13y10,x12y11/x13y11,x12y12/x13y12,x13y1/x14y1,x13y2/x14y2,x13y3/x14y3,x13y4/x14y4,x13y5/x14y5,x13y6/x14y6,x13y7/x14y7,x13y8/x14y8,x13y9/x14y9,x13y10/x14y10,x13y11/x14y11,x13y12/x14y12,x13y13/x14y13,x14y1/x15y1,x14y2/x15y2,x14y3/x15y3,x14y4/x15y4,x14y5/x15y5,x14y6/x15y6,x14y7/x15y7,x14y8/x15y8,x14y9/x15y9,x14y10/x15y10,x14y11/x15y11,x14y12/x15y12,x14y13/x15y13,x14y14/x15y14} {
        \draw[->] (\u) -- (\v);
      }
      
      \foreach \u/\v in {i1/x1y1,i2/x2y1,i3/x3y1,i4/x4y1,i5/x5y1,i6/x6y1,i7/x7y1,i8/x8y1,i9/x9y1,i10/x10y1,i11/x11y1,i12/x12y1,i13/x13y1,i14/x14y1,i15/x15y1}
      {
        \draw[->] (\u) -- (\v);
      }
      
      \foreach \u/\v in {x15y1/b1,x15y2/b2,x15y3/b3,x15y4/b4,x15y5/b5,x15y6/b6,x15y7/b7,x15y8/b8,x15y9/a9,x15y10/a10,x15y11/a11,x15y12/a12,x15y13/a13,x15y14/a14,x15y15/a15} 
      {
        \draw[->] (\u) -- (\v); 
      }
      \foreach \u/\v in {x1y1/x2y2,x2y2/x3y3,x3y3/x4y4,x4y4/x5y5,x5y5/x6y6,x6y6/x7y7,x7y7/x8y8,x8y8/x9y9,x9y9/x10y10,x10y10/x11y11,x11y11/x12y12,x12y12/x13y13,x13y13/x14y14,x14y14/x15y15,x15y15/a16} 
      {
        \draw[->] (\u) to[out=90,in=180] (\v);
      }
      \draw[->] (i0) to[out=90,in=180] (x1y1);
      \foreach \y/\v in {9/c1,10/c3,11/c5,12/c7,13/c9,14/c11,15/c13,16/c15} {
        \draw[->] (a\y) -- (26,\y) .. controls (27,\y) .. (\v);
      }
      \foreach \y/\v in {1/c1,2/c3,3/c5,4/c7,5/c9,6/c11,7/c13,8/c15} {
        \draw[->,red] (b\y) .. controls (27,\y) .. (\v);
      }
      \foreach \y in {1,2,...,16} {
        \node[term] (o\y) at (33,\y) {};
      }
      \foreach \x/\y in {1/1,1/2,3/3,3/4,5/5,5/6,7/7,7/8,9/9,9/10,11/11,11/12,13/13,13/14,15/15,15/16} 
      {
        \draw[->] (c\x) -- (o\y);
      }
      \begin{scope}[fill=blue,draw=blue]
        \foreach \n/\x/\y in {
          u1/18.5/13.5,u2/18.5/11.5,u3/18.5/9.5,u4/18.5/7.5,
          v1/21/11.5,v2/21/9.5,v3/21/7.5,v4/21/5.5,
          w1/23.5/9.5,w2/23.5/7.5,w3/23.5/5.5,w4/23.5/3.5} 
        {
          \node[splitter,blue] (\n) at (\x,\y) {};
        }
        \foreach \u/\v in {
          a16/u1,a15/u1,a14/u2,a13/u2,a12/u3,a11/u3,a10/u4,a9/u4}
        {
          \draw[->] (\u) -- (\v);
        }
        \foreach \u/\v in {
          u1/v1,u2/v2,u3/v3,u4/v4,u1/v2,u2/v1,u3/v4,u4/v3,
          v1/w1,v2/w2,v3/w3,v4/w4,v1/w3,v2/w4,v3/w1,v4/w2,
          w1/b8,w1/b7,w2/b6,w2/b5,w3/b4,w3/b3,w4/b2,w4/b1}
        {
          \draw[->] (\u) -- (\v);
        }
      \end{scope}

    \end{tikzpicture}
  \end{center}
  \caption{A $(2^k,2^k)$-balancer utilizing priorities, which contains
    significantly less fair splitters than the simple balancer. In
    this example, we set $k = 4$. Fair splitters are represented by
    full nodes. In each priority splitter, a small arrow indicates
    which arcs are prioritized: the outgoing arc pointed to by the
    arrow's head, and the incoming arc aligned with the origin of the
    arrow. On the right, the half-grid pushes up to $2^{k-1}$ units of
    flow to the top rows. In blue, a simple balancer of order
    $2^{k-1}$ proceeds to most of the balancing. The red arcs form a
    bottleneck, when the total throughput exceeds $2^{k-1}$. On the
    right, a last column of fair splitters distributes the flow to the
    outputs.}
  \label{fig:saturating-balancer}
\end{figure}
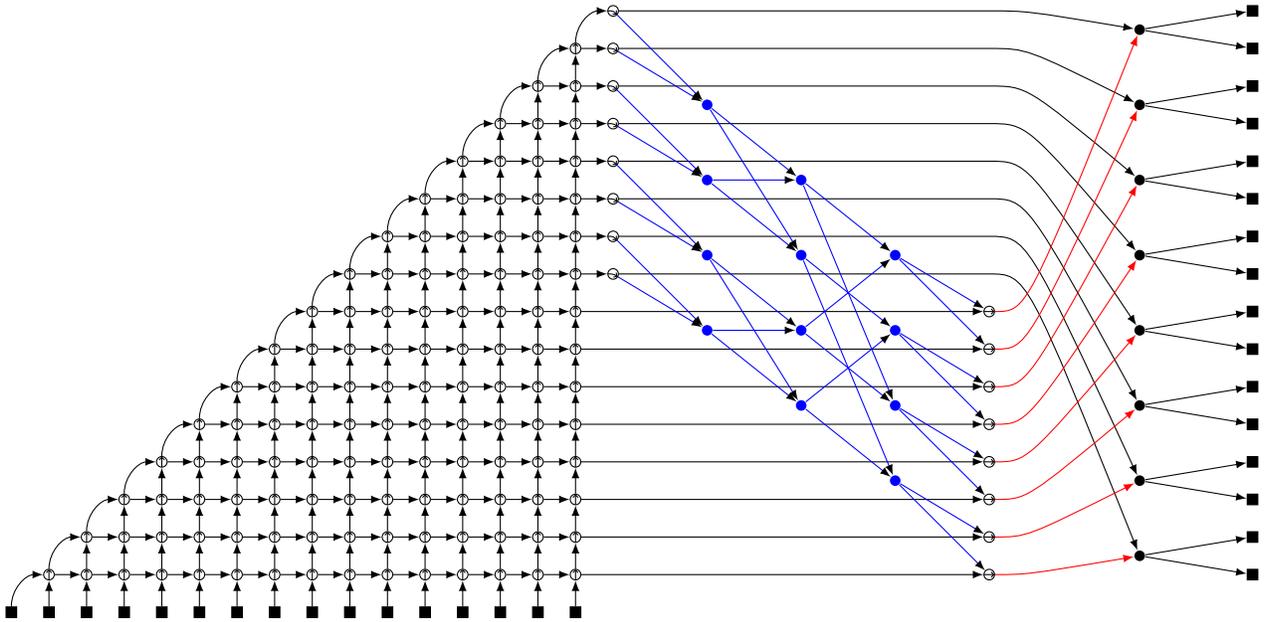

The saturating $(2^k,2^k)$-balancer is composed of three parts,
illustrated in \Cref{fig:saturating-balancer}. The first part contains
the input and a half-grid of priority splitters, whose output priority
are set to the same direction (vertical in the picture). This design
ensures that the flow is pushed as much as possible to the top of the
grid. Therefore the throughputs on the arcs leaving the half-grid,
from top to bottom, are $1,1,\ldots,1, t, 0, \ldots,0$, where
$t = c(I) - \lfloor c(I) \rfloor$. In particular, if
$c(I) \leq 2^{k-1}$, all the flow leaves the half grid from the top
$2^{k-1}$ arcs, else the $2^{k-1}$ receive $2^{k-1}$ units of flow.

The next part is made of a simple balancer of order $2^{k-1}$ from the
$2^{k-1}$ highest rows to the $2^{k-1}$ lowest rows. The last part
is a column of $2^{k-1}$ fair splitters, whose out-neighbors consist
in the network's outputs. From the outputs of the simple balancer to
those fair splitters, $2^{k-1}$ arcs serves as the main bottleneck of
this network.

The saturating balancer works under two possible regimes. When
$c(I) \leq 2^{k-1}$, all the flow is pushed in the simple balancer,
arrives balanced into the bottleneck arcs, and the is split evenly by
the last column of splitters toward the output. Then every arc is
fluid. The other regime happens when $c(I) > 2^{k-1}$. At
$c(I) = 2^{k-1}$, all the arcs in the simple balancer and in the
bottleneck reach throughput $1$. Increasing the input capacity above
$2^{k-1}$, the arcs of the simple balancer saturates. The excess flow
is pushed into the $2^{k-1}$ lowest rows, directly toward the
bottleneck arcs. Because the bottleneck is already full, the incoming
flow is pull backed by the simple balancer. At this point, the simple
balancer, which is fully saturated, acts a balancer on the pulled back
flow: the pulled back flow is sent back evenly to its input, from
which it proceeds on the horizontal arcs of the $2^{k-1}$ highest
rows, to the last column of splitters. Thus, the potential of each
fair splitter in the simple balancer is fully used: as long as
$c(I) < 2^{k-1}$ it splits the throughput fairly into its two outgoing
fluid arcs. When $c(I) > 2^{k-1}$, it splits the pulled-back
throughput evenly into its two incoming saturated arcs.

The saturating $(2^k,2^k)$-splitter contains
$2^{k-1} + S(2^{k-1}) = (k+1)2^{k-2}$ fair splitters, which is closer to
our lower bound of $k 2^{k-2}$ splitters from \Cref{thm:lower-bound}.
The difference is explained by the last column of fair splitters,
which are only used with fluid outgoing arcs. The total number of
splitters is $\Theta(2^{2k})$, dominated by the half-grid. However,
this part can be replaced by a network with the following property:
the total throughput on the $2^{k-1}$ highest leaving arcs must be
$\min\{c(I), 2^{k-1}\}$. Such a network can be defined with
$o(2^{2k})$ priority splitters, but we do not know whether one exists
with only $O(k 2^k)$ priority splitters.

\section{Perspectives}\label{sec:perspectives}

We formalized splitter networks and their steady-states, and presented
various load-balancing designs. The ability to design universal
balancers enables the simulation of networks with integral capacities:
each arc is replicated according to its capacity, and each splitter is
replaced by a universal balancer. A universal balancer is fair by the
balancing property, and maximizing by the unlimited-throughput
property, effectively generalizing splitters. Our definition of
splitter network can also be extended to support arc capacities
natively, with most of the proofs requiring only minor modifications.

Although our continuous model is convenient for modeling the expected
throughput of splitter networks, Factorio's belt systems operates
discretely. Therefore, the observed throughputs in Factorio's splitter
networks are only approximations of those theorized by our model.
Further investigation into the disparities between the discrete and
continuous splitter networks is necessary to accurately apply our
findings to Factorio.

Our lower bounds for the number of splitters in balancers have a
constant multiplicative gap across all designs, indicating they are
not tight. For simple balancers of order $k$, this gap is closed when
we forbid saturated arcs in the steady-state of the balancer.
Consequently, leveraging saturation is necessary to further reduce the
number of splitters in load-balancing networks. This is partially done
in \Cref{sec:saturating-balancer}, at the cost of introducing many
priority splitters. Furthermore, it is worth investigating stronger
lower bounds in the context of universal balancers.

Factorio allows splitters to be configured to prioritize either an
outgoing arc, or an incoming arc. Utilizing this feature, the
universal network described in~\cite{pocarski} achieves a
significantly smaller size compared to our design. Our technique still
establishes a lower bound on the number of fair splitters. In general,
what is the minimum size achievable for networks utilizing these more
general splitters? We proved several complexity results for the
problem of global throughput maximization, when we can choose which
arcs to prioritize in each splitter or a subset of those splitters.
However, the complexity of global throughput maximization when the
input and output capacities are arbitrary is not fully settled yet.

As a last series of questions, consider a network whose steady-state,
when all inputs and outputs have capacity 1, has no saturated arcs. If
the augmenting flow from any single input is uniformly distributed
across the outputs, then the network is a balancer. This provides a
polynomial-time procedure for deciding whether a network is a
balancer, subject to the absence of saturation. Is it feasible to
devise a general procedure to decide whether a splitter network is
balancing, throughput unlimited or universal?

\printbibliography

@misc{beltturingcomplete,
  author       = "{MatthaeusHarris}",
  title        = "{Factorio belts are Turing-complete}",
  howpublished = "\url{https://www.reddit.com/r/factorio/comments/lc25cx/factorio_belts_are_turing_complete/}"
}

@article{benevs1962rearrangeable,
  title={On rearrangeable three-stage connecting networks},
  author={Bene{\v{s}}, V{\'a}clav E},
  journal={The Bell System Technical Journal},
  volume={41},
  number={5},
  pages={1481--1492},
  year={1962},
  publisher={Nokia Bell Labs}
}

@article{benevs1964permutation,
  title={Permutation groups, complexes, and rearrangeable connecting networks},
  author={Bene{\v{s}}, V{\'a}clav E},
  journal={Bell System Technical Journal},
  volume={43},
  number={4},
  pages={1619--1640},
  year={1964},
  publisher={Wiley Online Library}
}

@inproceedings{bernstein2019decremental,
  title={Decremental strongly-connected components and single-source reachability in near-linear time},
  author={Bernstein, Aaron and Probst, Maximilian and Wulff-Nilsen, Christian},
  booktitle={Proceedings of the 51st Annual ACM SIGACT Symposium on Theory of Computing},
  pages={365--376},
  year={2019}
}

@misc{bgg,
  author="{BoardGameGeek}",
  title="{Boardgame category: transportation}",
  howpublished = "\url{https://boardgamegeek.com/boardgamecategory/1011/transportation}"
}

@inproceedings{boardman2021simulation,
  title={Simulation of Production and Inventory Control using the Computer Game Factorio},
  author={Boardman, Bonnie S. and Krejci, Caroline C.},
  booktitle={ASEE 2021 Gulf-Southwest Annual Conference},
  year={2021}
}

@inproceedings{covello2023using,
  title={Using graph theory to investigate the role of expertise on infrastructure evolution: A case study examining the game Factorio},
  author={Covello, Chase and Jung, Hyunjang and Watson, Bryan C.},
  booktitle={Conference on Systems Engineering Research},
  pages={297--311},
  year={2023},
  organization={Springer}
}

@article{COVIELLOGONZALEZ202098,
title = {Towards a theory of mixing graphs: A characterization of perfect mixability},
journal = {Theoretical Computer Science},
volume = {845},
pages = {98-121},
year = {2020},
issn = {0304-3975},
author = {Coviello Gonzalez, Miguel and Chrobak, Marek}
}

@article{dinits1973method,
  title={The method of scaling and transportation problems},
  author={Dinits, Yefim A.},
  journal={Studies in Discrete Mathematics, Moscow},
  pages={46--57},
  year={1973}
}

@article{duhan2019factory,
  title={Factory optimization using deep reinforcement learning AI},
  author={Duhan, Shivam and Zhang, Chengming and Jing, Wenyu and Li, Mingqi},
  year={2019},
  journal={Purdue Undergraduate Research Conference},
  volume={57}
}

@misc{factorio,
  author = "{Wube Software}",
  title = "{Factorio}",
  howpublished = "\url{https://www.factorio.com/}"
}

@misc{factoriosat,
  author       = "{R-O-C-K-E-T}",
  title        = "{factorio-SAT: Enhancing the Factorio experience with SAT solvers.}",
  howpublished = "\url{https://github.com/R-O-C-K-E-T/Factorio-SAT}"
}

@article{goldberg1988new,
  title={A new approach to the maximum-flow problem},
  author={Goldberg, Andrew V. and Tarjan, Robert E.},
  journal={Journal of the ACM (JACM)},
  volume={35},
  number={4},
  pages={921--940},
  year={1988},
  publisher={ACM New York, NY, USA}
}

@Inbook{greedoids,
author="Korte, Bernhard
and Schrader, Rainer
and Lov{\'a}sz, L{\'a}szl{\'o}",
title="Abstract Convexity --- Antimatroids",
bookTitle="Greedoids",
year="1991",
publisher="Springer Berlin Heidelberg",
address="Berlin, Heidelberg",
pages="19--43",
isbn="978-3-642-58191-5",
doi="10.1007/978-3-642-58191-5_3",
url="https://doi.org/10.1007/978-3-642-58191-5_3"
}

@misc{ketcheson,
  author="{Ketcheson, David}",
  title = "{Mathematics Stackexchange: Belt Balancer problem (Factorio)}",
  howpublished = "\url{https://math.stackexchange.com/questions/1775378/belt-balancer-problem-factorio}"
}

@inproceedings{knuth1976complexity,
  title={The complexity of nonuniform random number generation},
  booktitle={Algorithms and Complexity: New Directions and Recent Results},
  editor={JF Traub},
  author={Knuth, Donald E. and Yao, Andrew C.},
  year={1976},
  pages={357--428},
  publisher={Addison-Wesley}
}

@misc{verifactory,
  author       = "{Legnagi, Alessandro and Montini, Axel}",
  title        = "{VeriFactory}",
  howpublished = "\url{https://github.com/alegnagi/verifactory/}"
}

@phdthesis{leue2021verification,
  title={Verification of Factorio Belt Balancers using Petri Nets},
  author={Leue, Andre},
  year={2021},
  school={Bachelorarbeit, Darmstadt, Technische Universit{\"a}t Darmstadt}
}

@book{knuth1998art,
  title={The Art of Computer Programming: Sorting and Searching, Volume 3},
  author={Knuth, Donald E},
  isbn={9780321635785},
  year={1998},
  publisher={Pearson Education}
}

@article{meyers2009,
title = {Integer equal flows},
journal = {Operations Research Letters},
volume = {37},
number = {4},
pages = {245-249},
year = {2009},
issn = {0167-6377},
author = {Carol A. Meyers and Andreas S. Schulz}
}

@phdthesis{parmar2007integer,
  title={Integer programming approaches for equal-split network flow problems},
  author={Parmar, Amandeep},
  year={2007},
  school={Georgia Institute of Technology}
}

@article{patterson2023towards,
  title={Towards Automatic Design of Factorio Blueprints},
  author={Patterson, Sean and Espasa, Joan and Chang, Mun See and Hoffmann, Ruth},
  journal={arXiv preprint arXiv:2310.01505},
  year={2023}
}

@misc{pocarski,
  author       = "{pocarski}",
  title        = "{Universal 8-8: Perfectly Balanced, as All Things Should Be}",
  howpublished = "\url{https://web.archive.org/web/20230922022806/https://alt-f4.blog/ALTF4-27/}"
}

@inproceedings{reid2021factory,
  title={The factory must grow: automation in Factorio},
  author={Reid, Kenneth N. and Miralavy, Iliya and Kelly, Stephen and Banzhaf, Wolfgang and Gondro, Cedric},
  booktitle={Proceedings of the Genetic and Evolutionary Computation Conference Companion},
  pages={243--244},
  year={2021}
}

@inproceedings{srinathan2002theory,
  title={Theory of equal-flows in networks},
  author={Srinathan, K. and Goundan, Pranava R. and Kumar, MVN Ashwin and Nandakumar, R. and Rangan, C. Pandu},
  booktitle={Computing and Combinatorics: 8th Annual International Conference, COCOON 2002 Singapore, August 15--17, 2002 Proceedings 8},
  pages={514--524},
  year={2002},
  organization={Springer}
}

\end{document}